\documentclass[longauth]{aa}

\usepackage{graphicx}
\usepackage{txfonts}
\usepackage{lipsum}
\usepackage{subcaption}         
\usepackage{lscape}             
\usepackage{placeins}           

\usepackage{multirow}
\usepackage{relsize}
\usepackage{tikz}

\usepackage{hyperref}

\begin{document}

   \title{NOEMA\textsuperscript{3D}: A deep view of cold gas flows in a barred spiral galaxy at $z\sim1$}
   \titlerunning{ }


   \author{Stavros~Pastras\inst{1,2}
      \and Reinhard~Genzel\inst{1,3}
      \and Linda~J.~Tacconi\inst{1}
      \and Thorsten~Naab\inst{2}
      \and Natascha~M.~Förster~Schreiber\inst{1}
      \and Karl~Schuster\inst{4}
      \and Roberto~Neri\inst{4}
      \and Jianhang~Chen\inst{1}
      \and Giulia~Tozzi\inst{1}
      \and Jean-Baptiste~Jolly\inst{1}
      \and Letizia~Scaloni\inst{5,6}
      \and Capucine~Barfety\inst{1}
      \and Andreas~Burkert\inst{1,7}
      \and Yixian~Cao\inst{1}
      \and Françoise~Combes\inst{8,9}
      \and Ric~Davies\inst{1}
      \and Frank~Eisenhauer\inst{1,10}
      \and Juan~M.~Espejo~Salcedo\inst{1}
      \and Simon~Flesch\inst{1}
      \and Santiago~García-Burillo\inst{11}
      \and Rodrigo~Herrera-Camus\inst{12,13}
      \and Lilian~L.~Lee\inst{1}
      \and Minju~M.~Lee\inst{14,15}
      \and Daizhong~Liu\inst{16}
      \and Dieter~Lutz\inst{1}
      \and Giovanni~Mazzolari\inst{1}
      \and Amit~Nestor~Shachar\inst{17}
      \and Meghana~Pannikkote\inst{1}
      \and Eleonora~Parlanti\inst{18}
      \and Panos~A.~Patsis\inst{19}
      \and Sedona~H.~Price\inst{20}
      \and Claudia~Pulsoni\inst{1}
      \and Alvio~Renzini\inst{21}
      \and Taro~T.~Shimizu\inst{1}
      \and Amiel~Sternberg\inst{22,17,1}
      \and Eckhard~Sturm\inst{1}
      \and Stijn~Wuyts\inst{23}
      \and Hannah~Übler\inst{1}
          }
   
   \authorrunning{S. Pastras et al.}

   \institute{Max-Planck-Institut für Extraterrestrische Physik (MPE), Gießenbachstr. 1, D-85748 Garching, Germany 
         \and Max-Planck-Institut für Astrophysik (MPA), Karl-Schwarzschild-Str. 1, D-85748 Garching, Germany 
         \and Departments of Physics and Astronomy, University of California, Berkeley, CA 94720, USA 
         \and Institut de Radioastronomie Millimétrique (IRAM), 300 Rue de la Piscine, 38400 Saint-Martin-d’Hères, France 
         \and Department of Physics and Astronomy “Augusto Righi”, University of Bologna, Via Piero Gobetti 93/2, 40129 Bologna, Italy 
         \and INAF – Astrophysics and Space Science Observatory of Bologna, Via Piero Gobetti 93/3, 40129 Bologna, Italy 
         \and Universitäts-Sternwarte Ludwig-Maximilians-Universität (USM), Scheinerstr. 1, München, D-81679, Germany 
         \and Observatoire de Paris, LUX, CNRS, PSL Univ., Sorbonne University, Paris, France 
         \and College de France, 11 Pl. Marcelin Berthelot, 75231 Paris, France 
         \and Technical University of Munich, TUM School of Natural Sciences, Physics Department, 85747 Garching, Germany 
         \and Observatorio Astronómico Nacional (OAN-IGN)-Observatorio de Madrid, Alfonso XII, 3, 28014 Madrid, Spain 
         \and Departamento de Astronom\'{\i}a, Universidad de Concepción, Barrio Universitario, Concepción, Chile 
         \and Millenium Nucleus for Galaxies (MINGAL), Concepción, Chile 
         \and Cosmic Dawn Center (DAWN), Copenhagen, Denmark 
         \and DTU-Space, Technical University of Denmark, Elektrovej 327, DK2800 Kgs. Lyngby, Denmark 
         \and Purple Mountain Observatory, Chinese Academy of Sciences, 10 Yuanhua Road, Nanjing 210023, China 
         \and School of Physics and Astronomy, Tel Aviv University, Tel Aviv 69978, Israel 
         \and Scuola Normale Superiore, Piazza dei Cavalieri 7, I-56126 Pisa, Italy 
         \and Research Center for Astronomy and Applied Mathematics, Academy of Athens, Soranou Efessiou 4, 11527 Athens, Greece 
         \and Space Telescope Science Institute, 3700 San Martin Drive, Baltimore, MD 21218, USA 
         \and Osservatorio Astronomico di Padova, Vicolo dell’Osservatorio 5, Padova, I-35122, Italy 
         \and Centre for Computational Astrophysics, Flatiron Institute, 162 5th Avenue, New York, NY 10010, USA 
         \and Department of Physics, University of Bath, Claverton Down, Bath, BA2 7AY, UK 
             }

   \date{Received ...}

 

\abstract{We present a deep, high-resolution CO(4-3) IRAM-NOEMA observation of a main sequence, barred, spiral galaxy at $z\approx1.12$, with an on-source integration time of $\approx37$\,hours and a beam FWHM of $\approx$\,0\farcs3. We use the molecular gas data in conjunction with the available deep multi-band \textit{JWST} and \textit{HST} imaging, covering restframe UV to near-IR wavelengths, to quantitatively study the gas flows in the disk plane of this cosmic noon barred spiral. We find that this target is a massive ($\log (M_{\mathrm{baryons}}/M_\odot)\approx10.96$), baryon-dominated ($f_{\mathrm{dm}}(<R_e)=u^2_{\mathrm{circ,dm}}(R_e)/u^2_{\mathrm{circ}}(R_e)\sim4\%$), gas-rich ($f_{\mathrm{gas}}=M_{\mathrm{gas}}/(M_{\mathrm{\star}}+M_{\mathrm{gas}})\approx40\%$) disk, hosting a long ($a_{\mathrm{bar}}\approx4.2$\,kpc), strong ($Q_{\mathrm{b}}\approx0.37$), and fast ($\mathcal{R}=R_{\mathrm{CR}}/a_{\mathrm{bar}}\approx1.05$) bar, which rotates at an angular speed of $\Omega_{\mathrm{pattern}}\approx$\,50\,km/s/kpc. This bar is driving molecular gas inflows with a net inflow rate of $\dot{M}\sim30$\,M$_\odot$/yr, based on three estimates, which is of the same order as the galaxy-integrated star formation rate ($\mathrm{SFR}\approx36$\,M$_\odot$/yr). We additionally identify evidence of a well-defined dust lane shock at the northwestern side of the bar, with gas motions parallel to this feature, in agreement with expectations for an established bar-driven flow. Our study highlights the possible role of bars as key drivers of galaxy evolution for a significant fraction of cosmic noon galaxies, offering a detailed picture of well-defined, bar-driven inflows in a high-$z$ barred spiral.}


   \maketitle

\nolinenumbers

\section{Introduction}
\label{sec:introduction}

In the local Universe approximately two-thirds of spiral galaxies host galactic bars \citep{Eskridge_2000, Menendez-Delmestre_2007, Simmons_2014, Erwin_2018}. These non-axisymmetric structures shape the evolution of their host disks by driving large scale gas inflows toward the central regions \citep[e.g.,][]{Roberts_1979, Athanassoula_1992b, Kim_2012b, Sormani_2019a, Chown_2019, Pastras_2022, Yu_2022, Sormani_2023} and redistributing angular momentum from the inner to the outer parts of the disk and the dark matter halo \citep[e.g.,][]{Lynden-Bell_1972, Sellwood_1981, Debattista_1998, Debattista_2000, Athanassoula_2002b, Athanassoula_2003, Martinez-Valpuesta_2006}.\par

The role of bars as drivers of galaxy evolution has been extensively studied from both a theoretical and observational point of view. They have been shown to promote the formation of nuclear rings and spirals \citep[e.g.,][]{Piner_1995, Buta_1996, Regan_2003, Maciejewski_2004a, Maciejewski_2004b, Kim_2012a, Kim_2012b, Kolcu_2023, Sormani_2024, Pastras_2026a}, the building of central concentrations, i.e., bulges and nuclear disks, \citep[e.g.,][]{Gadotti_2015, Seo_2019, Gadotti_2020, Bittner_2020, Schinnerer_2023, Verwilghen_2024, Fraser-McKelvie_2025}, and possibly fueling active galactic nuclei (AGN) \citep[e.g.,][]{Shlosman_1990, Garcia-Burillo_2005, Silva-Lima_2022}, as well as enhancing or suppressing star formation in different regions (and evolutionary epochs) of their host disks \citep[e.g.,][]{Kim_2017, Wang_2020, Scaloni_2024}.\par

The effects of bar-driven inflows in the evolution of their hosts can be quantitatively assessed by constraining the gas inflow rates toward the central regions. In local studies, multiple techniques have been used to estimate these inflow rates, with typical inferred values of the order of a few M$_\odot$/yr \citep[e.g.,][]{Quillen_1995, Regan_1997, Sormani_2019a, Sormani_2023}, comparable to the typical star formation rates.\par

At high redshift, and more specifically at cosmic noon ($z\sim1-3$), the peak epoch of cosmic star formation \citep[e.g.,][]{Madau_2014}, recent studies with \textit{JWST} have revealed that a significant fraction of disk galaxies host bars ($f_{\mathrm{bar}}=N_{\mathrm{barred}}/N_{\mathrm{disk}}\approx10\%-20\%$) \citep{Le_Conte_2024, Guo_2025, Espejo_Salcedo_2025, Geron_2025, Le_Conte_2025}, with the true bar fraction being possibly even higher due to the limited bar detectability resulting from observational limitations \citep{Liang_2024}. This picture is also supported by high-resolution, cosmological simulations which reproduce the observed bar fraction \citep[e.g.,][]{Fragkoudi_2020, Zhao_2020, Fragkoudi_2025}, or even exceed it \citep[e.g.,][]{Rosas-Guevara_2022}; however, see also \citet{Kraljic_2012}. Thus, bars could also be responsible for shaping the evolution of a significant fraction of high-$z$ disk galaxies.\par

In recent years, multiple surveys targeting typical, main sequence galaxies at cosmic noon have shown that these galaxies are predominantly disks \citep[e.g.,][]{NMFS_2020} and significantly more gas-rich \citep[e.g.,][]{Tacconi_2018, Tacconi_2020, NMFS_2020} and turbulent \citep[e.g.,][]{Genzel_2006, NMFS_2009, Kassin_2012, Swinbank_2017, Uebler_2019, Wisnioski_2019} than their low-redshift counterparts. However, the number of studies focusing on the effects of bar-driven evolution in such conditions remains limited as: i) from a theoretical standpoint the large gas fractions make the proper treatment of the multiphase interstellar medium (ISM) very computationally expensive \citep[e.g.,][]{Naab_2017}, with relevant systematic studies carried out only recently \citep[e.g.,][]{Bland-Hawthorn_2024}, and ii) from an observational point of view these systems were not expected to exist in significant numbers, based on older \textit{HST} observations of restframe UV/optical wavelengths which pointed to the overall prevalence of irregular morphologies at $z>1$ \citep{Madau_1996, Abraham_1999, Sheth_2008, Lotz_2008, Lotz_2011, Melvin_2014, Madau_2014, Simmons_2014, Margalef-Bentabol_2022}.\par

Despite the recent identification of multiple barred galaxies at high redshift, given the long integration times needed for high-resolution ($\sim$\,1\,kpc) observations of the gas distribution and kinematics of normal $z\sim1-3$ star-forming galaxies, few detailed studies exist to date. Some notable examples are the study of a $z\sim2.2$ barred spiral (Q2343\_BX610) by \citet{Genzel_2023}, who identified rapid, large-scale inflows, the study of J0107a by \citet{Huang_2025}, the most massive high-$z$ barred spiral identified to date \citep{Huang_2023}, exhibiting characteristic signatures of bar flows and an estimated net inflow rate of the order of the SFR, two dusty star-forming galaxies (DSFGs) at $z\sim3.1$ \citep{Umehata_2025} and $z\sim3.8$ \citep{Amvrosiadis_2025}, as well as the identification of a disk bending wave in a $z\sim4.4$ barred spiral by \citet{Tsukui_2024}. In \citet{Pastras_2025b}, we presented a study of the residual velocities in a $z\sim1.5$ barred spiral, in which the signatures of circular motions agree well with a tailor-made simulation of a high-$z$, gas-rich barred spiral.\par

While these studies identify high-$z$ bar flows in overall agreement with theoretical predictions, the net effect of such flows, i.e., the net gas inflow rate, has been directly estimated in only a few cases \citep{Huang_2025, Jolly_2026}. NOEMA\textsuperscript{3D} \citep{Jolly_2026, Chen_2026}, an ambitious program targeting typical, massive, cosmic noon galaxies, provides us with a deep, high-resolution view of the molecular gas kinematics of high-$z$ normal and barred spirals, bringing new insights into their role in the context of galaxy evolution.\par

In this paper, we present a deep IRAM-NOEMA CO(4-3) observation of a $z\sim1$ main sequence, massive barred spiral, G4\_38232, observed as part of the NOEMA\textsuperscript{3D} survey. We exploit the NOEMA molecular gas data in addition to the available deep, multi-band \textit{HST} and \textit{JWST} imaging, to offer an in-depth study of this target, giving a detailed picture of the gas flows and estimates of the resulting net flow rates for this high-$z$ barred galaxy. Our study aims to constrain the possible role of a high-$z$ bar in shaping the evolution of its host disk at cosmic noon.\par

In Sect.~\ref{sec:observations} we present an overview of the observational data available for our target. In Sect.~\ref{sec:anatomyOfG438232} we apply an array of analyses to constrain important properties of G4\_38232. We make use of these properties in Sect.~\ref{sec:gasFlowsInG438232} to offer a quantitative view of the gas flows in the disk of the galaxy. In Sect.~\ref{sec:discussion} we discuss the results, implications and caveats of our analyses and finally in Sect.~\ref{sec:conclusions} we summarize our findings. Throughout this paper, we adopt the following typical cosmological parameters: $H_0=70$\,km/s/Mpc, $\Omega_m=0.3,$ and $\Omega_\Lambda=0.7$.\par

\section{Observations}
\label{sec:observations}

G4\_38232 (R.A.=14\textsuperscript{h}19\textsuperscript{m}48\textsuperscript{s}926, Dec.=+52\textdegree58$'$32$''$027; J2000) was observed with the IRAM-NOEMA interferometer as part of the NOEMA\textsuperscript{3D} survey (PIs: Genzel, Neri, Tacconi). The NOEMA observations, with a 12 antennas equivalent total on-source integration time of $\approx37$\,hrs, were obtained in the A (most extended) and C (intermediate) configurations. The calibration and imaging of the data was done using the \texttt{GILDAS}\footnote{\url{https://www.iram.fr/IRAMFR/GILDAS}} radioastronomical software \citep{GILDAS_2013}. Data products with the following three weightings were produced from the NOEMA visibilities: i) using a UV-taper of $800$\,m allowing for a better recovery of the extended emission, resulting in a resolution of $\approx$\,0\farcs46, ii) using natural weighting with an achieved resolution of $\approx$\,0\farcs32, and iii) using uniform weighting with a robust factor of $5$ \citep[see][for the definition]{GILDAS_2013} with a resolution of $\approx$\,0\farcs24. The achieved per channel sensitivity was $\approx0.12$\,mJy/beam, with the channel width of the data products being $20$\,MHz ($27.52$\,km/s). Finally, the total CO(4-3) flux of the two higher resolution data products was scaled to match that of the tapered one. We refer the reader to the following papers for a detailed overview of the data reduction process \citep{Jolly_2026, Chen_2026}.\par

This galaxy has also been observed with both \textit{HST} and \textit{JWST} with the multi-band available imaging ranging from restframe UV to NIR wavelengths. The \textit{HST} observations were carried out in the context of the Cosmic Assembly Near-infrared Deep Extragalactic Legacy Survey (CANDELS) \citep{Grogin_2011, Koekemoer_2011, van_der_Wel_2012}. The \textit{JWST} NIRCam data were taken as part of the Cosmic Evolution Early Release Science Survey (CEERS, Proposal ID: 1345, PI: Finkelstein, \citealt{Finkelstein_2025}), the JWST-legacy narrow-band survey of H$\alpha$ and [OIII] emitters in the epoch of reionization (Proposal ID: 2234, PI: Banados, \citealt{Banados_2021}), the Public Observation Pure Parallel Infrared Emission-Line Survey (POPPIES, Proposal ID: 5398, PI: Kartaltepe, \citealt{Kartaltepe_2024}), and the Slitless Areal Pure-Parallel High-Redshift Emission Survey (SAPPHIRES, Proposal ID: 6434, PI: Egami, \citealt{Sun_2025}). The MIRI imaging was obtained in the context of the MIRI EGS Galaxy and AGN Survey (Proposal ID: 3794, PI: Kirkpatrick, \citealt{Backhaus_2025}).\par

\begin{figure*}
    \centering
	\includegraphics[width=2.0\columnwidth]{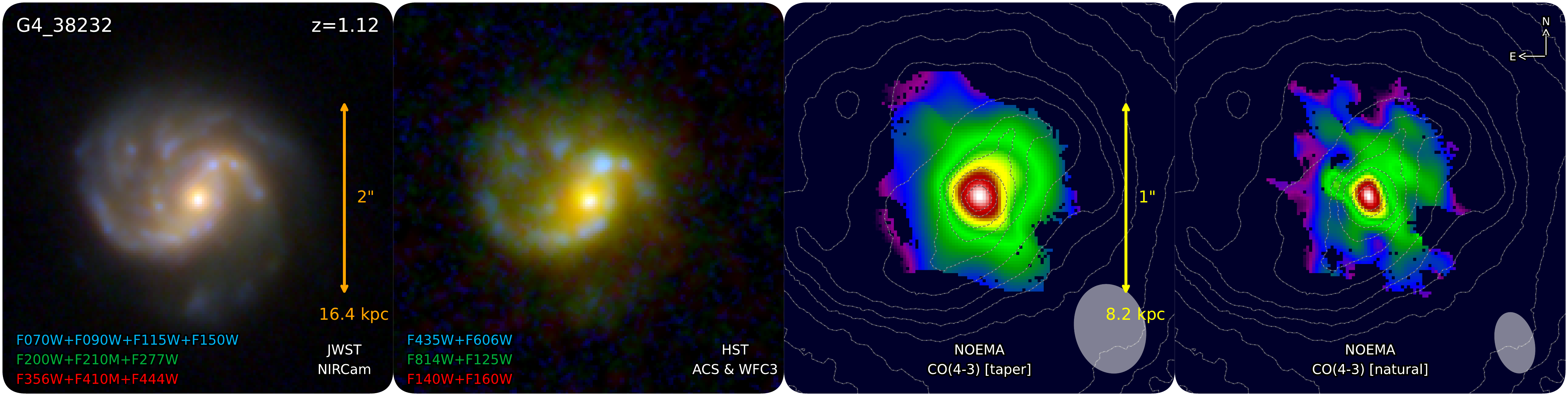}
    \caption{Imaging overview of G4\_38232: color composite images of the \textit{JWST} NIRCam 0.7-4.4\,$\mu m$ observed frame continuum (left) and the \textit{HST} ACS \& WFC3 0.43-1.6\,$\mu m$ (middle left), as well as NOEMA CO(4-3) line flux for the tapered (middle right) and naturally-weighted (right) data products with overlaid contours of the F444W continuum. In both \textit{HST} and \textit{JWST} continuum images a prominent bar is observed with spiral arms emanating from its ends, with the western spiral arm seemingly less extended compared to the eastern one. Additional substructure, in the form of fainter spiral arms at the northeastern side of G4\_38232, is identified in the more sensitive \textit{JWST} imaging. In both continuum images the galaxy appears asymmetric with more flux at the southwestern part of the bar and at the opposite, i.e., northeastern, side at larger radii. The CO(4-3) flux observed with NOEMA covers most of the bar region and has a centrally peaked distribution with higher values at the northwestern compared to the southeastern side.}
    \label{fig:imagingOfG4-38232}
\end{figure*}

In the case of the \textit{HST} ACS \& WFC3 and the \textit{JWST} MIRI data, we retrieved high quality reductions carried out using the \texttt{Grizli} \footnote{\url{https://github.com/gbrammer/grizli}} pipeline \citep{Brammer_2023}, from the DAWN JWST Archive (DJA). For the \textit{JWST} NIRCam data, we retrieved the uncalibrated data from the MAST archive, then reduced and combined them using a custom pipeline built around the \texttt{CrabToolkit} \footnote{\url{https://github.com/1054/Crab.Toolkit.JWST}}. We followed the standard \textit{JWST} reduction steps, with improved "snowball", "claw" and "wisp" removal through the use of the corresponding publicly available templates \footnote{\url{https://jwst-docs.stsci.edu/known-issues-with-jwst-data/nircam-known-issues/nircam-scattered-light-artifacts}}, application of astrometric corrections based on the CEERS photometric catalog \citep{Cox_2025}, as well as custom masking. The total on-source integration time for our NIRCam data reductions ranges from $\approx0.9$\,hrs (F410M) to $\approx6.0$\,hrs (F200W), with an average of $\approx3.6$\,hrs across all available filters. In all cases, the final products were astrometry corrected and background subtracted images with a pixel scale of $0.025$\,arcsec.\par

An overview of the available imaging for G4\_38232 is presented in Fig.~\ref{fig:imagingOfG4-38232}. The deep CO(4-3) observations combined with the available deep multi-band imaging allow for a detailed quantitative study of the molecular gas flows in this $z\sim1$ target and their correlation with the prominent bar structure. This enables the assessment of the possible role of bar-driven flows in the evolution of a typical, main sequence barred spiral at high redshift.\par

\section{Anatomy of G4\_38232}
\label{sec:anatomyOfG438232}

In this section we use the available data for G4\_38232 to constrain its properties. We first focus on the properties that would be relevant for any disk galaxy, such as its orientation, mass distribution and kinematic information, and then apply more refined analyses to constrain the properties of the bar, namely, its orientation, length, strength and pattern speed.\par

\subsection{Morphology}
\label{sec:morphology}

In the multi-band imaging of Fig.~\ref{fig:imagingOfG4-38232}, we identify a prominent central bar structure and a clear but asymmetric spiral structure at larger radii. Specifically, at the northwestern side of the galaxy the spiral arm emanating from the end of the bar stops abruptly and the whole southwestern side appears less bright in stellar continuum than its northeastern counterpart. In the \textit{HST} (ACS \& WFC3) imaging, the observed spiral structure consists of two arms, emanating from the ends of the bar. In the more sensitive and higher-resolution \textit{JWST} (NIRCam) images, additional substructure, in the form of more spiral arms, can be discerned in the northeastern part of the galaxy. Additionally, a visual inspection of the overall observed morphology indicates that G4\_38232 is probably quite face-on, with a low inclination angle.\par

The barred morphology of this target was first identified and studied by \citet{Guo_2023}, who presented it as part of six cases of barred spirals at $z\sim1.1-2.3$ identified through the quantitative analysis of \textit{JWST} NIRCam imaging probing the restframe near-IR continuum. Here, we use the combined photometric and kinematics data to derive robust new measurements of the morphological parameters.\par

\subsubsection{Elliptical isophotes; the effects of non-axisymmetric structures}
\label{sec:ellipticalIsophotes}

\begin{figure}
    \centering
	\includegraphics[width=\columnwidth]{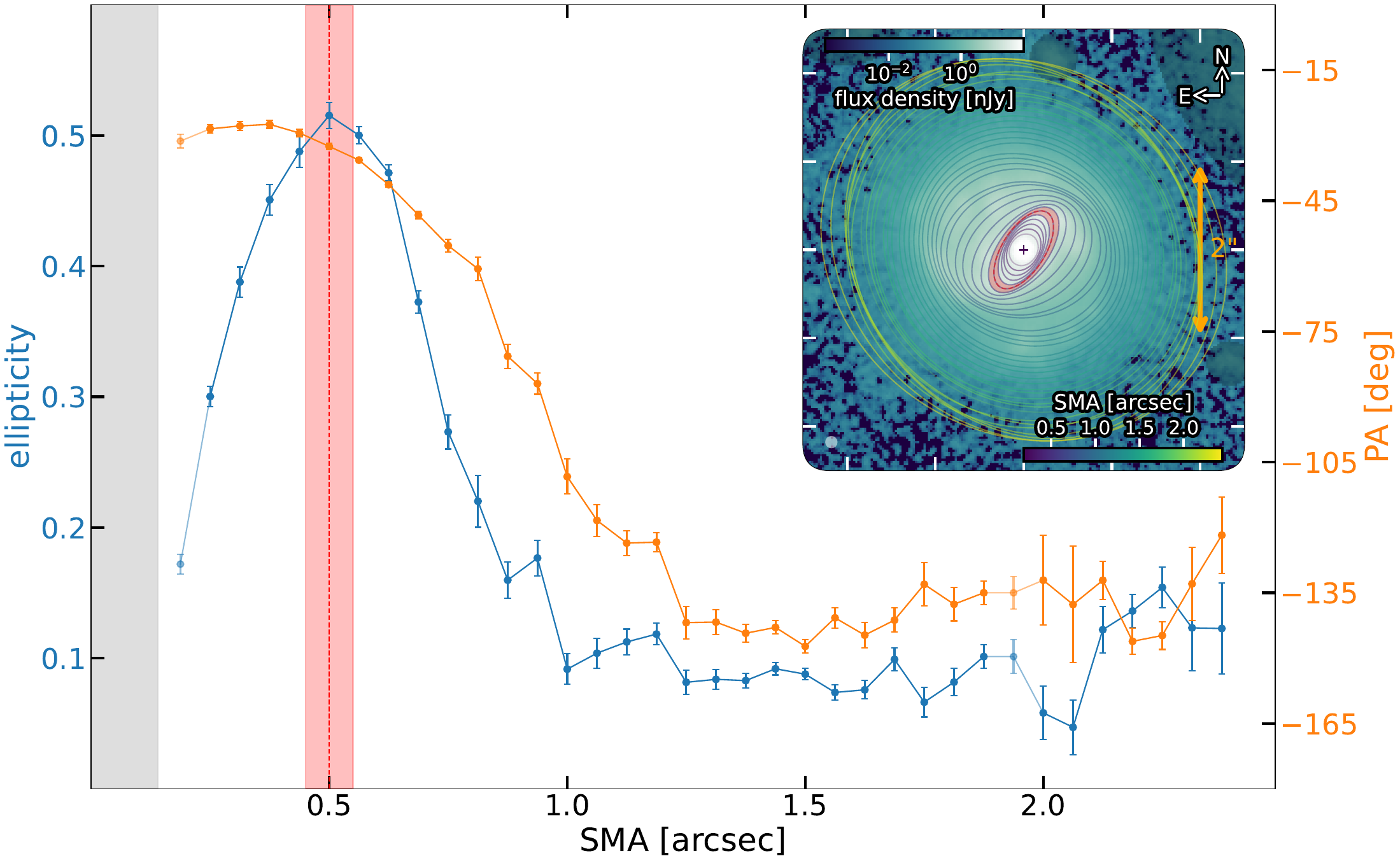}
    \caption{Overview of the elliptical isophote fitting of the NIRCam F444W image: isophote ellipticity and position angle profiles (main plot) and the F444W continuum with overlaid contours of the elliptical isophotes (inset). The color of each overlaid isophote indicates its semimajor axis (SMA). While the outer isophotes do not provide us with reliable information for the disk orientation (as they are elongated in a different direction than the major kinematic axis and consequently they do not trace the axisymmetric part of the disk), the bar orientation and radius can be constrained using the peak of the ellipticity profile.}
    \label{fig:ellipticalIsophoteFitting}
\end{figure}

Since the orientation of the disk, as well as the orientation and length of the bar are needed for our analyses, we carried out an elliptical isophote analysis, similar to that presented in \citet{Guo_2023}, focusing on constraining these properties. We determined the center of G4\_38232 by fitting a 2D Gaussian to the central mass concentration, in other words the bulge, observed in the reddest NIRCam image (F444W). We assume that the center of the fitted Gaussian indicates the center of G4\_38232, which we use throughout the rest of our analyses. With this center in hand, we used the respective routines of \texttt{Photutils} \citep{Bradley_2024} to fit elliptical isophotes to the reddest NIRCam image (F444W). We used linear steps in the radial direction, with changes of 0\farcs0625 in the semimajor axis of the isophotes. The centers of these isophotes were fixed to the previously determined center of G4\_38232. We also implemented two iterations of sigma-clipping, each time masking the pixels with values deviating by more than $2\times\sigma$ from the mean value along each elliptical path. The fitted elliptical isophotes, in addition to the profiles of their ellipticities ($e=1-b/a$, where $a$ and $b$ are the semimajor and semiminor axis, respectively) and position angles (PAs) as a function of their semimajor axes are presented in Fig.~\ref{fig:ellipticalIsophoteFitting}.\par

The outer isophotes are elongated along the northeastern to southwestern direction, that is with a PA of $\sim45$\textdegree. This orientation is very different from the kinematically inferred PA (see Sect.~\ref{sec:molecularGasKinematics}), thus, we can not assume that in this case the outer isophotes trace the outer parts of an axisymmetric disk. They are significantly affected by the spiral structure at large radii. This outer spiral structure can extend out to the outer Lindblad resonance (OLR), a region typically elongated perpendicular to the bar \citep[e.g.][]{Buta_1996}, meaning that the observed offset between the outer isophotes and the apparent PA of the bar is probably expected. At any rate, we will use the kinematic information to estimate the orientation parameters of the disk.\par

Both the ellipticity and position angle profiles display the characteristic signatures of the presence of a bar; the ellipticity climbs to a peak at $\approx$\,0\farcs5 before dropping at $\approx$\,0\farcs625, while the position angle stays almost constant throughout the central parts of the galaxy, out to $\approx$\,0\farcs4375. We derive the bar length using the peak of the observed ellipticity profile, i.e., $a_{\mathrm{bar,proj}}\approx$\,0\farcs5, and the corresponding value for its PA, namely, $\mathrm{PA}_{\mathrm{bar}}\approx32.5$\textdegree~(e.g., \citet{Wozniak_1991, Liang_2024}, see also \citealt{Erwin_2003}).\par

We infer an in-plane semimajor axis of $a_{\mathrm{bar}}\approx4.2$\,kpc, assuming that the bar is very thin \citep[i.e.,][Eq.~D.5]{Pastras_2025b} and adopting the dynamically inferred disk $\mathrm{PA}\approx-66$\textdegree~and $i\approx20$\textdegree~(see Sect.~\ref{sec:dynamicalForwardModeling}). Given the modest disk inclination, this value is expected to be in excellent agreement with the proper deprojection of the elliptical isophote \citep{Gadotti_2007}. Following \citet{Cuomo_2019a}, we derive two additional estimates: i) the semimajor axis of the first isophote with a position angle difference $\vert\Delta\mathrm{PA}\vert\geq5$\textdegree~from the isophote with the maximum ellipticity, and ii) the bar-interbar contrast from the Fourier decomposition of Sect.~\ref{sec:potentialEstimation} \citep[e.g.,][]{Aguerri_2000}. The resulting estimates of the in-plane bar semimajor axis were $a_{\mathrm{bar,\Delta PA}}\approx5.2$\,kpc and $a_{\mathrm{bar,FD}}\approx4.9$\,kpc, respectively, in both cases larger than our adopted value. However, the apparent bi-symmetric spiral structure emanating from the ends of the bar can result in an overestimation of its true length. This possibility seems to be confirmed through the visual inspection of the deprojected galaxy morphology presented in Figs.~\ref{fig:baryonSDPotentialAndTorque}~and~\ref{fig:imagingOverviewDeprojected}.\par

We find that G4\_38232 hosts an extended bar, at the high-end of the respective distributions of both projected \citep{Guo_2025, Le_Conte_2025} and deprojected \citep{Le_Conte_2025} bar lengths at $z>1$.\par

\subsubsection{Curve of growth; an extended exponential disk}
\label{sec:curveOfGrowthFitting}

After deriving an estimate for the disk inclination and position angle through the two-dimensional modeling of the lower resolution molecular gas kinematics presented in Sect.~\ref{sec:dynamicalForwardModeling}, we used the curve of growth of the continuum in the F444W filter and the stellar mass map derived through the SED fitting of the \textit{JWST} NIRCam and \textit{HST} ACS imaging (see Appendix~\ref{sec:SEDFitting} and G. Tozzi et al., in prep.) to constrain the effective radius and S\'ersic index of the disk and bulge-to-total ratio of the galaxy.\par

\begin{figure}
    \centering
	\includegraphics[width=\columnwidth]{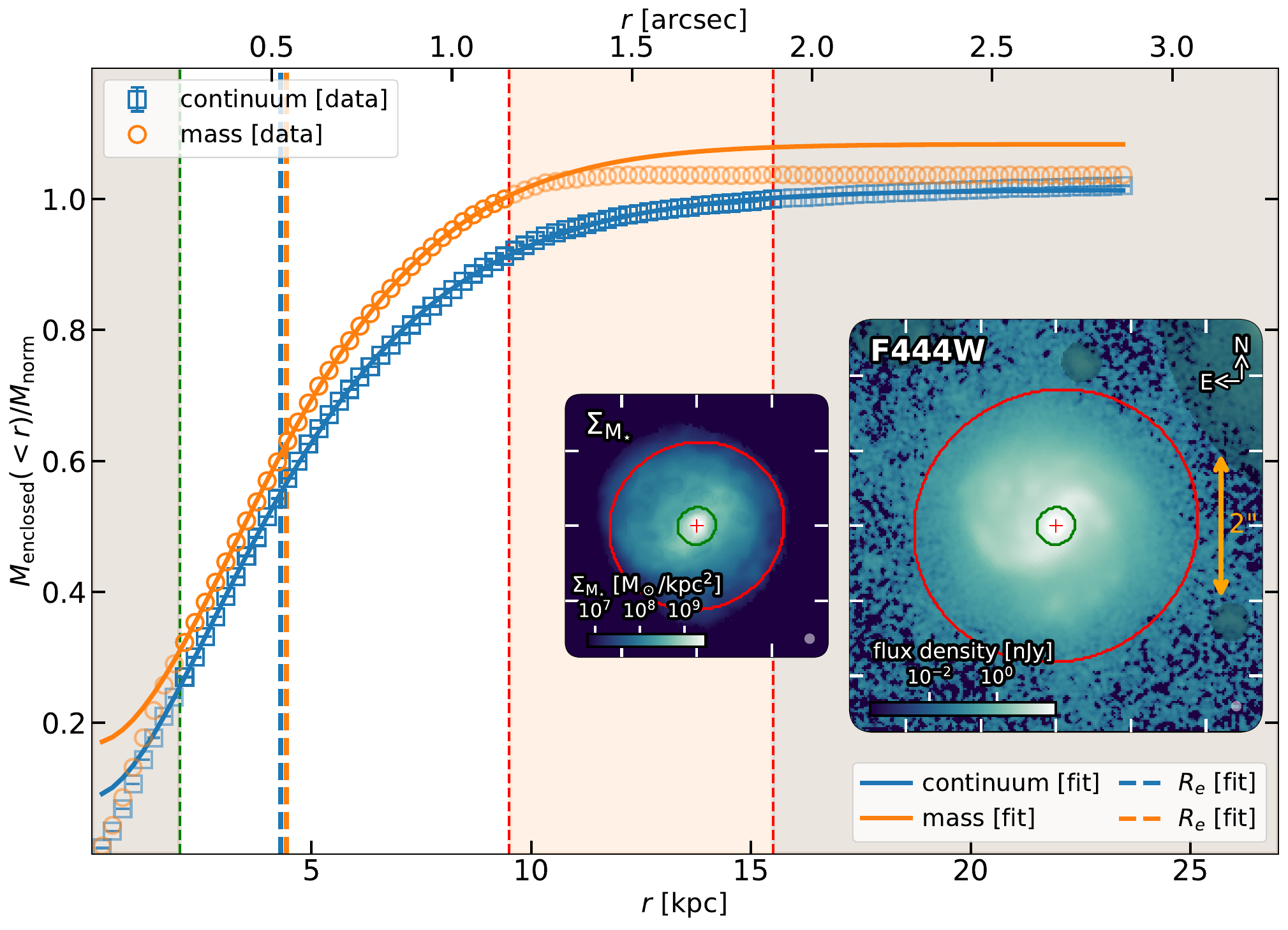}
    \caption{Overview of the fit to the curve of growth of the NIRCam F444W image and stellar mass map: normalized cumulative flux / mass profiles (main plot) and the F444W continuum (right inset) and stellar mass (left inset) maps. The blue squares (orange circles) indicate the cumulative flux (mass) profile extracted using elliptical annuli of increasing semimajor axes while the fitted profile is indicated with a solid blue (orange) line. The corresponding effective radius is also given with a dashed blue (orange) vertical line. The inner (outer) radius of the region considered in the fit is indicated with a green (red) dashed vertical line in the main plot and a solid green (red) ellipse in the insets. The discrepancies between the fitted profiles in the central region stem from the presence of a central mass concentration (smoothed out by the observational PSF in the data). In the outer regions, the discrepancies are the result of the identified clumps in the outer disk regions in the F444W flux and the limited extent of the stellar mass map.}
    \label{fig:curveOfGrowthFitting}
\end{figure}

We extracted the F444W flux and stellar mass profiles using elliptical annuli of increasing semimajor axes. We covered the whole extent of our images, i.e., a field of view (FOV) of 6\farcs0$\times$6\farcs0, with $100$, $\approx$\,0\farcs03 thick annuli. The resulting curves of growth are shown in Fig.~\ref{fig:curveOfGrowthFitting}.\par

We fit each curve of growth with the sum of the cumulative mass of a S\'ersic profile (with a total mass $M$, S\'ersic index $n_s$, and effective radius $R_e$) and an additional central mass concentration $M_0$. The presence of a nuclear disk, bulge, or even a composite of the two is commonly observed in local barred spirals \citep[e.g.,][]{Gadotti_2015, Gadotti_2020, Bittner_2020, de_Sa_Freitas_2023a, de_Sa_Freitas_2023b, de_Sa_Freitas_2025, Fraser-McKelvie_2025}. The ages of nuclear disks hint towards their presence already at cosmic noon \citep{Gadotti_2015, de_Sa_Freitas_2023a, de_Sa_Freitas_2025}, with evidence of such structures uncovered by recent observations at $z>1$ \citep{Le_Conte_2026}. Interestingly, in the case of G4\_38232, a central mass concentration with a low S\'ersic index (see Appendix~\ref{sec:morphologicalForwardModeling} and \citealt{Chen_2026}) and high star-formation rate (see Appendix~\ref{sec:SEDFitting} and G. Tozzi et al., in prep.) is tentatively identified through the forward modeling and resolved SED fitting of the available \textit{HST} ACS and \textit{JWST} NIRCam data, hinting towards the possible presence of a nuclear disk.\par

In our fits of the curve of growth, we excluded the very central region ($r\approx$\,2\,kpc), since given the PSF of the NIRCam F444W filter ($\approx$\,0\farcs15) the contribution of the central bulge would affect the observed profile. Thus, we limited our fits to larger radii, where its contribution to the observed surface density is negligible meaning that its effect on the curve of growth can be adequately described by the addition of the central mass concentration $M_0$. We also imposed an upper limit to the radial extent of the fit, in order to avoid contributions from clump-like structures identified in the outer disk of G4\_38232, as well as a lower redshift galaxy (G4\_38651, $z\approx1.062$) at the northwestern edge of the FOV.\par

Our fits show that our target hosts a nearly exponential ($n_s\approx0.9\pm0.1$), extended ($R_e\approx4.3\pm0.2$\,kpc) disk, with a modest central bulge ($B/T\approx0.16\pm0.07$). There is a very good overall agreement between the results from the fits to the F444W continuum and the stellar mass map, based on the differences of which we estimated the uncertainties of the fitted parameters, since the formal fitting uncertainties in each case were even lower.\par

Considering that the F444W and stellar mass surface density profiles could offer a better constraint on the S\'ersic index of the disk, we additionally fit these surface density profiles with a S\'ersic profile in a similar fashion to our curve of growth analysis, with recovered parameters in excellent agreement with our previous results. An overview of the fits is presented in Appendix~\ref{sec:surfaceDensityProfileFitting}.\par

\subsection{Molecular gas kinematics}
\label{sec:molecularGasKinematics}

We extracted maps of the observed line of sight (LOS) velocity, velocity dispersion and CO(4-3) flux, by fitting our data spaxel-by-spaxel with a single Gaussian. We estimated the uncertainty of the recovered parameters for each fit, by performing $100$ Monte Carlo simulations following a similar procedure as described by \citet{NMFS_2009}. For each simulation, the spectrum of each spaxel is refitted after being perturbed at the noise level, estimated as the standard deviation of the residuals in the part excluding the region within $\sim2\times$FWHM of the originally fitted line. The standard deviation of the recovered parameters from these Monte Carlo simulations was used as the uncertainty of the recovered parameters by our initial fits.\par

With the line centroid, dispersion and flux in hand, we assessed the reliability of the CO(4-3) line detection for each pixel, initially through a S/N threshold (S/N$\geq$3) considering the peak amplitude of the fitted line as signal and then requiring that the CO flux of the pixel is well constrained, i.e., $F/\delta F\geq1$. Additionally, since spatially isolated pixels could still be the result of noise-introduced artifacts we masked our flux, velocity and velocity dispersion maps using a Friends-of-Friends (FoF) algorithm, starting from the center of the galaxy and masking pixels with no valid immediate neighbors. We visually assessed that no coherent substructures were masked through this process. Finally, we additionally manually masked unreliable regions of the resulting maps, rejecting pixels with values that are either extreme or inconsistent with their surrounding regions.\par

\begin{figure}
    \centering
	\includegraphics[width=\columnwidth]{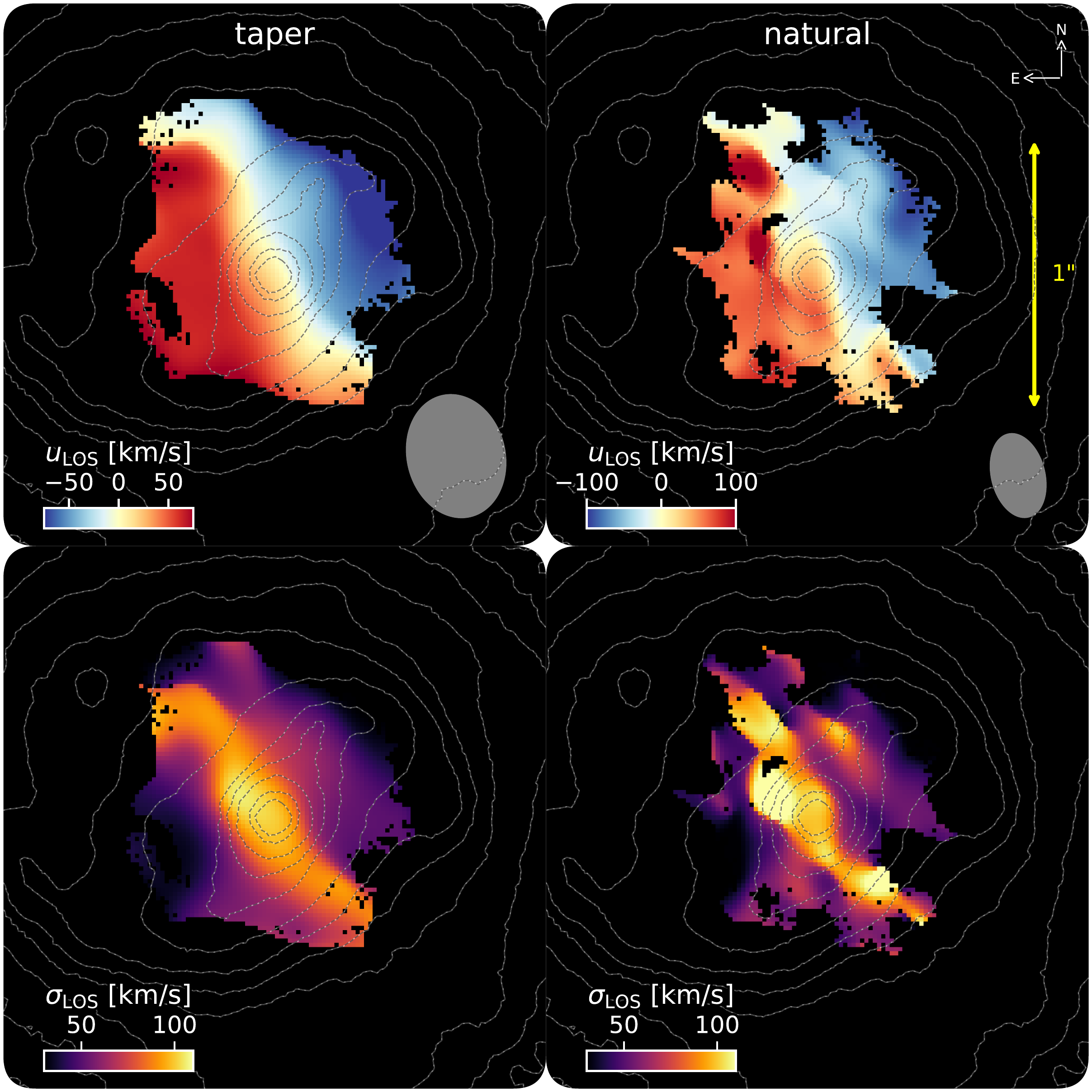}
    \caption{Velocity (top) and velocity dispersion (bottom) for the tapered (left) and naturally-weighted (right) data with overlaid contours of the reddest (F444W) NIRCam image. In the tapered data, a clear rotation pattern is identified, with deviations from this regular, axisymmetric rotation becoming obvious in the higher resolution naturally-weighted data. In both weightings, an elevated velocity dispersion along the kinematic minor axis is observed, in agreement with expectations for a thick disk in combination with beam-smearing, given the orientation of the disk of this galaxy.}
    \label{fig:velocityAndVelocityDispersion}
\end{figure}

The velocity and velocity dispersion maps for the tapered and naturally-weighted data are presented in Fig.~\ref{fig:velocityAndVelocityDispersion}. In the case of the robust data, following the same masking procedure results in a very limited region of reliably constrained kinematics, thus we chose not to include them in our two-dimensional, spatially resolved analyses. In both of our two other data products, we identify a clear rotation pattern, with a seemingly almost unperturbed velocity field in the inner regions of the tapered data and clear deviations from circular motions in the naturally-weighted data. The apparent zero velocity curve and velocity gradient in the former helps constrain the PA of the disk of the galaxy, while the detailed structure in the latter allows for the study of the in-plane noncircular motions.\par

\subsubsection{Dynamical forward modeling; a turbulent baryon dominated disk}
\label{sec:dynamicalForwardModeling}

We used the forward modeling tool \texttt{DysmalPy} \footnote{\url{https://github.com/dysmalpy/dysmalpy},\\\url{https://www.mpe.mpg.de/resources/IR/DYSMALPY/}} \citep{Davies_2004a, Davies_2004b, Cresci_2009, Davies_2011, Wuyts_S_2016, Lang_2017, Price_2021, Lee_2025}, to fit the major axis kinematics of G4\_38232 and constrain its mass distribution. This goal has been shown to be achievable through this one-dimensional fitting method \citep{Price_2021}. Additionally, while tangential noncircular motions can affect the major axis kinematics, the radial flows, in which we are particularly interested, do not contribute to the LOS velocities in this region. Moreover, the larger radial extent we are able to probe through the extraction of the major axis kinematics using apertures of increasing size as a function of radius, allows for more robust constraints on the mass distribution of the galaxy, with the contribution of the dark matter halo being an important example \citep{Wuyts_S_2016, Lang_2017, Genzel_2017, Genzel_2020, Price_2021, Bouche_2022, Puglisi_2023, Nestor_Shachar_2023}.\par

\begin{figure}
    \centering
	\includegraphics[width=\columnwidth]{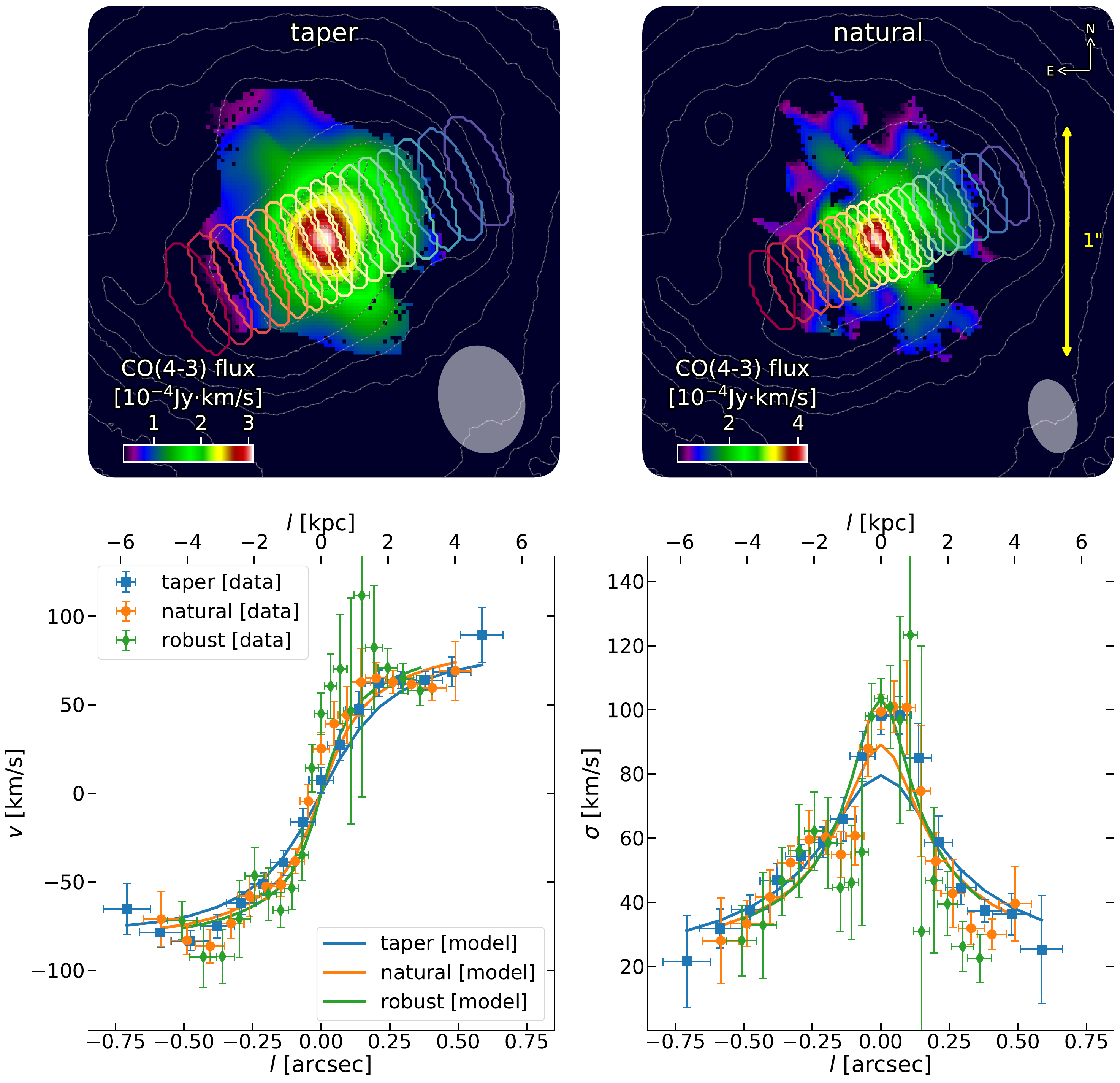}
    \caption{CO(4-3) flux maps (top), as well as velocity (bottom left) and velocity dispersion (bottom right) profiles for the tapered (blue), naturally-weighted (orange) and robust (green) data, along with the profiles extracted from the beam-convolved, axisymmetric model fit by \texttt{DysmalPy}. The apertures used in the extraction of the profiles along the major axis of the disk are superimposed on the CO(4-3) flux maps, color-coded with respect to the relative distance along the major axis. Both the observed rotation curve and velocity dispersion profile can be well reproduced in all cases, with only minor discrepancies between the data and the model, such as the velocity at the central positive side of the rotation curve for the robust data or central velocity dispersion peak for the tapered data.}
    \label{fig:DysmalPySlitProfiles}
\end{figure}

Our model comprises a baryonic disk and a bulge, embedded in a dark matter halo. The baryonic disk is modeled as a flattened spheroid \citep{Noordermeer_2008} with a vertical to horizontal ratio of $q_0=0.2$, in agreement we observed values for edge-on galaxies at cosmic noon \citep{Tsukui_2025}, and a S\'ersic index of $n_{\mathrm{s,disk}}=0.9$, the bulge as an axisymmetric, deprojected S\'ersic profile with $n_{\mathrm{s,bulge}}=4$ and an effective radius of $R_{\mathrm{eff,bulge}}=1$\,kpc, and the dark matter halo as a \citep*{Navarro_1996} (NFW) profile, with a concentration parameter of $c\approx5.85$ derived through the average relation $c\sim10.9\times(1+z)^{-0.83}$ presented in \citet{Genzel_2020} for a redshift of $z\approx1.12$.\par

Our model includes a velocity dispersion component, $\sigma_0$, that is isotropic in the radial and tangential directions. In the vertical direction, we implemented a dispersion component required to account for the thickness of the disk, following $\sigma_z(r)=(z_0/r)\times u_{\mathrm{rot}}(r)$ \citep[][Eq.~1b]{Genzel_2008}, where $z_0$ is the scale-height of the disk and $u_{\mathrm{rot}}$ the rotational velocity. As a result, the velocity dispersion of our model along the LOS was $\sigma_{\mathrm{LOS}}=\sqrt{\left(\sigma_0\sin i\right)^2+\left(\sigma_z\cos i\right)^2}$. This was necessary for reproducing the central peak of the velocity dispersion profile, the values of which are too large to be explained by beam smearing alone. Since the isotropic, in-plane component is the one contributing to the velocity dispersion in the radial direction, this component is used in the application of the pressure support correction implemented in \citep{Price_2022} (see also \citealt{Burkert_2010}).\par

We started by fitting the two-dimensional velocity and velocity dispersion maps of the tapered data to constrain the position angle and inclination of the disk of G4\_38232. The degeneracy between the inclination and the total baryonic mass could be alleviated by the shape of the spider diagram, for example, consider the interdependence of the inclination, $i$, and projected azimuthal angle, $\phi$, in the LOS contributions of in-plane tangential motions in Eq.~E.29 of \citealt{Pastras_2025b}. The vertical component of the velocity dispersion, $\sigma_z$, the contribution of which depends on the inclination as well as the in-plane rotation, could also help alleviate the aforementioned degeneracy. Thus, we used the lower resolution data, in which the contributions from non-circular motions are expected to be smoothed out, to derive a refined estimate of the orientation parameters, which are crucial for the next parts of our analysis.\par

To increase the reliability of our results, we applied a more conservative mask to the tapered velocity and velocity dispersion maps, adopting a stricter flux-detection criterion of $F/\delta F\geq3$. This removed the asymmetric features previously visible at the northeastern and southwestern edges of the maps. Finally, since the constant velocity dispersion component is expected to be more accurately recovered from the respective profile along the major axis \citep[e.g.,][]{Genzel_2017, Price_2021}, we adopted a Gaussian prior of $\sigma_{0}\approx20\pm10$\,km/s, a value derived from the outer radii of the profiles, corrected for the line spread function (LSF), and assuming a roughly isotropic ($\sigma_z\sim\sigma_0$) dispersion at large radii.\par

Following \citet{Lee_2025}, we used \texttt{DysmalPy} and the \texttt{emcee} sampler, in this case with $200$ burn-in steps followed by $500$ iterations for $560$ walkers. We also repeated our fits using the nested sampler \texttt{dynesty} with results in excellent agreement with those of our \texttt{emcee} runs. After multiple fitting iterations our best estimates for these two parameters are $\mathrm{PA}\approx-66$\textdegree$\pm9$\textdegree~and $i\approx20$\textdegree$\pm5$\textdegree. We adopted these values and then followed the methodology of \citet{Genzel_2023} of fitting its major axis kinematics, fixing the $\mathrm{PA}$ and $i$.\par

We started by extracting the kinematics along the major axis using elliptical apertures, elongated perpendicular to it (see Fig.~\ref{fig:DysmalPySlitProfiles}). The semimajor axis (width -- perpendicular to the PA axis) of our apertures is fixed so that it matches that of our beam, while the semiminor axis (length -- along the PA axis) was increasing following a cosine function from $1/4$ (in the center) to $1/2$ (in the outer edges) of the semimajor axis of our beam. This way, we managed to probe in detail the central, higher S/N, parts of our data sets, while at the same time enabling the detection of our CO(4-3) line at the outer, lower S/N, regions of the disk.\par

After the extraction, we fit a single Gaussian to the spectrum of each aperture recovering the flux, velocity and velocity dispersion as well as the respective uncertainties from Monte Carlo simulations. We then masked the spectra with unreliable line detections, confirming the results through visual inspection of the Gaussian fits.\par

\begin{figure*}
    \centering
	\includegraphics[width=2.0\columnwidth]{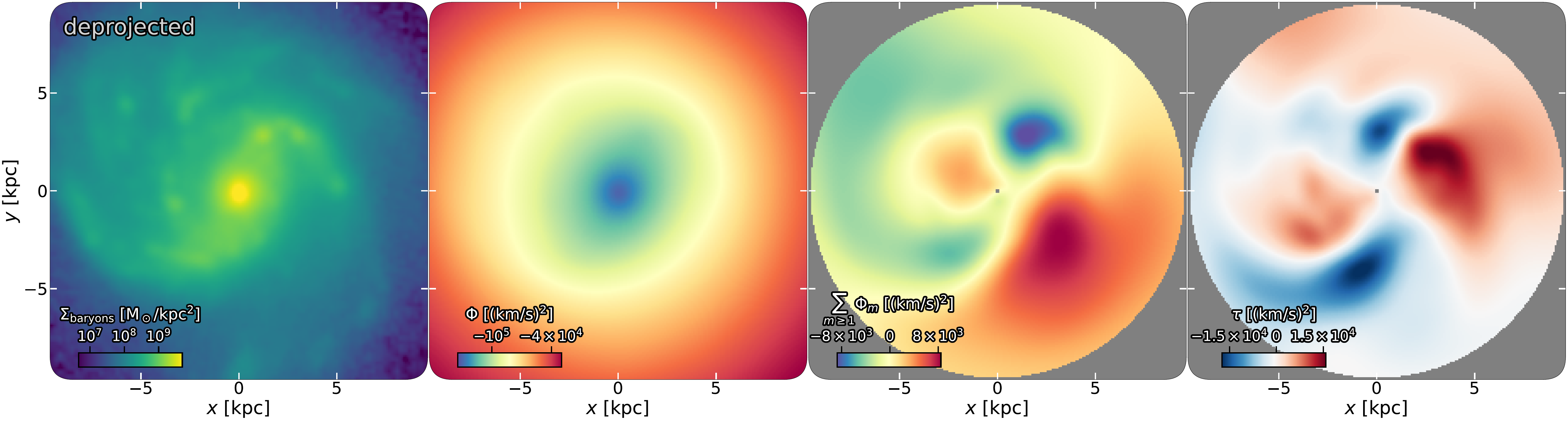}
    \caption{Estimate of the deprojected baryonic surface density (left), gravitational potential at the disk midplane (middle left), non-axisymmetric component of the potential (middle right) and torque per unit mass due to this non-axisymmetric component (right). A bisymmetric pattern, typical of a bar, is identified in both of the two latter figures in addition to an overall asymmetry. As expected, the torques are negative (positive) in front of the local minima (maxima) of the potential in the azimuthal direction and with respect to the rotation of the galaxy (counter-clockwise).}
    \label{fig:baryonSDPotentialAndTorque}
\end{figure*}

Next, these velocity and velocity dispersion profiles were fit using the forward modeling of an axisymmetric, rotating disk using \texttt{DysmalPy}. All three of our data products were fit concurrently, using the same configuration as for our two-dimensional fits, with the fitted quantities being the total baryonic mass of the galaxy, $M_{\mathrm{baryons}}$, the isotropic, in-plane component of the intrinsic velocity dispersion of the disk material, $\sigma_0$, the dark matter fraction within the effective radius of the disk, $f_{\mathrm{dm}}(<R_e)=u^2_{\mathrm{circ,dm}}(R_e)/[u^2_{\mathrm{circ,dm}}(R_e)+u^2_{\mathrm{circ,baryons}}(R_e)]$, and the systematic velocity, $u_{\mathrm{sys}}$. 
We used flat priors for all fitted parameters, except for the intrinsic, in-plane velocity dispersion for which we adopted a Gaussian prior centered at the LSF-corrected value measured at the disk outskirts. Additionally, we fixed the values of the bulge-to-total ratio, $B/T$, and effective radius of the disk, $R_e$, to the photometrically-derived estimates presented in Sect.~\ref{sec:morphology}.\par

In our \texttt{DysmalPy} modeling, we fit the observed velocity and velocity dispersion profiles but excluded the flux profiles from the fit, since they appear asymmetric and, given the well-constrained photometric light profiles of G4\_38232, we considered that fitting the CO flux would introduce more uncertainties. At each iteration of the fitting process, a model of the mass distribution of the fitted components is realized and a four-dimensional hypercube is produced. This cube is collapsed along the LOS, and then convolved with the Point-Spread function (PSF) and LSF of each of our data products. For each resulting cube, the spectra along the major axis are extracted using the same apertures used for the extraction from the observational data, and fit using a single Gaussian, the properties (centroid, $\mu$, and dispersion, $\sigma$) of which are compared to those fitted to our observations. The best-fit profiles are presented in Fig.~\ref{fig:DysmalPySlitProfiles} and the respective constrained properties for G4\_38232 in Tab.~\ref{tab:galaxyProperties}.\par

The velocity and velocity dispersion profiles of G4\_38232 are generally well reproduced. However, the identified central peak in the velocity dispersion profile still exceeds the predictions of our fitted model for the tapered and naturally-weighted data. This could be evidence that the observed velocity dispersion profiles can not be modeled in detail using a purely axisymmetric rotating disk, possibly indicating strong noncircular motions near the center of this target, the presence of a super-massive black hole (SMBH), or, less likely given that our observations trace molecular gas \citep[e.g.,][]{Herrera-Camus_2019, Barfety_2025}, the possible contribution of an AGN. The possibility of strong noncircular motions in the central regions is also supported by the asymmetric and steeply rising central part of the observed rotation curve.\par

We find that G4\_38232 is a massive ($\log(M_{\mathrm{baryons}}/\mathrm{M_\odot})\approx10.96$), baryon dominated galaxy, with an inner dark matter fraction of $f_{\mathrm{dm}}(<R_e)\approx4$\%. Its dynamically-constrained baryonic mass is in good agreement with the sum of its integrated stellar ($\log(M_{\mathrm{\star}}/\mathrm{M_\odot})\approx10.82$) and gas ($\log(M_{\mathrm{gas}}/\mathrm{M_\odot})\approx10.65$) mass. The former was derived through the integrated SED fitting of all available imaging (\citet[][Appendix~A]{Jolly_2026}, \citealt[][Sect.~2.2.2]{Chen_2026}), while the latter using the integrated, within a radius of $\sim2.5$\,$R_e$, CO(4-3) flux from the tapered data product, following \citet[][Eq.~2]{Tacconi_2020} and assuming a CO(1-0) conversion factor of $a_{\mathrm{CO~1}}=4.36$\,M$_\odot$/[K(km/s)pc$^2$]. Its disk is gas-rich ($f_{\mathrm{gas}}\approx40$\%) and turbulent $\sigma_0\approx20$\,km/s, with a gas Toomre Q parameter of $Q_{\mathrm{gas}}\sim0.3$, estimated following \citet{Genzel_2014a} and assuming a flat rotation curve, meaning that its gas disk is gravitationally unstable \citep{Behrendt_2015}. This could be a possible explanation for the apparent clumpy morphology of the restframe UV/optical images of the galaxy.\par

\subsection{Potential estimation; an asymmetric bar}
\label{sec:potentialEstimation}

Focusing on the properties of the bar, we estimated the potential in the midplane of G4\_38232. The baryonic part of this potential corresponds to the sum of the contributions of the stellar and gaseous components. In G4\_38232 the molecular gas amounts to $\approx40\%$ of the mass of the disk, while the contribution of atomic and ionized gas can be neglected because the bulk of the gas budget in the central parts of star-forming galaxies is predominantly in the molecular state \citep[e.g.,][]{Leroy_2009, Saintonge_2016, Tacconi_2020}. We assumed that the distribution of gas follows that of the stars and estimated the projected baryonic surface density of the disk using the F200W NIRCam image. While this filter is not the reddest available and could be affected by extinction, we still chose it based on its better spatial resolution compared to longer-wavelength filters, e.g., $\sim2\times$ better than F444W, and the fact that it still probes restframe NIR wavelengths. Additionally, given the high gas fraction of the disk, a modest contribution from young stars to the observed F200W continuum emission might alleviate possible discrepancies between the true stellar and gaseous mass distributions; with young stars preferentially tracing the regions of a high gas concentration a more significant contribution from these objects could help better trace the true potential of the disk. Since these factors can introduce significant uncertainties in our analyses, we verified that our main results remain unaffected for different estimates of the baryonic surface density.\par

With the baryonic surface density map in hand, we followed the methodology of \citet{Quillen_1994} to estimate the baryonic potential in the midplane of the disk. In summary: i) we deprojected the baryonic surface density following the procedure in (S. Pastras et al, in prep.), realizing the grid of the face-on image, projecting its pixels on the sky plane, sampling the original projected surface density distribution and finally correcting for the difference between the area of the projected and deprojected pixels by multiplying the values of the latter by $\cos i$, ii) produced a convolution function following \citet[][Eq.~3.3]{Quillen_1994}, effectively assuming that the baryonic mass is spread in the vertical direction following a squared hyperbolic secant function, $\rho_{z}(z)=\left[1/\left(4z_0\right)\right]\mathrm{sech}^2\left[z/(2z_0)\right]$, with $z_0$ being the scale-height of the disk model of Sect.~\ref{sec:dynamicalForwardModeling}, thus accounting for the disk thickness, and iii) used the \citet{Hohl_1969} method to compute the baryonic potential in the disk midplane, by convolving this function with the deprojected baryonic surface density.\par

The total potential in the midplane of the disk also includes the contribution of the dark matter distribution. Thus, we computed the potential of the NFW profile of the \texttt{DysmalPy} model of Sect.~\ref{sec:dynamicalForwardModeling} and added it to the axisymmetric component of the baryonic potential. The resulting total potential at the midplane of the disk is presented in Fig.~\ref{fig:baryonSDPotentialAndTorque}.\par

In this derived potential, a bisymmetric perturbation appears, in the form of an elongated pattern along the same direction as the bar identified in the deprojected baryonic surface density. This is a typical feature for barred galaxies, in which the potential minimum (in the azimuthal direction) is found along the bar. This non-axisymmetric feature can be quantified through the Fourier decomposition of the potential, and the study of the azimuthal $m\geq1$ modes. A qualitative overview of the resulting non-axisymmetric part of the potential is presented in the third panel of Fig.~\ref{fig:baryonSDPotentialAndTorque}, with an apparent bisymmetric feature, namely, lower (higher) values along the major (minor) axis of the bar, and additionally an $m=1$ perturbation at larger radii highlighting the lopsided nature of the outskirts of the galaxy.\par

\begin{figure}
    \centering
	\includegraphics[width=\columnwidth]{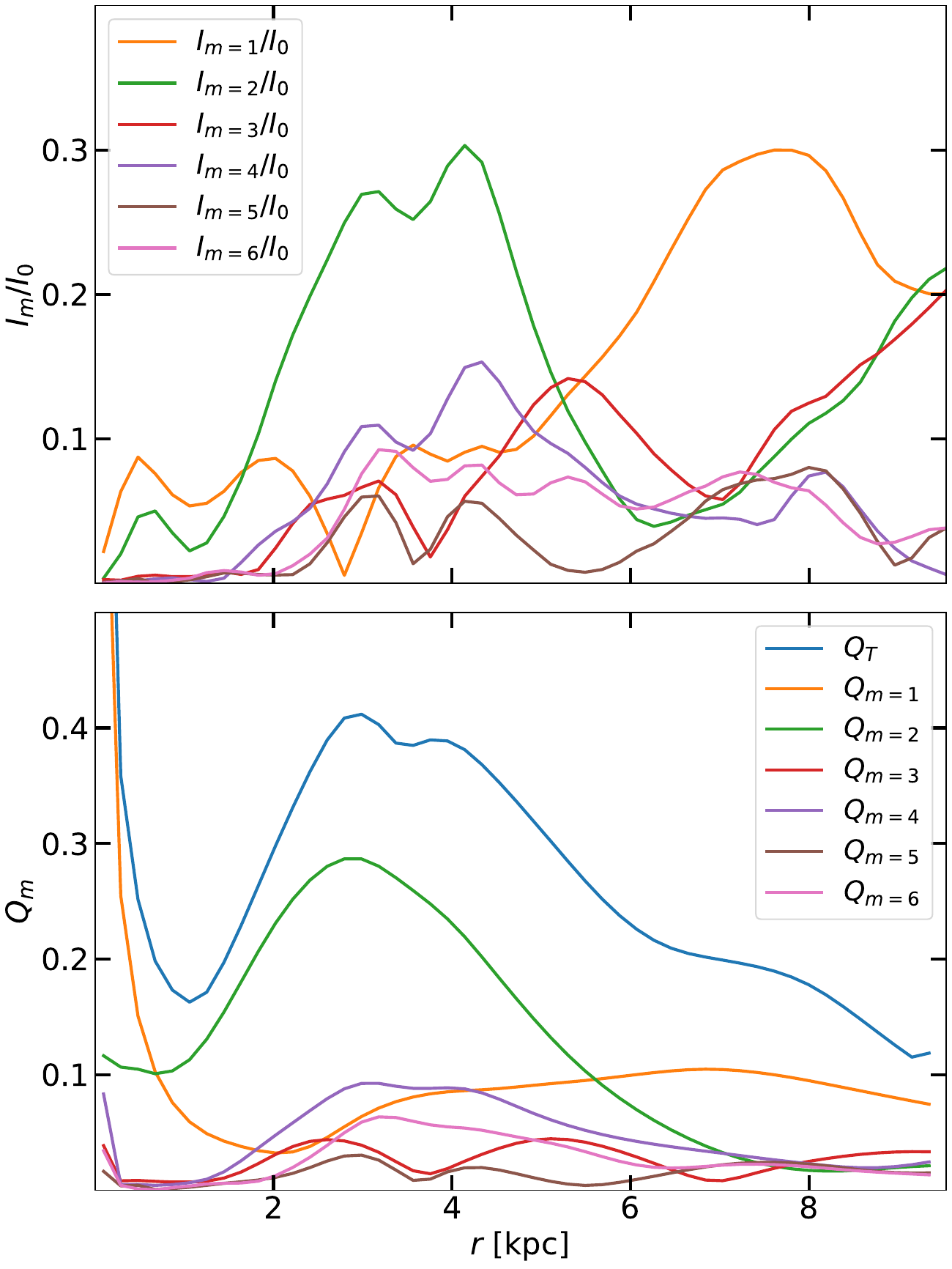}
    \caption{Mode amplitudes from the Fourier decomposition of the deprojected baryonic surface density of G4\_38232 (top) and measure of the strength of the non-axisymmetric potential perturbation (bottom). A clear peak of the $m=2$ mode amplitude is identified in the Fourier decomposition plot with an equivalent peak in the strength of the $m=2$ potential perturbation, quantified as the ratio of the azimuthally maximum tangential force from each mode over the radial force from the axisymmetric part of the potential \citep{Combes_1981}. At large radii, the $m=1$ mode is dominant in both cases.}
    \label{fig:non-axisymmetricPerturbation}
\end{figure}

A quantification of the strength of these non-axisymmetric features is presented in Fig.~\ref{fig:non-axisymmetricPerturbation}. The top panel shows the decomposition of the baryonic surface density used in the derivation of the potential estimate. We find a dominant amplitude in the normalized $m=2$ mode, characteristic of a bar structure, followed by a prominent $m=1$ amplitude at the region beyond the bar radius, stemming from the asymmetry in the mass distribution of the disk. This asymmetry could be a consequence of ongoing gas accretion \citep{Bournaud_2005, Jog_2009}, which is known to take place at high-$z$ \citep[e.g.,][]{Keres_2005, Dekel_2009, van_de_Voort_2011, Nelson_2013}, as well as evidence of efficient outwards angular momentum transport \citep{Saha_2014}.\par

\begin{figure}
    \centering
	\includegraphics[width=\columnwidth]{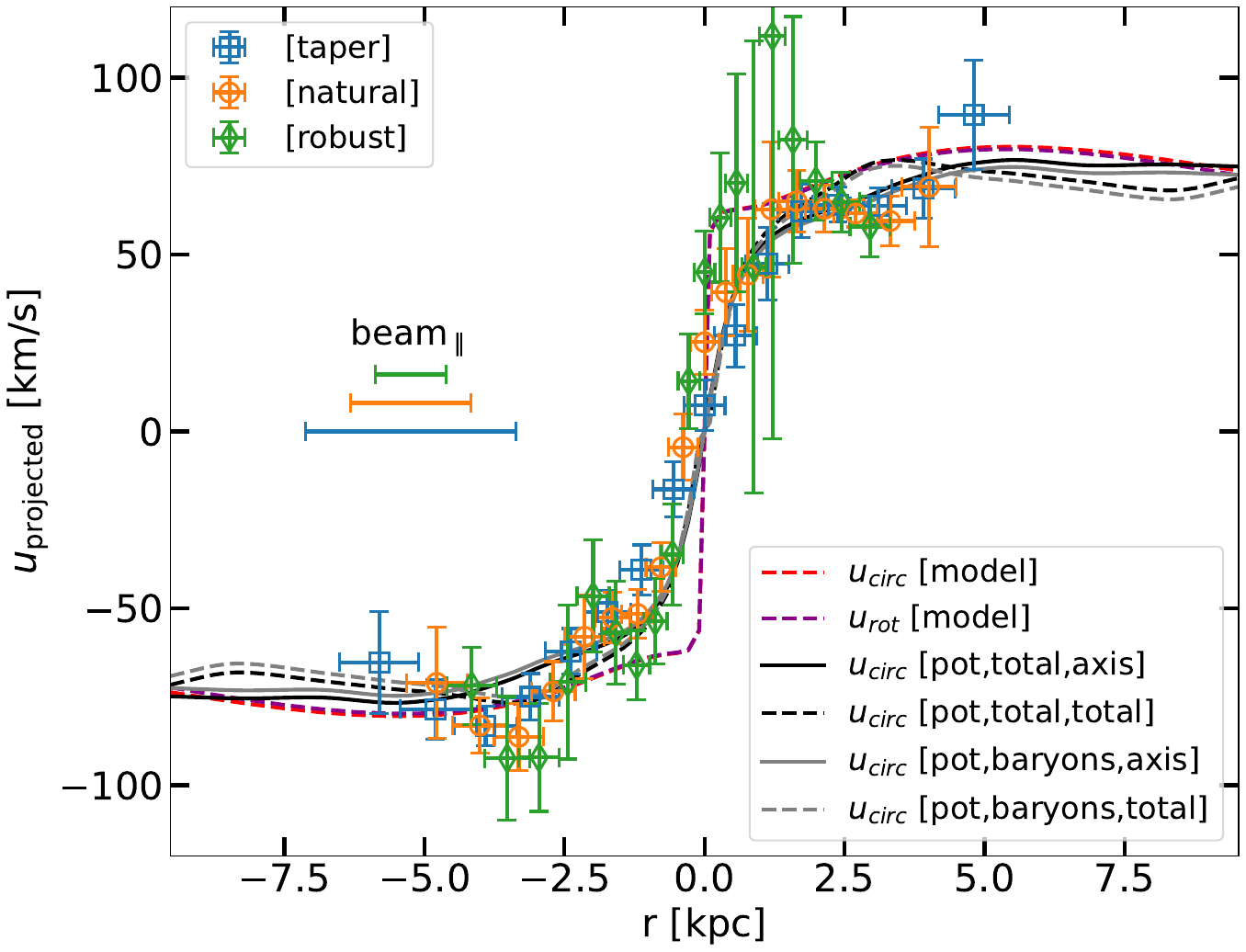}
    \caption{Expected rotation curve of G4\_38232 based on the estimated potential at the disk midplane. The projected intrinsic circular (red) and rotational (purple) velocity of the \texttt{DysmalPy} model, without beam-smearing, and the likewise projected circular velocity stemming from the axisymmetric (solid) and total (dashed) estimated baryonic (gray) and total (black) potentials. The observed rotation is plotted for our tapered (blue), naturally-weighted (orange) and robust (green) data. We find that the projected velocity of the fitted model can be very well reproduced by the circular velocity of the baryonic potential, leaving little-to-no room for a significant contribution from an additional component, such as the dark matter.}
    \label{fig:potentialRotationCurve}
\end{figure}

The perturbations to the baryonic mass distribution are also reflected in the potential. A quantification of the strength of the potential perturbations can be achieved through the comparison of the maximum tangential force stemming from each $m\geq1$ Fourier mode, to the respective radial force of the axisymmetric component, through the computation of the $Q_m$ parameter as a function of radius \citep{Combes_1981}. The strength of the total non-axisymmetric perturbation can be also assessed in a similar fashion, this time taking into account the total of the $m\geq1$ terms \citep{Combes_1981}. As indicated in the bottom panel of Fig,~\ref{fig:non-axisymmetricPerturbation}, the strongest non-axisymmetric potential perturbation stems from the $m=2$ mode, as expected for a barred spiral. Beyond the bar region, $a_{\mathrm{bar}}\approx4.2$\,kpc, the contribution of the bisymmetric mode steadily decreases and is eventually surpassed by that of the lopsidedness ($m=1$) of the outer disk.\par

We estimated the strength of the bar by computing the typically used parameter $Q_b$ \citep{Buta_2001}. This parameter is based on the strength of the non-axisymmetric perturbation and its value for G4\_38232 is $Q_b\approx0.37$ (see Appendix~\ref{sec:barStrengthEstimate}). A comparison with similar measurements for cosmic noon or lower redshift galaxies \citep{Kim_2021, Kalita_2026} indicates that G4\_38232 hosts a strong bar, while based on the local classification scheme of \citet{Buta_2001} this target is a class 4 (out of 6) barred galaxy.\par

With the evidence of a very limited dark matter content from our \texttt{DysmalPy} forward modeling, we assessed the agreement between the observed rotational profile of G4\_38232 and the one inferred from the estimated baryonic and total potentials. This comparison is motivated by the fact that the dark matter content of our target is lower than that of typical, massive galaxies, namely, $f_{\mathrm{dm}}(<R_e)\approx38\%$ at $z\sim1$ \citep{Wuyts_S_2016, Genzel_2017, Genzel_2020, Bouche_2022, Price_2021, Nestor_Shachar_2023}. We computed the circular velocity due to the axisymmetric part of these estimated potentials, projected it on the sky plane and compared it to the observed rotation and the projected expectation based on our axisymmetric \texttt{DysmalPy} model. The comparison is shown in Fig.~\ref{fig:potentialRotationCurve}.\par

We find a very good agreement between the two types of rotational profiles. However, in the innermost regions, the PSF of the \textit{JWST} NIRCam image used to derive the potential limits the maximum steepness of the rotation curve that can be recovered. Thus, the rotation curve inferred directly from the potential deviates slightly from our \texttt{DysmalPy} model. At larger radii, however, the agreement between the two is apparent, also in the case of the circular velocity with the contributions of all non-axisymmetric terms of the potential taken into account (e.g., in the negative part of the rotation curve). While, such an agreement is, in principle, expected given the scaling of the surface density distribution used in deriving the potential, the observed stellar continuum offers a more direct insight; it may deviate from the S\'ersic profile assumed in our parametric model and it explicitly includes the contributions of non-axisymmetric features, in this case, the bar, spiral arms and clumps.\par

\subsection{Bar pattern speed; a fast bar at high redshift}
\label{sec:barPatternSpeed}

\begin{figure*}
    \centering
	\includegraphics[width=2.0\columnwidth]{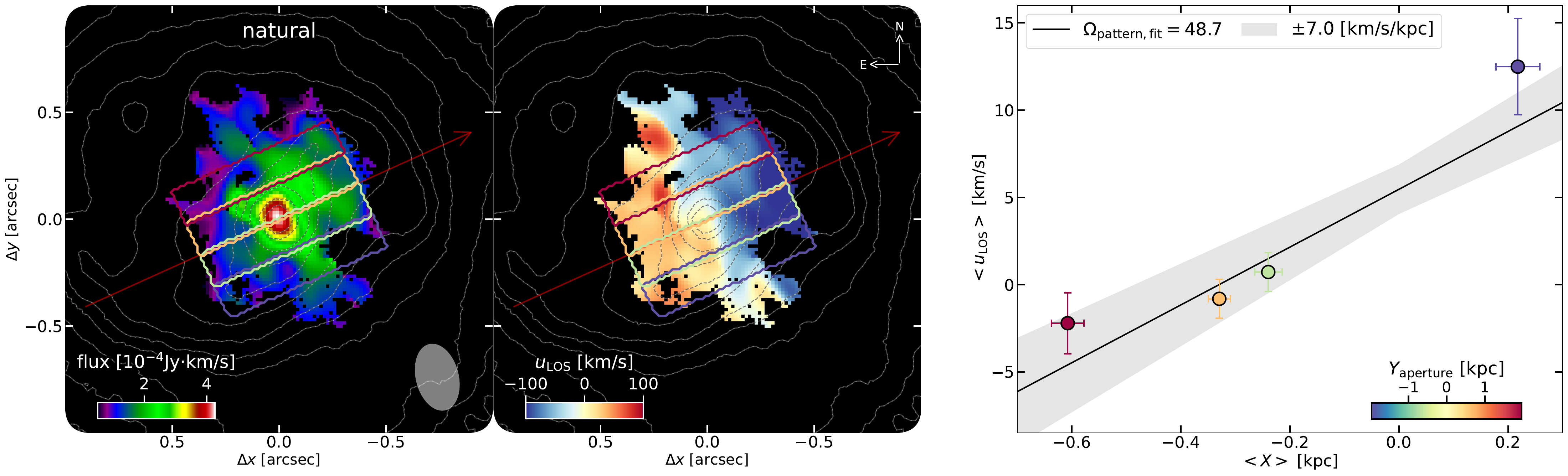}
    \caption{Overview of the application of the \citep*{Tremaine_1984} method: the slits overlaid on the CO(4-3) flux (left) and LOS velocity (middle) maps along with the flux-weighted LOS velocity as a function of the flux-weighted position along the slits. The color-coding of the slits is with respect to the projected offset from the disk major axis (plotted with a red arrow). The linear relation between the flux-weighted LOS velocity and position indicate the presence of a coherent solid-body rotating pattern, i.e., the bar, the pattern speed of which is determined by the slope of the fitted line plotted in black. The evolution with respect to the offset from the major axis is coherent, even for the central slits, which could be affected by the kinematics of the central regions.}
    \label{fig:TremaineWeinbergMethod}
\end{figure*}

An important property of bars is the pattern speed, namely, the angular velocity with which the bar structure rotates in its host disk. Its value is connected with the presence and loci of resonances which determine to a great extent the dynamics of both the stellar and gaseous components \citep[e.g.,][]{Athanassoula_1992a, Athanassoula_1992b, Wada_1994, Buta_1996, Bournaud_2002, Kim_2012b}.\par

The bar pattern speed is additionally by itself interesting since it is related to a disagreement between simulations and observations, even in the local Universe. While bars in most cosmological simulations appear to be slowly rotating (\citet{Algorry_2017, Peschken_2019, Habibi_2024}, however, see also \citet{Fragkoudi_2021, Merrow_2026} who find fast bars in the AURIGA simulations \citealt{Grand_2017}), direct observational measurements of bar pattern speeds, indicate a prevalence of fast-rotating bars (\citet{Aguerri_2003, Corsini_2011, Aguerri_2015, Cuomo_2019b, Cuomo_2020}, however, see also \citet{Geron_2023} who find slow bars in $\approx2/3$ of their sample). The pattern speed of a bar, is usually quantified through $\mathcal{R}=R_{\mathrm{CR}}/a_{\mathrm{bar}}$, with $R_{\mathrm{CR}}$ being the corotation radius and $a_{\mathrm{bar}}$ the semimajor axis of the bar. Since the potentially bar supporting $x1$ orbits can extend up to the corotation radius (with $x1$ becoming elongated perpendicular to the bar at even larger radii) \citep{Contopoulos_1980a}, the value of $R$ is expected to be $\mathcal{R}\geq1$, with the threshold value separating slow and fast bars being $\mathcal{R}\approx1.4$ (a fast bar has $\mathcal{R}<1.4$, while a slow one $\mathcal{R}>1.4$) \citep{Debattista_2000}.\par

In the course of its evolution, a bar is expected to exchange angular momentum with the dark matter and outer disk (losing angular momentum) and slow down \citep[e.g.,][]{Debattista_1998, Debattista_2000, Athanassoula_2003}. Thus, the typical theoretical expectation would be that local bars are slowly rotating, having had a lot of time to interact with their host dark matter halo and having lost a fraction of their initial angular momentum. In contrast, observational evidence show that local bars are fast-rotating, with $\mathcal{R}<1.4$, and in some cases even $\mathcal{R}<1$ considered to be short-lived configurations termed as ``ultra-fast'' bars \citep[e.g.,][]{Aguerri_2015, Guo_2019}.\par

Recent theoretical works have updated this picture, possibly reconciling the differences between observations and simulations. In the case of idealized simulations, a significant gas component has been shown to disrupt the mechanism through which the angular momentum of the bar is absorbed by the dark matter halo, allowing for a bar to remain fast \citep{Beane_2023}. Earlier studies have also identified the role of a high gas fraction in preventing the bar from slowing down \citep{Athanassoula_2003, Athanassoula_2014}. Additionally, in state-of-the-art, high-resolution, cosmological, zoom-in simulations, bars have been shown to remain fast until local times, as a result of the baryon-dominated nature of their host disks \citep{Fragkoudi_2021, Merrow_2026}.\par

At higher redshifts measurements of bar pattern speeds are scarce. The highest redshift direct pattern speed measurement to date using the Tremaine-Weinberg method, is that of the very massive J0107a at $z\approx2.5$, which was shown to host a fast bar by \citet{Huang_2025}. Indirect measurements of how fast bars are in intermediate redshifts were carried out by \citet{Perez_2012}, who estimated $\mathcal{R}$ in $44$ barred spirals at $z\leq0.8$ using the loci of the outer rings. From the fact that their presence is expected in the region of the OLR \citep{Buta_1996} and assuming a flat rotation curve, these authors found all cases to be compatible with fast rotating bars.\par

\subsubsection{Tremaine-Weinberg method; direct measurement}
\label{sec:tremaineWeinbergMethod}

The \citet*{Tremaine_1984} method is the only direct and the most commonly used method to constrain bar pattern speeds \citep[e.g.,][]{Merrifield_1995}. It works by identifying the signatures of a solid-body-like rotating pattern, in the surface density-weighted LOS velocities and positions in slits parallel to the major axis of the galaxy. The assumptions made in deriving the Tremaine-Weinberg method formulation, or in other words the requirements for the application of this method, are the following: i) the disk must be flat (i.e., not warping), ii) the pattern must have a well-defined angular velocity (pattern speed), and iii) the tracer must obey the continuity equation \citep{Tremaine_1984}.\par

To apply the Tremaine-Weinberg method to G4\_38232, we used the naturally-weighted data product, taking advantage of the better spatial resolution. We used 4 slits oriented parallel to the PA of the galaxy, each 0\farcs8 wide and $(1/2)\times\mathrm{FWHM}_{\mathrm{beam}}$ thick. We computed the mass-weighted average LOS velocity and position along each slit, in the form of the flux-weighted average LOS velocity and projected position along the major axis of the pixels of our data product which fell within the bounds of the slit. An overview of the used slits and resulting positions and LOS velocities is presented in Fig.~\ref{fig:TremaineWeinbergMethod}.\par

We find that the data points from our $4$ slits appear to follow a linear relation. Thus, there appears to be a well-defined pattern speed detected in our data. We constrain the pattern speed of the bar by fitting a first degree polynomial to these data points, resulting in an estimate of $\Omega_{\mathrm{pattern}}\approx49\pm7$\,km/s/kpc. The uncertainty in this pattern speed estimate reflects only the formal fitting uncertainty, while the real uncertainty considering also the errors in the adopted disk position angle and inclination are expected to be quite significant \citep[e.g.,][]{Corsini_2011}. Nevertheless, this is a direct estimate of the angular velocity with which the pattern (bar) rotates.\par

While this method should in principle be applied to the stellar component of the disk, it is being routinely applied to gaseous tracers, namely, molecular (CO) and ionized ($\mathrm{H\alpha}$) gas observations. Applications to such tracers, can result in artificially larger pattern speeds as consequence of false signal due to the limited physical coverage or clumpy nature of the tracer \citep{Williams_2021}. Furthermore, the Tremaine-Weinberg method is expected to be most efficient in the cases of bars with an observed PA close to (but not coinciding with) the PA of the host disk, and moderately inclined (but not face-on) galaxies \citep[e.g.,][]{Debattista_2003, Corsini_2011, Borodina_2023}.\par

At high redshift, \citet{Roshan_2025} showed that the Tremaine-Weinberg method can accurately measure the pattern speed of a bar provided that: i) the projected length of the bar along the disk minor axis allows for at least three independent slits, and ii) the extent of the data allows for a slit length of $\sim2.5$ times the projected length of the bar along the disk major axis. In our case, none of these two requirements are satisfied. In addition, our CO data trace molecular gas in a highly star-forming galaxy (with a depletion timescale of $t_{\mathrm{depl}}=M_{\mathrm{gas}}/\mathrm{SFR}\approx1.2$\,Gyr), thus, the continuity equation is not strictly obeyed. Finally, the possibly clumpy nature of our tracer, as indicated by the available UV/optical imaging, could introduce additional uncertainties in the recovered pattern speed. Thus, the systematic uncertainties in this pattern speed estimate are significant.\par

\subsubsection{Hydrodynamical features matching; indirect measurement}
\label{sec:hydrodynamicFeaturesMatching}

Since the bar is expected to rotate as a fixed pattern in the plane of the galactic disk, an indirect measurement of the pattern speed of the bar can be made through the comparison of hydrodynamical features in the response of the gaseous disk to the barred potential of galaxy. This method has been successfully applied in local galaxies, through the use of gas response models to observationally derived galactic potentials \citep[e.g.,][]{Laine_1996, Laine_1998, Rautiainen_2008, Patsis_2009}, or even to typical, analytical bar potentials (e.g., to the simple and general $\Phi_b(R,\phi,t)=\Phi(R)\cos\left[2(\phi-\Omega_pt)\right]$ by \citet{Lin_2013} for the local barred galaxy NGC~1097). Additionally, the matching of stellar morphologies has also been used in a similar manner \citep[e.g.,][]{Rautiainen_2008, Patsis_2009, Kalapotharakos_2010b}.\par

In order to constrain the bar pattern speed of G4\_38232 using this method, we ran a series of isothermal gas responses to the estimated potential of this target, and determined the pattern speed in which the resulting gas morphologies better resembled the observed morphologies in the short-wavelength, higher-resolution \textit{JWST} NIRCam images. To that end, we used the adaptive mesh, hydrodynamical code \texttt{RAMSES} \citep{Teyssier_2002}, with a custom patch based on that of \citet{Pastras_2022}, enabling the introduction of a user defined potential in the form of a set of Fourier mode amplitudes, however in our case interpolated with cubic splines, in a similar fashion to \citet{Kalapotharakos_2010b}. For each simulation, the non-axisymmetric part of this potential was rotated at a constant angular speed, $\Omega_{\mathrm{pattern}}$, and the resulting gas response was evaluated.\par

\begin{sidewaysfigure*}
    \centering
    \includegraphics[width=0.714\columnwidth]{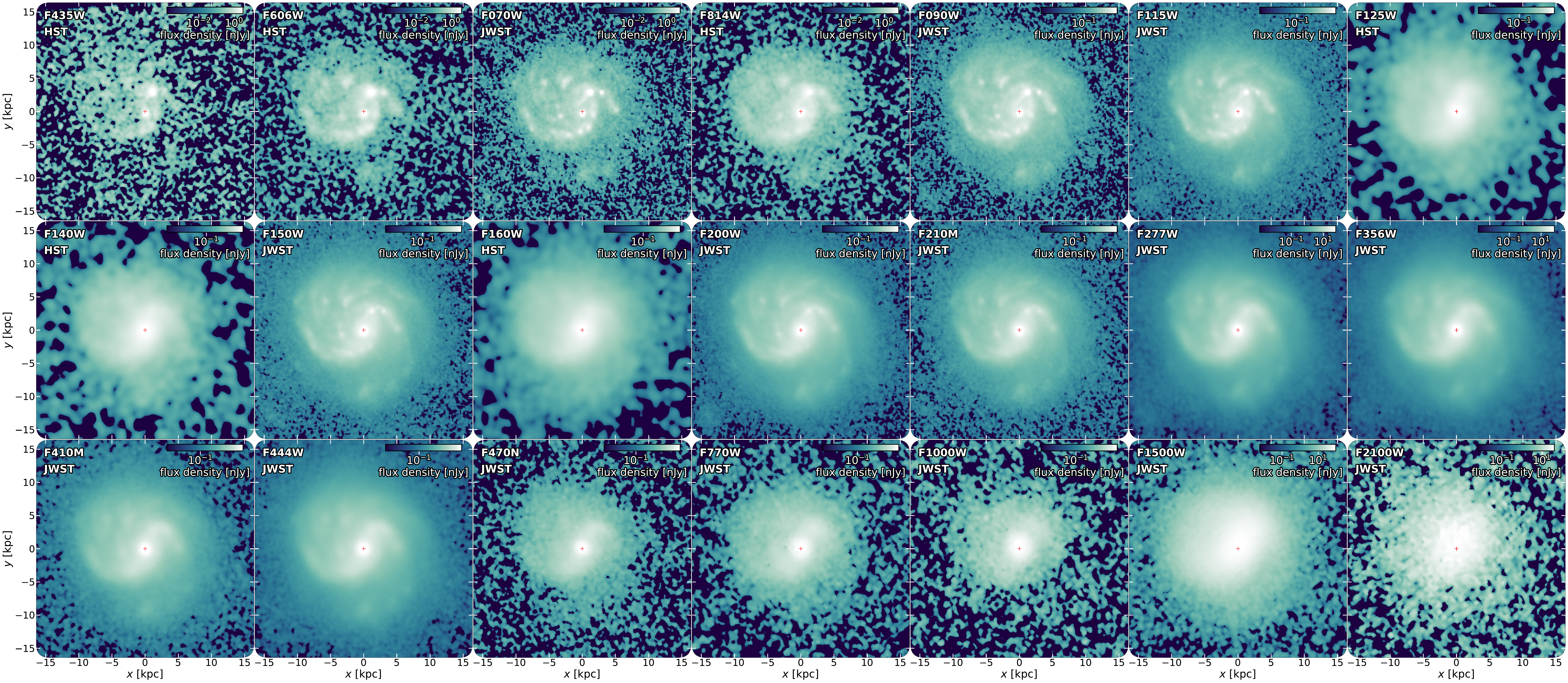}
    \caption{Deprojected imaging overview, using all available imaging bands, with data from \textit{HST} (ACS/WFC3) and \textit{JWST} (NIRCam and MIRI).}
    \label{fig:imagingOverviewDeprojected}\vspace{10pt}
    \includegraphics[width=\columnwidth]{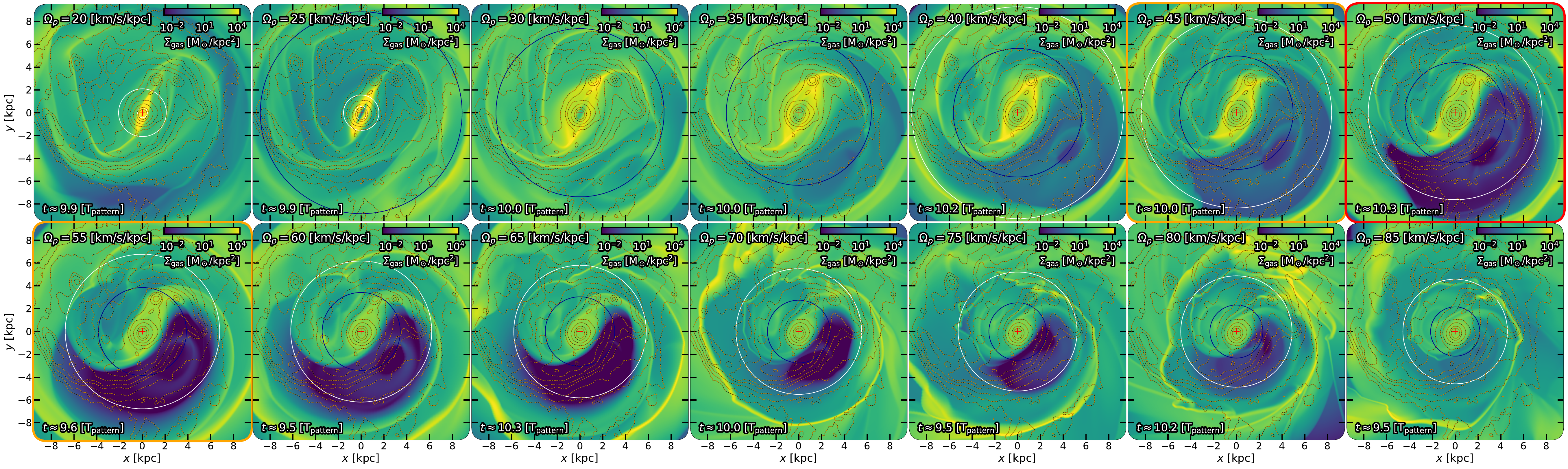}
    \caption{Gas surface density maps of the isothermal responses to the estimated potential in the midplane of G4\_38232 for pattern speeds $20\leq\Omega_{\mathrm{pattern}}\leq85$\,km/s/kpc, with overlaid contours of the NIRCam F150W image, probing the restframe optical continuum and within the bandwidth of which $\mathrm{H\alpha}$ falls. White circles mark the radii of the inner and outer Lindblad resonance (if they exist), derived through frequency analysis on the axisymmetric potential, while the dark blue circle indicates the corotation radius. In the comparison between the observed and simulated morphologies, emphasis is put in the agreement in and around the bar region, in which the dynamics are expected to be mostly affected by the barred perturbation. The simulation with the best-matching pattern speed is marked with a red outline (top rightmost panel) and those within the estimated uncertainty marked with an orange one.}
    \label{fig:isothermalResponsesGas}
\end{sidewaysfigure*}

The initial conditions of our models comprised an isothermal gaseous disk with homogeneous density, in rotational equilibrium in the presence of the axisymmetric part of the potential. We set the total mass of the gaseous disk to the gas mass inferred from the observed CO(4-3) flux. However, we note that in such isothermal responses this value serves as a simple scaling factor since the hydrodynamical equations are invariant under rescaling of the initial gas density \citep[e.g.,][]{Sormani_2023}. The non-axisymmetric Fourier terms of the potential were introduced gradually in the course of three pattern rotations, $T_{\mathrm{pattern}}=2\pi/\Omega_{\mathrm{pattern}}$, to avoid abruptly disturbing the gaseous distribution of the disk. The isothermal responses were run for at least another $\approx10$ pattern rotations after the introduction of the full potential.\par

We present an overview of all available multi-band imaging for G4\_38232 in Fig.~\ref{fig:imagingOverviewDeprojected}, alongside the gas distributions of the isothermal response snapshots after $\approx7$ pattern rotations with the full potential in Fig.~\ref{fig:isothermalResponsesGas} (see also Fig.~\ref{fig:isothermalResponsesStars} for the respective stellar responses). In order to facilitate the comparison of our response models with the observed morphologies we have overlaid the contours of the NIRCam F150W filter on top of Fig.~\ref{fig:isothermalResponsesGas}. Given the optical restframe wavelengths (tracing young stellar populations) probed by this filter (with $\mathrm{H\alpha}$ falling within its bandwidth) and the excellent spatial resolution, this image serves as an apt comparison to our isothermal gas models.\par

In this comparison emphasis is put on the features near the region of the bar, where the dynamics are dominated by the contribution of the barred component of the potential. While the non-axisymmetric contributions of the spiral arms are also imprinted in the estimated potential, it is not clear whether the pattern speed of the spiral pattern is the same as that of the bar \citep[e.g., as is the case for ``chaotic spirals''][]{Romero-Gomez_2006, Patsis_2006, Romero-Gomez_2007, Athanassoula_2009a, Athanassoula_2012}, or even if there is a well-defined pattern speed for the arms in the first place \citep[e.g.,][]{Toomre_1981, Dobbs_2014}. Thus, the features we aim to reproduce are mainly those that stem from the dynamics of the bar itself.\par

With these in mind and looking at the morphologies of the response models of Fig.~\ref{fig:isothermalResponsesGas}, we identify a clear pattern; while the models with low pattern speeds ($\Omega_{\mathrm{pattern}}<40$\,km/s/kpc) produce bar-like response features that are more extended than the observed bar region, high pattern speed models ($\Omega_{\mathrm{pattern}}>60$\,km/s/kpc) result in a very compact bar region, which does not provide a good match to the observed morphologies. The intermediate pattern speeds, however, do result in reasonable responses; in these cases the extent (and shape) of the bar region matches that observed in F150W. Additionally, in the models with pattern speeds closer to $\Omega_{\mathrm{pattern}}\approx50$\,km/s/kpc, we find that the regions close to the bar from which the spiral arms emanate (one at the top right and two at the bottom left side of the bar) closely resemble the respective observed regions.\par

Based on the good correspondence between the observed and simulated features, we conclude that the pattern speed of the bar must be $\Omega_{\mathrm{pattern}}\approx50\pm5$\,km/s/kpc. This value is in excellent agreement with the estimate of the Tremaine-Weinberg method. While each method comes with its own caveats, this agreement suggests that the pattern speed of our galaxy is relatively well constrained to a value of $\Omega_{\mathrm{pattern}}\approx50$\,km/s/kpc.\par

Given the rotational profile constrained through forward modeling in Sect.~\ref{sec:dynamicalForwardModeling}, this bar pattern speed results in the corotation radius being at $R_{\mathrm{CR}}\approx4.4$\,kpc and consequently $\mathcal{R}=R_{\mathrm{CR}}/a_{\mathrm{bar}}\approx1.05$, indicating that our bar is a fast-rotating one ($\mathcal{R}<1.4$). This is in agreement with theoretical expectations for gas-rich \citep[e.g.][]{Athanassoula_2003, Athanassoula_2014, Beane_2023} and baryon dominated \citep[e.g.][]{Fragkoudi_2021, Merrow_2026} disks, as is the case for this galaxy.\par

A frequency analysis of the estimated potential for G4\_38232, i.e., a comparison between $\Omega(r)-\kappa(r)/2$, with $\kappa(r)$ being the epicyclic frequency of the estimated potential, and $\Omega_{\mathrm{pattern}}$, suggests the absence of an inner Lindblad resonance (ILR) for the constrained pattern speed. However, there is morphological evidence supporting the presence of an ILR; the tentatively identified bar lane at the northern leading side of the bar (see Sect.~\ref{sec:gasFunnelingParallelToTheBarLanes}) is off-centered with respect to the major axis of the bar, a configuration associated with a significant extent of the $x_2$ and $x_3$ families of periodic orbits \citep{Athanassoula_1992b, Patsis_2000}, the existence of which means that there is an ILR. Consequently, and modulo the caveats of identifying resonances through frequency analyses of the axisymmetric potential in the presence of perturbations with significant amplitudes, these conflicting indications highlight the uncertainty in the shape of the estimated potential at its very central regions.\par

It is also worth noting that the presence of a central SMBH with $M_{\mathrm{BH}}\sim10^8$\,M$_\odot$ (consistent with the relation between the stellar mass and black hole mass for local galaxies \citealt{Reines_2015, Greene_2020}), would also result in the presence of an ILR, at the same time having essentially no effect on the observed kinematics given the observational limitations. Thus, in principle, the absence of an ILR for the estimated potential and pattern speed could serve as an indirect indication of the presence of a central SMBH. However, the overall uncertainties make such a discussion speculative.\par

\subsection{Summary of the properties of G4\_38232}
\label{sec:summaryOfProperties}

\begin{table*}
    \setlength{\tabcolsep}{1.75pt}
	\centering
	\caption{Global galaxy properties inferred for G4\_38232.}
	\label{tab:galaxyProperties}
	\begin{tabular}{cccccccccccccc} 
		\hline
        \multirow{2}{*}{$z$} & SFR & $\mathrm{PA}_{\mathrm{disk}}$ & $\mathrm{i}_{\mathrm{disk}}$ & $R_{\mathrm{e}}$ & $u_{\mathrm{circ}}(R_{\mathrm{e}})$ & $\sigma_0$ & \multirow{2}{*}{$\log\left(\frac{M_{\mathrm{\star}}}{M_\odot}\right)$} & \multirow{2}{*}{$\log\left(\frac{M_{\mathrm{gas}}}{M_\odot}\right)$} & \multirow{2}{*}{$f_{\mathrm{gas}}$} & \multirow{2}{*}{$\log\left(\frac{M_{\mathrm{baryons}}}{M_\odot}\right)$} & \multirow{2}{*}{$B/T$} & \multirow{2}{*}{$f_{\mathrm{DM}}(<R_{\mathrm{e}})$} & \multirow{2}{*}{$Q_{\mathrm{gas}}$} \\
        & [M$_\odot$/yr] & [deg] & [deg] & [kpc] & [km/s] & [km/s] & & & & & & & \\
		\hline
		1.1159 & 36$\pm$15 & -66$\pm$9 & 20$\pm$5 & 4.3$\pm$0.2 & 234 & 20 & 10.82$\pm$0.11 & 10.65$\pm$0.05 & 0.4$\pm$0.07 & 10.96$\pm$0.1 & 0.16$\pm$0.07 & 0.04$^{+0.22}_{-0.03}$ & 0.3 \\
		\hline
	\end{tabular}
    \tablefoot{Redshift (column 1), star-formation rate (column 2), disk position angle, inclination and effective radius (columns 3-5), circular velocity at a disk effective radius (column 6), velocity dispersion (column 7), logarithm of the stellar and molecular gas mass, derived from integrated SED fitting and the CO(4-3) flux, respectively, (columns 8 and 9), gas fraction (column 10), logarithm of the dynamically inferred baryonic mass (column 11), bulge-to-total ratio (column 12), dark matter fraction within a disk effective radius (column 13) and gas Toomre Q stability parameter (column 14).}
\end{table*}

\begin{table}
    \setlength{\tabcolsep}{1.25pt}
	\centering
	\caption{Bar related properties constrained for G4\_38232.}
	\label{tab:barProperties}
    \begin{tabular}{cccccccc} 
		\hline
        $\mathrm{PA}_{\mathrm{bar}}$ & $a_{\mathrm{bar}}$ & $\Omega_{\mathrm{pattern}}$ & $R_{\mathrm{CR}}$ & $\mathcal{R}$ & \multirow{2}{*}{$Q_{\mathrm{b}}$} & \multirow{2}{*}{$R_{\mathrm{ELN}}$} & $f_{\mathrm{disk}}$ \\
        \hspace{0pt}[deg] & [kpc] & [km/s/kpc] & [kpc] & ($=R_\mathrm{CR}/a_{\mathrm{bar}}$) & & & $(<2.2R_d)$ \\
		\hline
		-32.5 & 4.2 & 50 & 4.4 & 1.05 & 0.37 & 0.7 & 0.8 \\
		\hline
	\end{tabular}
    \tablefoot{Position angle of the bar (column 1), in-plane bar semimajor axis (column 2), bar pattern speed (column 3), corotation radius (column 4) and ratio of the corotation radius over the semimajor axis of the bar (column 5), bar strength (column 6), and values of the ELN (column 7) and disk fraction (column 8) bar instability criteria.}
\end{table}

We have constrained an array of properties for G4\_38232, ranging from basic ones typically constrained for high-$z$ disks to some specifically relevant for barred cases. We summarize these properties in Tab.~\ref{tab:galaxyProperties} and Tab.~\ref{tab:barProperties}, respectively.\par

Our main conclusions about G4\_38232, are the following:

\begin{itemize}
    \item Our target is a massive ($\log(M_{\mathrm{baryons}}/\mathrm{M}_\odot)\approx10.96$), gas-rich ($f_{\mathrm{gas}}\approx40\%$), main sequence galaxy at $z\approx1.12$. It is highly star-forming, with a $\mathrm{SFR}\approx36$\,M$_\odot$/yr \citep{Jolly_2026, Chen_2026}. In its inner regions, it is particularly baryon dominated, with $f_{\mathrm{dm}}(<R_e)\approx4\%$ and in the center we identify indications of the presence of a modestly massive bulge ($B/T\approx0.16$). Its gaseous disk is unstable \citep{Behrendt_2015}, as indicated by the gas Toomre Q parameter of $Q_{\mathrm{gas}}\sim0.3$ and its clumpy restframe UV/optical morphology.
    \item Morphologically, it hosts a prominent bar structure in the central regions and spiral structure at larger radii. In the northeastern leading side of the bar, we identify a straight-line, dust lane shock, in correspondence with local barred spirals, in which dust lanes are routinely found at the leading sides of bars \citep[e.g.,][]{Athanassoula_1992b, Patsis_2000, Stuber_2023, Sormani_2023, Schinnerer_2023}.
    \item The bar of this galaxy is long ($a_{\mathrm{bar}}\approx4.2$\,kpc), as supported by the bar length distribution of cosmic noon barred spirals \citep{Guo_2025, Le_Conte_2025}. It rotates with an angular velocity of $\Omega_{\mathrm{pattern}}\approx50$\,km/s/kpc, placing the corotation radius at $R_{\mathrm{CR}}\approx4.4$\,kpc, meaning it is fast-rotating ($\mathcal{R}\approx1.05$), in agreement with expectations for gas-rich \citep{Athanassoula_2003, Athanassoula_2014, Williams_2021, Beane_2023}, baryon-dominated \citep[e.g.,][]{Athanassoula_2003, Fragkoudi_2021, Merrow_2026} disks.
    \item The low inner dark matter content of G4\_38232 \citep[e.g.,][]{Efstathiou_1982, Widrow_2008, Bland-Hawthorn_2023, Bland-Hawthorn_2024}, in addition to the modest central mass concentration \citep{Athanassoula_2005b, Saha_2018, Kataria_2018}, means that the bar could be the result of the intrinsic bar instability of the disk. Indeed, the ELN criterion \citep*{Efstathiou_1982} with a value of $R_{\mathrm{ELN}}\approx0.7$ corroborates this conclusion ($R_{\mathrm{ELN}}<1.1$) (however, see \citealt{Athanassoula_2008, Izquierdo-Villalba_2022, Romeo_2023, Bland-Hawthorn_2023, Ghosh_2023, Frosst_2026}), and so does the disk fraction criterion with $f_{\mathrm{disk}}(<2.2R_d)\approx0.8$ \citep{Widrow_2008, Fujii_2018, Bland-Hawthorn_2023, Bland-Hawthorn_2024, Frosst_2026}. However, the possibility of tidal triggering can not be excluded.
\end{itemize}

\section{Gas flows in G4\_38232}
\label{sec:gasFlowsInG438232}

We used the properties of G4\_38232 derived in the previous section to quantitatively study the gas flows in its disk. Taking advantage of the exquisite IRAM-NOEMA molecular gas data and the deep, multi-band imaging, we used three independent analyses to estimate the net gas flow rates in this high-$z$ barred galaxy.\par

\subsection{Residual velocities; signatures of noncircular motions}
\label{sec:residualVelocitites}

\begin{figure}
    \centering
	\includegraphics[width=\columnwidth]{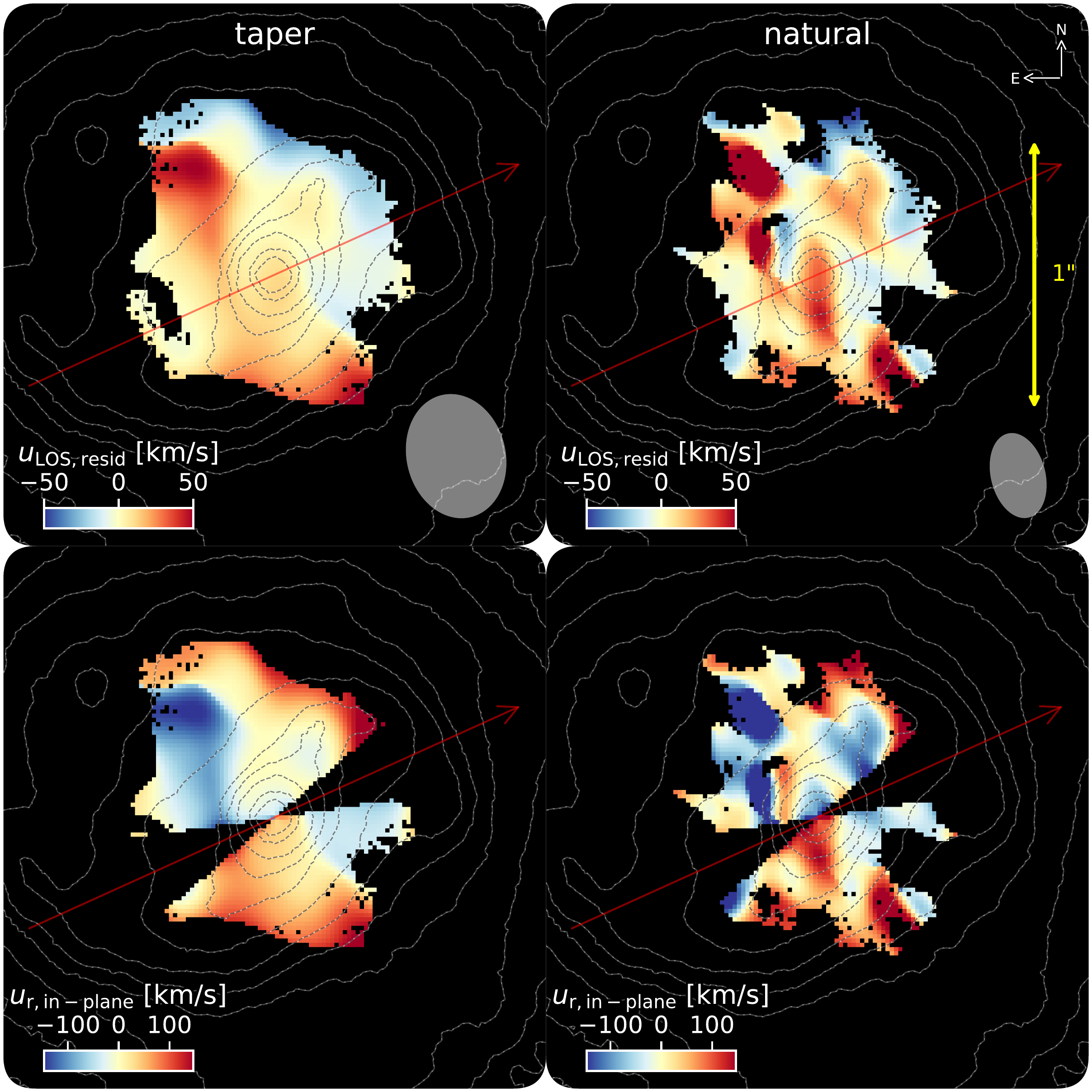}
    \caption{Observed residual LOS velocity (top) and inferred in-plane radial velocity (bottom) maps for the tapered (left) and naturally-weighted (right) data with overlaid contours of the reddest (F444W) NIRCam image. The red arrow indicates the disk major axis. The residual velocity maps are the result of the subtraction of our beam-smeared model of the axisymmetric rotation from the observed LOS velocity maps, while the inferred in-plane radial velocities are derived assuming that these residuals stem exclusively from in-plane radial motions. A comparison of the observed residual patterns with a simulation of a gas-rich, barred spiral (see Appendix~\ref{sec:simulationMockObservedVelocityResiduals}) offers insights on the possible nature of the in-plane noncircular motions. These motions are in broad agreement with bar-driven flows, with some additional features, and diluted by observational effects.}
    \label{fig:residualVelocityMaps}
\end{figure}

Our dynamical modeling of the major axis kinematics of G4\_38232 has provided us with a model of its rotation. With this model in hand, we applied the \citet{Genzel_2023} methodology to unveil the signatures of noncircular motions, namely, we produced beam convolved models of the fitted axisymmetric rotation and subtracted them from the observed velocity maps, resulting in the residual maps presented in Fig.~\ref{fig:residualVelocityMaps}.\par

Considering that the spiral arms are trailing \citep[e.g.][]{Contopoulos_1971, Toomre_1981, Romero-Gomez_2006, Romero-Gomez_2007, Athanassoula_2009a, Athanassoula_2012, Patsis_2006}, we conclude that G4\_38232 rotates counter-clockwise in the plane of the sky. Thus, since the northwestern part of the galaxy is approaching and the southeastern one is receding (see Fig.~\ref{fig:velocityAndVelocityDispersion}), its northeastern (southwestern) side is the near (far) side.\par

Considering the features present in both the tapered and naturally-weighted data, we can interpret the observed flow patterns through the comparison with the high-resolution simulation of a gas-rich barred spiral presented in Appendix~\ref{sec:simulationMockObservedVelocityResiduals}. Based on this model, the most prominent features and respective possible origin of the observed LOS velocity residuals are: i) the positive residuals at the northwestern side of the bar, possibly stemming from negative in-plane radial motions (local inflows) and/or slower rotation (shocks) at the leading side of the bar (smeared by the observational PSF), ii) the positive residuals at the southern central part of the galaxy, likely due to positive in-plane radial velocities (local outflows), and iii) a strongly positive residual pattern along the arm at the northeastern part of the FOV, possibly due to in-plane negative radial motions (local inflows). It is worth mentioning that the expected negative residual pattern at the southern leading part of the bar is not observed, possibly due to a weaker shock with relatively mild negative radial velocities and/or shocks (see also Sect.~\ref{sec:gasFunnelingParallelToTheBarLanes}). Alternatively, a significant amount of gas overshooting the central region, having been funneled towards it along the northern bar lane, could be the source of the positive residuals in the southern central part of the galaxy as well as the clumpy morphology in this region due to its collision with gas along the southern bar lane \citep[e.g.,][]{Sormani_2019b, Hatchfield_2021}.\par

We find that the observed residual patterns are in broad agreement with expectations for a barred spiral, albeit with some deviations from this expected picture. A top-level interpretation of these residual patterns is possible through the assumption of purely radial noncircular motions and the subsequent computation of the in-plane radial velocities. In general, deviations from axisymmetric rotation are not purely in the radial direction \citep[e.g.,][]{Warner_1973, van_der_Kruit_1978, Pastras_2025b}, but such a computation could provide an intuitive qualitative interpretation of the in-plane flows. Despite the possibility of considerable deviations in the azimuthal direction, the dominance of the contributions of radial motions in the regions away from the major axis can be a reasonable assumption (see e.g., Fig.~\ref{fig:simulationIdealizedVelocityResiduals} of Appendix~\ref{sec:simulationMockObservedVelocityResiduals}). In the bottom panels of Fig.~\ref{fig:residualVelocityMaps} we plot the resulting in-plane radial velocity maps, following \citet[][Eq.~E.33]{Pastras_2025b} for purely radial in-plane motions, i.e., a radial to tangential ratio of $u_{\mathrm{r}}/du_{\mathrm{\theta}}\to\infty$.\par

In the regions close to the major axis, small deviations in the azimuthal direction can result in significant LOS velocity residuals. In the same areas, the contributions of radial motions amount only to a small fraction of their in-plane values, possibly hindering the detection of the respective signatures in the residuals. To improve the robustness of our results, we masked the regions close to the major axis in which the signatures of in-plane radial motions would be $\leq10\%$ of their in-plane amplitudes, considering any residuals in these areas less likely to stem from such motions.\par

From the in-plane radial velocity maps we derived the first estimate of the gas flow rate for G4\_38232. We started by estimating the molecular gas mass using the CO(4-3) flux of each pixel with \citet[][Eq.~2]{Tacconi_2020}. We split the galaxy into annuli with a projected width of $(1/8)\times\mathrm{FWHM}_{\mathrm{beam}}$, and then used the inferred in-plane radial velocities and gas masses to compute the gas flow rate in each annulus, as:

\begin{equation}
    \dot{M}_{\mathrm{net}}=\frac{M\bar{u}_{\mathrm{r}}}{\Delta r}=\frac{1}{\Delta r}\left(\sum_{i=1}^{N_{\mathrm{cells}}}{m_{\mathrm{i}}u_{\mathrm{r},{\mathrm{i}}}}\right),
        \label{eq:residulasRadialFlowRate}
\end{equation}

where $i$ is the index of each pixel falling within the bounds of the annulus, $m_{\mathrm{i}}$ and $u_{\mathrm{r},{\mathrm{i}}}$ the gas mass and radial velocity of said pixel, respectively, and $\Delta r$ the radial size of the annulus. Having estimated the radial mass flow rate profile, we estimated the net flow rate as a function of radius by averaging the contributions of all annuli from the center of the galaxy up to each target radius. Since the inflow rate at each radius is an expression of the mass per unit time crossing the boundary of a cylinder with said radius, in a steady-state picture and from a purely hydrodynamical perspective, that is neglecting star-formation, for a certain amount of mass flow through the bounds of this cylinder to occur, an equal amount of mass must flow through every similar enclosed cylinder, up to the very center; otherwise there will be either an accumulation of gas or a depletion of the gas reservoir at some radius, invalidating the initial steady-state hypothesis. Thus, the inflow rate at the bounds of each annulus can be estimated through the averaging of the flow rates of all annuli enclosed within it.\par

\begin{figure}
    \centering
	\includegraphics[width=\columnwidth]{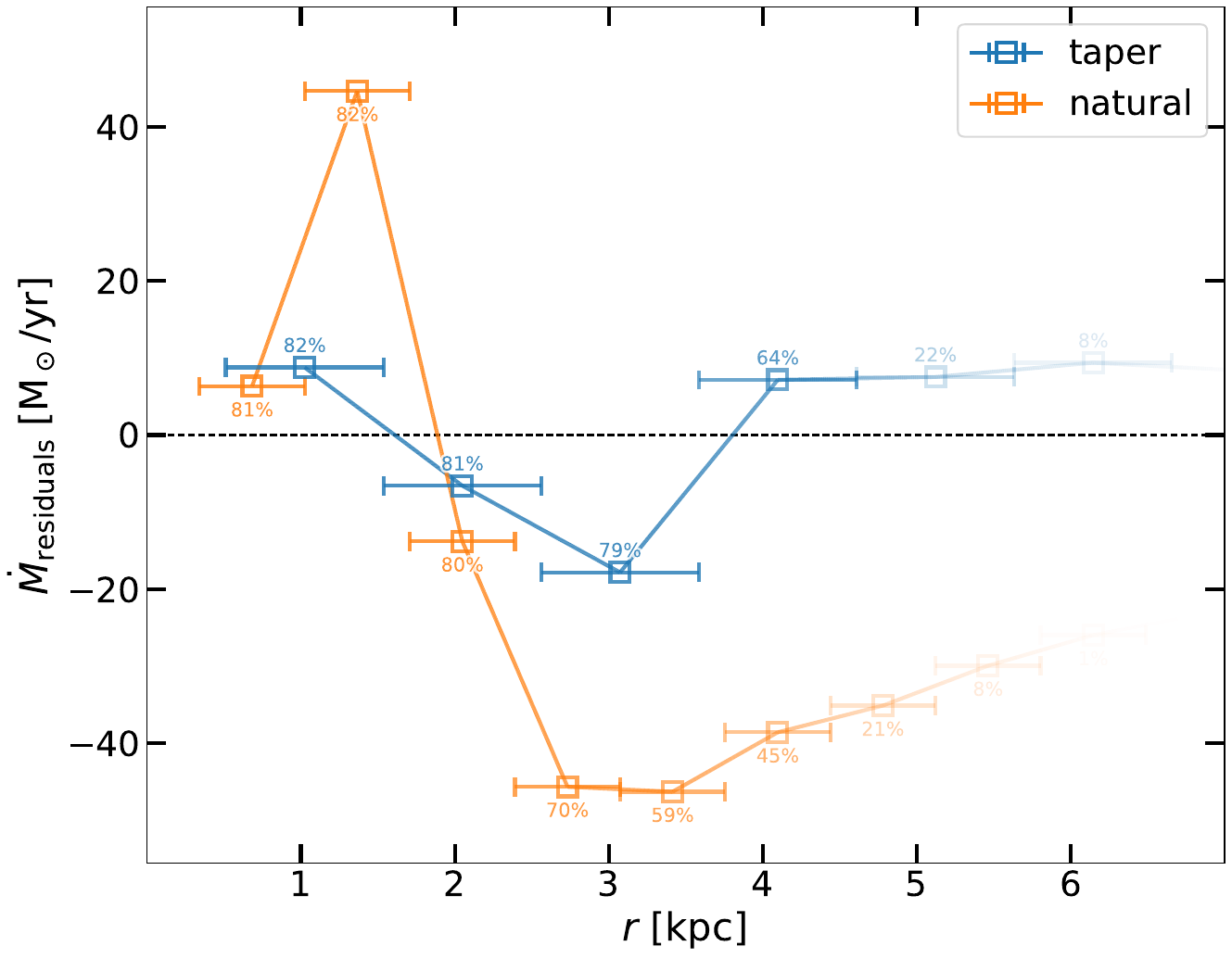}
    \caption{Radial gas flow rates based on the inferred in-plane radial velocity maps, assuming all LOS velocity residuals stem from in-plane radial motions. The fraction of the pixels in each elliptical annulus for which the in-plane radial velocity and molecular gas mass can be estimated is given on top of each data point. In the central regions, we find a net outflow in both of our data sets, succeeded by negative net radial flows at larger radii within the bar region. At even larger radii, our higher resolution data offer evidence of significant net inflows, while the result for our lower resolution data product corresponds to either no net gas transport or even modest net outflows. The fraction of pixels with meaningful information limits the extent up to which the estimated flows are meaningful, to $\approx4$\,kpc.}
    \label{fig:residualVelocityRadialFlowRates}
\end{figure}

We present the estimated net radial gas flow rates in Fig.~\ref{fig:residualVelocityRadialFlowRates}. Starting from the central regions, we first identify a net outflow, which turns into a net inflow at larger radii. After reaching a minimum value at a radius comparable to the bar radius, the net flow rate increases (becomes less negative). Given the declining pixel coverage fraction in the outer annuli, the radial flow rates can be constrained up to $\approx4$\,kpc, that is roughly the radius of the bar. In addition, we identify larger variations in the case of our higher resolution data compared to our tapered data. This is expected, since the more detailed view in the higher-resolution data enable the better recovery of the underlying noncircular motions.\par

All in all, according to this first estimate of the net gas flow rate and considering the values constrained for both data products, the net flow rate within the bar region is approximately $\dot{M}_{\mathrm{net}}\approx20-45$\,M$_\odot$/yr, of a similar order as the galaxy-integrated $\mathrm{SFR}\approx36$\,M$_\odot$/yr. In the central region, the net outflow identified in both data products could be an indication of the stalling of the inflowing gas streams due to the presence of an unresolved central structure, such as a nuclear ring or leading nuclear spirals \citep[e.g.,][]{Regan_2003, Combes_2002, Maciejewski_2004a, Maciejewski_2004b, Combes_2022}. It could also be the effect of a contribution from a possible AGN outflow, something less probable as such features are not expected to be prominent in molecular gas kinematics \citep[e.g.,][]{Herrera-Camus_2019, Barfety_2025}.\par

In the outer regions, in the case of our higher resolution data product the net gas flow remains significant ($\dot{M}_{\mathrm{net,natural}}\approx40$\,M$_\odot$/yr), while a modest net outflow is identified in our lower resolution data. Based on the higher resolution, we assume the former scenario to be more accurate. However, a picture in which high angular momentum gas flows outwards in the disk outskirts, enabling the damping of angular momentum and contraction of the gaseous component in the inner regions, can not be ruled out. This interpretation is also compatible with the prominent influence of the bar driving local outflows at large radii, and specifically beyond the corotation region, with the pattern \citep[e.g., spiral arms attached to the ends of the bar][]{Romero-Gomez_2006, Romero-Gomez_2007, Athanassoula_2009a, Athanassoula_2012} rotating faster than the underlying material of the disk.\par

With respect to this first methodology of estimating the net gas flow rate, we note the following caveats; i) we cannot recover the in-plane radial velocities in the masked regions along the major axis, so the contribution of these regions is implicitly neglected, ii) even for the regions in which an in-plane radial velocity value can be estimated, our assumption of purely radial noncircular motions is a simplistic approximation, and iii) the finite sensitivity of our observations leads to the incomplete (in the azimuthal sense) recovery of the molecular gas distribution, possibly biasing the inferred flow rates in two ways: creating a spurious net flow with some regions of the gaseous distribution being ignored, and making rates artificially decline in the outer regions due to the division by an increasing number of annuli which, however, offer no significant contributions.\par

\subsection{Torque modeling; bar-driven inflows}
\label{sec:torqueModeling}

We followed the methodology of \citep{Garcia-Burillo_2005} to estimate the mass flow rate driven by the torques of the non-axisymmetric component of the estimated potential for G4\_38232.\par

We first deprojected the gas mass distribution, in the similar fashion as we did in Sect.~\ref{sec:hydrodynamicFeaturesMatching} for the stellar continuum. We then used the non-axisymmetric Fourier modes of the potential to compute the torques per unit mass at the centers of the pixels of the deprojected gas surface density maps ($\vec{\tau}=-\vec{r}\times\vec{\nabla}\Phi$). An overview of the computed torques is presented in the rightmost panel of Fig.~\ref{fig:baryonSDPotentialAndTorque}.\par

In this torque map, we identify a quadrupole pattern of negative torques ahead and positive behind, with respect to the rotation of the galaxy, the local potential minima in the azimuthal direction. These torques enable the driving of gas inflows by the bar, offering a mechanism through which the gas can lose angular momentum and migrate toward the central regions. This mechanism, which has been studied from both a theoretical \citep[e.g.,][]{Wada_1994, Combes_2002} as well as an observational perspective \citep[e.g.,][]{Garcia-Burillo_2005, Audibert_2019}, can be used to evaluate the efficiency with which inflows can be driven by torques from non-axisymmetric features.\par

In the deprojected plane, we split the galaxy into circular annuli, each $\Delta r\approx0.5$\,kpc wide (corresponding to the PSF of the F200W NIRCam filter used in the estimation of the potential), and computed the mass-weighted average, in the azimuthal sense, torque per unit mass, $\tau$, as a function of radius, $r$. We used the rotational profile of our \texttt{DysmalPy} model to estimate the in-plane rotational velocity of the gas, $u_{\mathrm{rot}}$. We assessed the efficiency of the absorption of angular momentum by the stellar component by computing the relative change in the azimuthal angular momentum of the gas component, $\Delta L/L$, in one rotation period, $T_{\mathrm{rot}}$, as follows \citep{Garcia-Burillo_2005}:

\begin{equation}
    \frac{\Delta L}{L}\bigg\vert_{\mathrm{rot}}=\frac{\tau\times T_{\mathrm{rot}}}{r\times u_{\mathrm{rot}}}=\frac{2\pi\tau}{u_{\mathrm{rot}}^2}
        \label{eq:angularMomentumTransferEfficiency}
\end{equation}

Assuming that at each infinitesimally small annulus the rate of change in the relative angular momentum will be equal to the relative flow rate, i.e., a decrease (increase) of $x$\% in the angular momentum will result in $x$\% of the mass to inflow (outflow), we can compute the flow rate using the following expression \citep{Garcia-Burillo_2005}:

\begin{equation}
    \frac{dM}{dt}=\mathlarger{\mathlarger{\sum}}\frac{d}{dt}\left(\frac{dM}{dr}\right)\Delta r=\mathlarger{\mathlarger{\sum}}\left(\frac{\Delta L}{L}\bigg\vert_{\mathrm{rot}}\frac{1}{T_{\mathrm{rot}}}\right)\left(2\pi r\Sigma_{\mathrm{gas}}\right)\Delta r
        \label{eq:torqueModelingMassFlowRate}
\end{equation}

\begin{figure}
    \centering
	\includegraphics[width=\columnwidth]{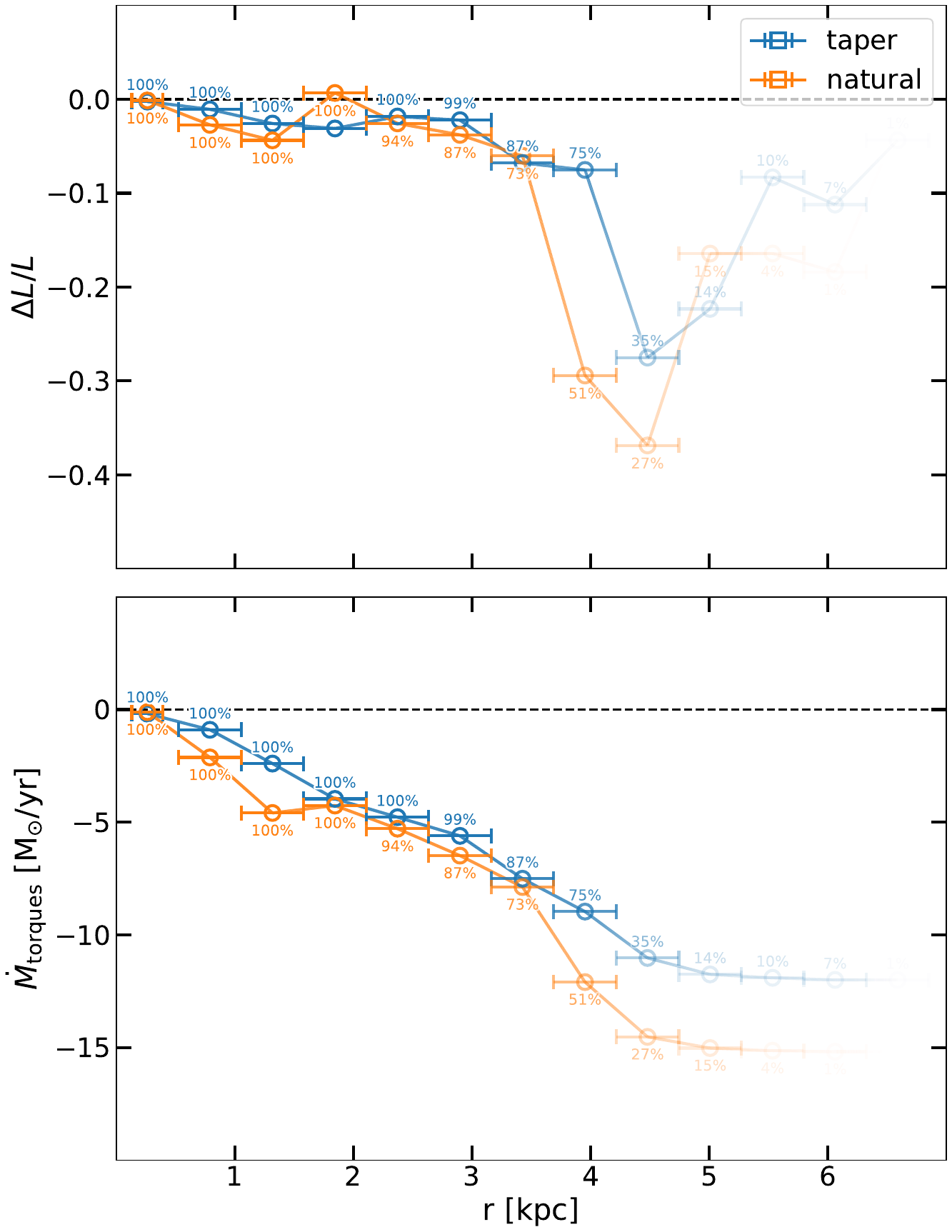}
    \caption{Angular momentum loss efficiency (top) and mass inflow rate (bottom) profiles for the tapered and naturally-weighted data. We find that gas loses angular momentum in the bar region, with a maximum angular momentum loss efficiency near the end of the bar, enabling the efficient gas transport toward the center, as indicated by the increasingly negative values of the mass flow rate. Beyond the bar region the efficiency of angular momentum loss decreases. However, this trend is based on a limited number of pixels with a modest amount of flux (as reflected in the roughly constant mass flow rate).}
    \label{fig:torqueModeling}
\end{figure}

The resulting angular momentum transfer efficiency, $\Delta L/L$, and mass flow rate, $\dot{M}_{\mathrm{torques}}$, are presented in Fig.~\ref{fig:torqueModeling}. We find that the angular momentum of the gas is efficiently lost in the outer bar region, with a lower efficiency identified in the very inner and outer parts of the galaxy. The whole extent of the disk is dominated by inflows, with modest values in the very central regions which become more significant moving toward the outer bar region and eventually converge to $\dot{M}\sim15$\,M$_\odot$/yr ($10$\,M$_\odot$/yr) for our naturally-weighted (tapered) data.\par

In the outer bar region, the torques exerted by the potential appear to be efficient in causing the loss of angular momentum of the gas, resulting in inflows toward the galactic center, highlighting the role of the bar as their possible main driver. However, in the central regions, this efficiency is decreased meaning that another mechanism, such as the dynamical friction or viscous torques \citep[e.g.,][]{Garcia-Burillo_2005}, could be responsible for funneling the gas toward the very center. On the other hand, if the torques from the stellar potential are able to efficiently drive inflows at small distances from the center, even higher-resolution observations would be required for the identification of morphologies that could possibly be responsible for such flows, e.g., trailing spirals \citep[e.g.,][]{Audibert_2019, Combes_2022}, and the quantification of their effect on the central gas flows.\par

In the outer disk, the observed lopsidedness could also promote gas inflows. The $m=1$ mode has been shown to be the most efficient one in driving outwards angular momentum transfer at large radii \citep{Saha_2014}. In the same regions, the lopsidedness itself could also be the result of gas accretion \citep{Bournaud_2005, Jog_2009}. Thus, the large $m=1$ amplitude could be the dominant mechanism driving gas inflows in the outskirts of the disk of this galaxy.\par

The following caveats should be considered in the interpretation of our torque modeling results: i) in the very central regions the resolution of our observations might be inadequate to properly resolve the gas mass distribution; ii) in the central regions the estimation of the baryonic potential could be inaccurate due to the implicit assumption of a flattened bulge, since a refined treatment was not applied to it in the deprojection; and iii) in the outer regions, the limited coverage of our molecular gas observations would introduce uncertainties in the constrained angular momentum transfer efficiency, since we account for only part, in the azimuthal sense, of the torques per unit mass, and consequently of the radial flow rate profile.\par

\subsection{Gas funneling parallel to the bar lanes; a well established dust lane shock?}
\label{sec:gasFunnelingParallelToTheBarLanes}

\begin{figure*}
    \centering
	\includegraphics[width=2.0\columnwidth]{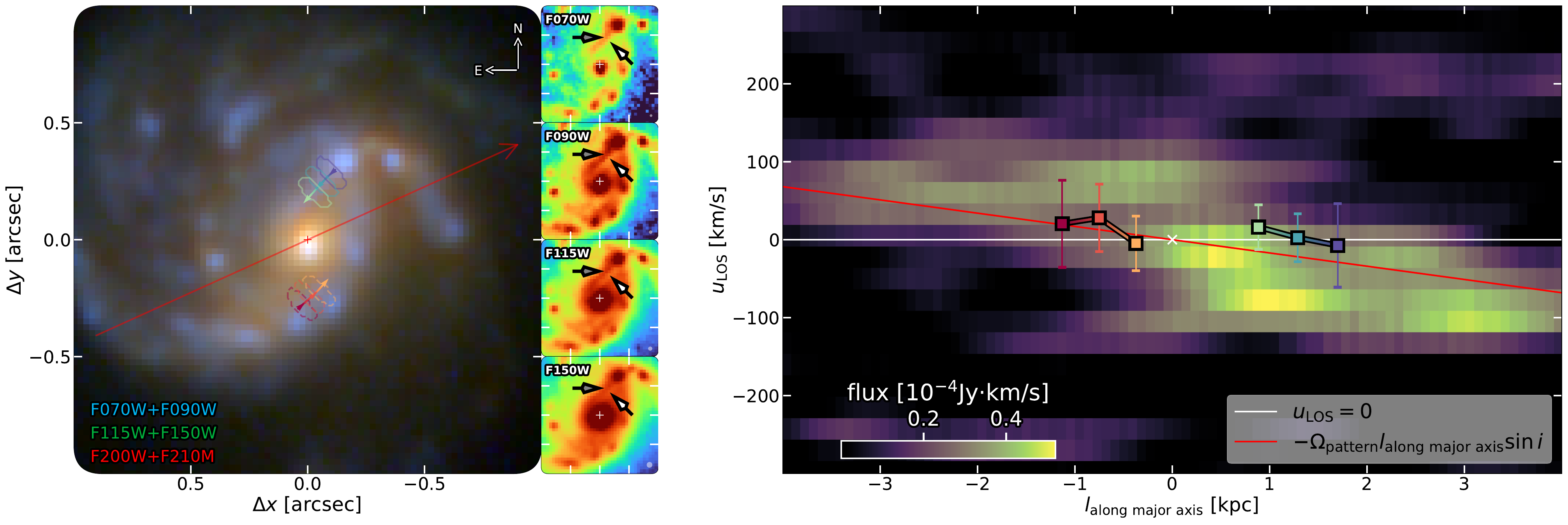}
    \caption{Color composite image in the short-wavelength \textit{JWST} NIRCam bands (left), the four shortest wavelength NIRCam images (F070W, F090W, F115W, F150W) (middle), and PV diagram of an 1\farcs0$\times$1\farcs0 region parallel to the disk major axis (right). A dust lane is tentatively identified at the north side of the bar, in the form of a dark straight-line feature (white arrow), succeeded by a straight line of an increase in the restframe UV/optical continuum ahead of it (gray arrow), with respect to the direction of rotation. In the PV diagram, the values of the pixels within $\approx1.5\times\text{beam}$ from the disk major axis are integrated, with the rotation of the galaxy clearly identified in the resulting pattern. The solid red line indicates the LOS velocity contribution of the pattern (bar) rotation. Given the orientation of the galaxy and the bar lanes, LOS velocities above (below) this line for $l_{\mathrm{along~dust~lanes}}>0$ ($<0$) (as is the case for the northern dust lane), are compatible with the presence of flows along the shocks toward the center of the galaxy. In the opposite case (as is the case for the middle bin along the southern expected shock), the inferred flows would be nonphysical under the assumption of a flow purely parallel to the shock, so this assumption is not justified.}
    \label{fig:apertureAlongDustLanesAndPVDiagram}
\end{figure*}

We have established that the flows in G4\_38232 are in agreement with expectations for barred galaxies. We have further estimated the net flow rates, uncovering evidence of significant net radial inflows, with a total inflow rate of the same order as the star formation rate. Building on these results, and motivated by evidence in the literature pointing towards typical bar flows in high-$z$ barred galaxies \citep[e.g.,][]{Genzel_2023, Amvrosiadis_2025, Umehata_2025, Huang_2025, Pastras_2025b}, we take a yet deeper look into the gas flows in G4\_38232, assuming that expectations from theoretical works \citep[e.g.,][]{Athanassoula_1992b, Patsis_2000, Kim_2012a} and observations in the local Universe \citep[e.g.,][]{Regan_1997, Sormani_2023} are also valid (at least in part) for this high-$z$ barred spiral.\par

We examine the planar gas motions under the assumption that in the frame corotating with the bar, gas at the leading side of the straight-line shocks (dust lanes) flows parallel to them. In this picture, gas rotates in an elliptical-like manner in the bar region, runs into the straight-line shock formed ahead, with respect to the sense of rotation, of the local minimum of the potential (semimajor axis of the bar), gets shocked, and consequently flows parallel to the dust lane identified at the location of the shock \citep[e.g.,][]{Athanassoula_1992b}.\par

A color composite image obtained by combining the available short wavelength NIRCam filters (i.e., F070W, F090W, F115W, F150W, F200W, and F210M) reveals a redder straight-line morphology at the inner edge of a blue linear feature at the central northern side of G4\_38232 (see Fig.~\ref{fig:apertureAlongDustLanesAndPVDiagram}). This feature, at the leading side of the bar, appears to be a well-defined dust lane shock. Hints of a similar feature can be identified at the symmetric southern part of the bar. However, this morphology is overshone by the clumpy nature of this region, leaving its northern counterpart as the single prominent such structure in the galaxy. In fact, the morphology of the northern part of the bar is very reminiscent of typical, local-like, gas morphologies identified in simulation models. For example, in this region, the apparent increased density at the trailing side of the bar could be evidence of the presence of a "smudge", i.e., an enhanced density region identified in models of gas flows in barred potentials \citep{Patsis_2000, Kim_2012a} and molecular gas observations \citep[e.g.,][]{Regan_1995}.\par

Motivated by the identification of this straight-line feature, we chose to investigate the kinematics parallel to it following the methodology presented in \citet{Regan_1997, Sormani_2023}. We assume that in the frame corotating with the bar the direction of the flow is parallel to the bar lanes, and in that frame gas flows into the bar lane, gets shocked and consequently gets funneled toward the central regions \citep[e.g.,][]{Athanassoula_1992b, Regan_1997, Kim_2012a, Sormani_2023}. Under this assumption, after the subtraction of the LOS contribution of the bar pattern rotation, the residual LOS velocities are expected to stem exclusively from motions parallel to the shock. Following the formulation of \citet[][Appendix~E]{Pastras_2025b}, and specifically, with $\psi$ indicating the angle with respect to the major axis of the disk in the face-on orientation and $i$ the fiducial inclination ($i:=i_{\mathrm{fiducial}}$), we can express the LOS velocities at the loci of the bar lanes as the sum of the two aforementioned contributions ($u_{\mathrm{LOS}}=u_{\mathrm{pattern,LOS}}+u_{\mathrm{\parallel~bar~lane,LOS}}$). The first is the LOS contribution of a tangential rotation. The second stems from a flow at an angle with the major axis, with only its component perpendicular to the major axis contributing along the LOS. Thus, we can express the total LOS velocity as:\par

\begin{equation}
    u_{\mathrm{LOS}}=-\Omega_{\mathrm{pattern}}r\cos\psi\sin i-u_{\mathrm{\parallel~bar~lane}}\sin\psi_{\mathrm{\parallel~bar~lane}}\sin i,
        \label{eq:velocityParallelToTheDustLaneInPlane}
\end{equation}

with $r\cos\psi=l_{\mathrm{\parallel~major~axis}}$ being the projection of the point along the dust lane on the major axis of the disk, and $\psi_{\mathrm{\parallel~bar~lane}}$ the angle between the tangent to the bar lane and the major axis in the face-on orientation. Substituting this angle with its projected counterpart, $\phi_{\mathrm{\parallel~bar~lane}}$, using \citet[][Eq.~E.23]{Pastras_2025b}, and solving for $u_{\mathrm{\parallel~bar~lane}}$, the in-plane velocity parallel to the bar lane, we reach the following relation, in the plane projected on the sky:

\begin{align}
    \begin{split}
        u_{\mathrm{\parallel~bar~lane}}=-&\frac{u_{\mathrm{LOS}}+\Omega_{\mathrm{pattern}}l_{\mathrm{\parallel~major~axis}}\sin i}{\tan i\sin\phi_{\mathrm{\parallel~bar~lane}}}~\times\\
        &\left[\frac{\sin^2\phi_{\mathrm{\parallel~bar~lane}}}{\cos^2i}+\cos^2\phi_{\mathrm{\parallel~bar~lane}}\right]^{1/2}
    \end{split}.\label{eq:velocityParallelToTheDustLaneProjected}
\end{align}

This is the equivalent to the equation presented in \citet{Regan_1997} and \citet[][Eq.~5]{Sormani_2023}, but following our own conventions, with the resemblance between these expressions being quite apparent.\par

After establishing a method to derive the in-plane velocities, we used the JWST imaging to draw a path in-front of and parallel to this dust-lane-like feature, starting from the edge of the apparent bar lane and up to the northern edge of the central region. We extracted spectra from the higher resolution naturally-weighted data cube along this path using three elliptical apertures, each $\approx$\,0\farcs16 ($1/2$ the major axis of our beam) wide and $\approx$\,0\farcs05 long. We followed a similar procedure for the southern expected bar lane. The size of the apertures is well below the observational PSF, however, we still use them since the higher surface density along the shocks should increase the CO flux contribution from the targeted region, compared to its surroundings.\par

We fit the extracted spectra with a Gauss-Hermite (GH) series to account for the asymmetry of the line profiles \citep{van_der_Marel_1993}. Since the S/N ratio of the data does not allow for the fitting of an $h_4$ correction, we truncated the series after the $h_3$ term, so that our normalized fitting function was:

\begin{equation}
    \mathrm{f}(y)=\frac{e^{-\frac{y^2}{2}}}{\sigma\sqrt{2\pi}}\left[1+\frac{h_3}{\sqrt{6}}\left(2\sqrt{2}y^3-3\sqrt{2}y\right)\right], \mbox{ with } y=\frac{x-\mu}{\sigma}.\label{eq:GaussHermiteSeries}
\end{equation}

\begin{figure}
    \centering
	\includegraphics[width=\columnwidth]{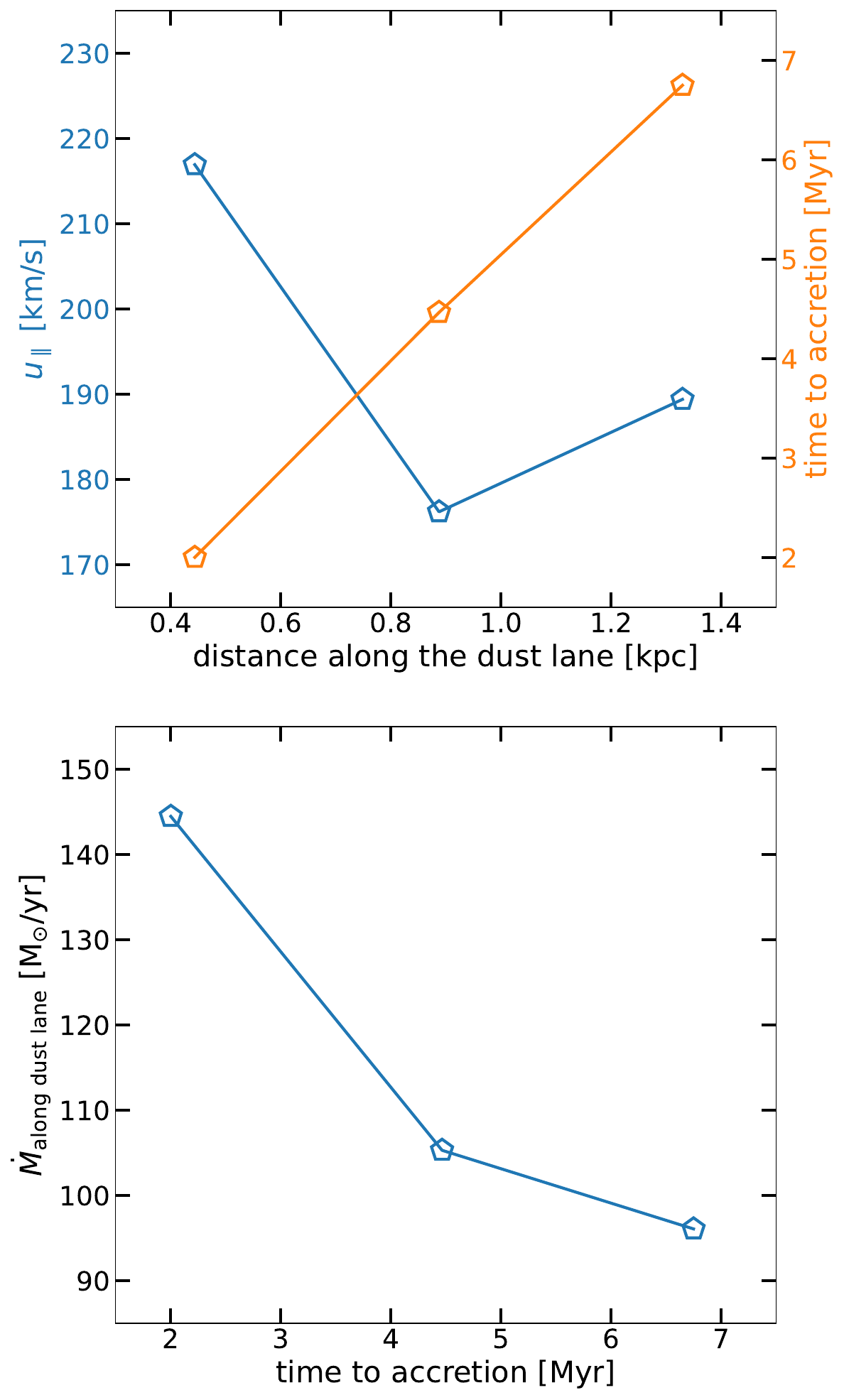}
    \caption{In-plane velocity parallel to the northern dust lane and time for the material of each aperture to reach the base of the dust lane (top) along with the inferred mass flow rate (bottom). We find a median flow of $\approx100$\,M$_\odot$/yr at a velocity of $\approx200$\,km/s. Assuming an accretion efficiency of $\approx30\%$, the resulting net inflow rate due to the northern bar lane is estimated at $\sim30$\,M$_\odot$/yr.}
    \label{fig:inflowAlongDustLane}
\end{figure}

The in-plane velocity of the gas is assumed to be parallel to the dust-lane in the frame corotating with the bar. We assess to what extent this assumption is justified by considering a PV-diagram along the major axis of the disk, covering the bar region. In the second panel of Fig.~\ref{fig:apertureAlongDustLanesAndPVDiagram}, we present the PV-diagram covering a region of $\approx$\,1\farcs0 ($\approx8.2$\,kpc) in length and width, sampled with $37$ pixels along the PA, and a velocity range $\pm300$\,km/s around the systemic velocity of G4\_38232. In the same figure, we plot the LOS contribution of the rotation of the bar, which has to be subtracted for the computation of the velocity parallel to the bar lanes (see Eq.~\ref{eq:velocityParallelToTheDustLaneProjected}).\par

Given the orientation of the bar lanes, the angle with the disk major axis, $\phi_{\mathrm{\parallel~bar~lane}}$, has a negative (positive) sine for the northern (southern) dust lane. Thus, LOS velocities, $u_{\mathrm{LOS}}$, with higher (lower) values than the contribution of the bar rotation at the northern (southern) part of the bar are required for the flow to be toward the center of the galaxy. Fig.~\ref{fig:apertureAlongDustLanesAndPVDiagram} provides an intuitive way to assess whether our assumptions of flows parallel to the dust lanes could be valid since this can only be the case when the inferred in-plane velocities parallel to the shocks are toward the central regions. We find that our assumption could be valid for the northern dust lane shock, while in the case of the southern the inferred in-plane velocity under this assumption is not physical (either approximately zero or parallel to the shock towards the outer regions). In simulations it is possible to find flows at the loci of the bar lanes that do not drive gas inwards, but such occurrences are rare, brief and arise due to the interaction of the material flowing parallel to the bar lane with gas originating from that on the opposite side of the galaxy, overshooting the central regions, and slumming into the inflowing material \citep[e.g.,][]{Regan_1999, Sormani_2019b}.\par

In G4\_38232, while the northern dust lane shock is well-defined, its expected southern counterpart, if present, is clearly much less prominent. Additionally, in front of the southern possible dust lane, we identify a small number of clumps. Thus, while at the northern part we have a clear bar lane, with gas flowing parallel to it in an as-expected configuration, at the southern part, the flow is more spiral like. Gas flows into the region of enhanced density, gets compressed, but instead of undergoing a strong shock and getting funneled directly towards the central regions, it continues to flow downstream in an in-spiraling manner, with clumps resulting from the aforementioned compression found downstream of the shock. The PV diagram of Fig.~\ref{fig:apertureAlongDustLanesAndPVDiagram} supports this interpretation, given that the LOS velocities at the location of the southern shock are not consistently lower, but in some cases in excess to that of the pattern rotation, with gas been able to overtake the bar pattern in that region.\par

Focusing on the northern dust lane shock, we used Eq.~\ref{eq:velocityParallelToTheDustLaneProjected} to compute the in-plane velocities parallel to it, using the first moment of the fitted GH series, $u_{\mathrm{LOS}}=\mu+\sqrt{3}\sigma h_3$, as the observed velocity centroid. We then used the deprojected length of each aperture along the dust lane, $dl_{\mathrm{\parallel~bar~lane,in-plane}}$, to compute the time required for its material to cross it, $dt_{\mathrm{cross}}$. We also derived the gas mass enclosed in each aperture along the bar lane, $dM_{\mathrm{gas}}$, using the CO(4-3) flux enclosed within its bounds with \citet[][Eq.~2]{Tacconi_2020}. The gas flow rate at each position along the bar lane can be computed as:

\begin{equation}
    \dot{M}_{\mathrm{\parallel~bar~lane}}=\frac{dM_{\mathrm{gas}}}{dt_{\mathrm{cross}}}=\frac{dM_{\mathrm{gas}}\times u_{\mathrm{\parallel~bar~lane}}}{dl_{\mathrm{\parallel~bar~lane,in-plane}}}.
        \label{eq:flowRateAlongDustLane}
\end{equation}

The estimation of the time of accretion in the central region is also interesting. Following \citet{Sormani_2023}, considering that for a gas element to reach the base of the dust lane, in this case the base of the aperture closest to the center, it has to go through all of the intermediate regions, the time required is:

\begin{equation}
    t_{\mathrm{accretion}}=\sum dt_{\mathrm{cross}}=\sum\left(dl_{\mathrm{\parallel~bar~lane,in-plane}}/u_{\mathrm{\parallel~bar~lane}}\right).
        \label{eq:timeOfAccretionAlongTheDustLane}
\end{equation}

The results from this analysis are presented in Fig.~\ref{fig:inflowAlongDustLane}. The assumed direction of the dust lane shock goes toward the center of the galaxy, thus the inferred velocity and flow rate parallel to it are positive when gas is funneled toward the center, corresponding to inflows. We find a consistent gas flow along the outer regions of the dust lane toward the center. Near the central regions the inferred flow velocity and gas flow rate increases, an effect also observed in local galaxies \citep[e.g.,][]{Sormani_2023}. In our case, this effect could also be a consequence of observational limitations, since due to the PSF the kinematics of the regions of the shock closer to the galactic center can be affected by contamination from the central regions. Additionally, our assumption of purely parallel flows breaks down as we approach the central regions and a possible nuclear ring.\par

The median inferred gas flow rate is $\approx100$\,M$_\odot$/yr, which is $\sim3\times$ larger compared to the integrated SFR of G4\_38232 ($\mathrm{SFR}\approx36$\,M$_\odot$/yr). However, while this amount of gas is moving along the dust lane, only part of it will be accreted into the inner regions at the first passage. Simulations show that, on average, roughly $30\%$ of the material flowing parallel to the bar lanes will get trapped in the central region \citep{Regan_1997, Hatchfield_2021}, while the rest will overshoot it and flow into the dust lane of the opposite side of the galaxy \citep{Regan_1999, Sormani_2019b, Whitmore_2023}. Thus, the time-averaged net gas inflow rate is estimated at $\approx30$\,M$_\odot$/yr over a period of $\approx7$\,Myr.\par

\begin{figure*}
    \centering
	\includegraphics[width=2.0\columnwidth]{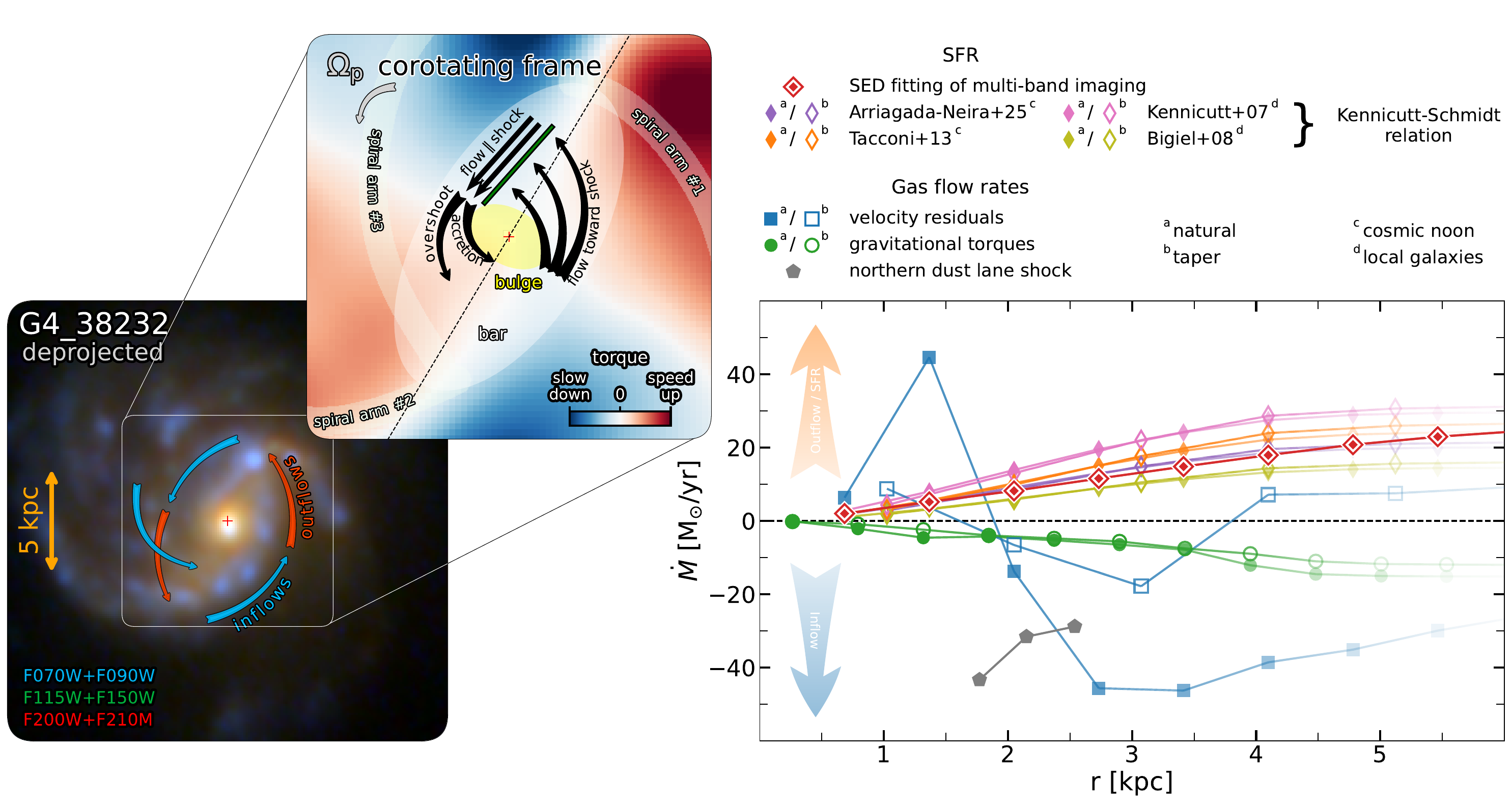}
    \caption{Overview of inferred gas flow patterns in the disk of G4\_38232 (left) and resulting gas flow rate profiles (right) determined through three different methods. Our residual analysis and torque modeling results can be considered reliable within the bar region ($r\leq a_{\mathrm{bar}}\approx4.2$\,kpc), while at larger radii the limited azimuthal coverage hinders the derivation of reliable measurements of the molecular gas flow rate. The inflow rate based on our analysis of the flows along the dust lanes considers the contribution of only one bar lane and an accretion efficiency of $\approx30\%$. Overall, we find a clear dominance of inflows with higher inflow rates inferred from the naturally-weighted data product. Our best estimate for the net inflow rate in the bar region is $\dot{M}\sim30$\,M$_\odot$/yr, that is of the order of the SFR in the same region. The dominant driver of gas flows in G4\_38232 is the bar, exerting torques on the gas, leading to the formation of the shocks at its leading side and the subsequent funneling of gas toward the central regions parallel to the northern bar lane. A schematic view of these flows in the frame corotating with the bar is shown at the top middle panel.}
    \label{fig:gasFlowOverview}
\end{figure*}

\section{Discussion}
\label{sec:discussion}

In our study of G4\_38232, we quantitatively examine the in-plane molecular gas flows using three different analyses. In all cases, we find an integrated negative net gas flow rate, i.e., inflow, with a value of a few tens of M$_\odot$/yr, a similar order as the SFR, indicating efficient funneling of gas toward the central regions. The values resulting from each one of our analyses are the following: i) based on the observed velocity residuals assuming only in-plane, radial, noncircular motions ($\dot{M}\approx20-45$\,M$_\odot$/yr), ii) based on the torques exerted on the molecular gas by the galactic potential in the midplane of the disk ($\dot{M}\approx10-15$\,M$_\odot$/yr), and iii) through the study of the gas flows along the well-defined northern dust lane shock ($\dot{M}\approx30$\,M$_\odot$/yr due to only this shock).\par

In addition to these integrated values, we have derived profiles of the radial gas flow, an overview of which is shown in Fig.~\ref{fig:gasFlowOverview}. We find an efficient funneling of gas toward the central regions along the bar but this efficiency declines in the innermost regions of the galaxy. While this can be the effect of the resolution of our observations, hindering the accurate study of such flows at small radii, (or artificially amplified inflows at larger radii due to the limited azimuthal coverage) it could also be a sign of less efficient funneling in the nuclear regions, with further gas transport possibly impeded by the presence of nuclear rings or spirals, as seen in local barred cases \citep[e.g.,][]{Combes_2002, Regan_2003, Maciejewski_2004a, Maciejewski_2004b, Combes_2022, Pastras_2026a}.\par

Our flow rate profiles are a measure of the rate at which gas flows into (or out of) a region starting from the center of the galaxy and ending at a certain radius. The SFR that takes place within this region serves as an intuitive baseline for comparison. If the gas inflow rate exceeds the SFR, there will be an accumulation of gas in this region, if it is lower the region will become quenched, otherwise, if the two rates are on the same order, a steady state is achieved, with the efficient replenishment of gas that is converted into stars. We derived two estimates of the cumulative SFR profile: i) using the SFR map resulting from resolved SED fitting (see Appendix~\ref{sec:SEDFitting}), and ii) through the application of the Kennicutt-Schmidt (KS) relation \citep{Schmidt_1959, Kennicutt_1998}, connecting the molecular gas mass and the SFR, to our molecular gas data. For the latter, we used the slopes and normalizations constrained by galaxy-integrated and spatially-resolved studies of cosmic noon \citep{Tacconi_2013, Arriagada-Neira_2025} and local \citep{Bigiel_2008, Kennicutt_2007} galaxies \citep[see also, e.g.,][]{Kennicutt_1998, Schuster_2007, Genzel_2010, Genzel_2013}. We find that the SFR estimate derived from SED fitting is in broad agreement with the KS-based ones, for all considered slopes and normalizations. Comparing these estimates with the respective gas flow rates, we find that, overall, the estimated flow rate into each region is of the same order as the SFR taking place within it. Thus, G4\_38232 appears to be in an overall steady state, with the gas lost in star formation being replenished through planar inflows.\par

A comparison between the estimated flow rates in local barred galaxies and those constrained for G4\_38232 shows that for this galaxy, the inflow rates are approximately one order of magnitude higher than those found in local cases \citep[e.g.,][]{Quillen_1995, Regan_1997, Sormani_2019a, Wu_2021, Di_Teodoro_2021, Sormani_2023}. Thus, our work shows that in the inner parts of cosmic noon disk galaxies, non-axisymmetric structures (in this case the bar) could play a dominant role in driving gas flows toward the central regions (see also \citealt{Genzel_2023, Huang_2025, Jolly_2026}, C. Pulsoni et al, in prep.). In the outer parts, flows from the circumgalactic medium (CGM) in the form of either cold gas streams or cooling flows have been suggested as a viable way of replenishing the gas reservoir of the disk with pristine gas \citep[e.g.,][]{Keres_2005, Dekel_2009, van_de_Voort_2011, Combes_2014, Dutta_Chowdhury_2024}. From this point on, bar or spiral driven gas flows can take over, dominating the radial transport in the inner regions.\par

Our study also offers compelling evidence of the presence of kinematically well-defined bar lanes at high redshift. Gas flows in agreement with theoretical expectations in the presence of a bar have already been identified in a handful of barred spirals at cosmic noon \citep{Genzel_2023, Amvrosiadis_2025, Umehata_2025, Huang_2025, Pastras_2025b}. Our study takes advantage of the exquisite IRAM-NOEMA CO data and \textit{JWST} multi-band imaging to offer a yet deeper view of these flows. In G4\_38232 we uncover evidence of gas flowing parallel to an apparent dust lane shock. This orderly gas streaming could serve as a benchmark for simulations of high-$z$-like, turbulent, gas-rich barred galaxies \citep[e.g.,][]{Bland-Hawthorn_2024}; successful models should, at least in some cases, reproduce straight shock fronts accompanied by downstream gas streaming parallel to them.\par

Moreover, it is worth mentioning that the velocities parallel to the northern bar lane are very significant, similarly to those constrained for local barred galaxies \citep[e.g.,][]{Regan_1997, Sormani_2023}. It is known that these velocities at which the gas is funneled toward the central regions correlate with the properties of the bar \citep[e.g.,][]{Athanassoula_1992b}. However, in G4\_38232 the resulting net flow rates are almost one order of magnitude higher than those constrained in a similar fashion for local cases \citep{Regan_1997, Sormani_2023}, due to the higher gas content of its disk. This is expected to apply in general to main sequence, barred spirals at cosmic noon given the their increased gas fractions \citep{Tacconi_2018, Tacconi_2020}. Thus, the effects of bar flows in such gas-rich conditions are considerably amplified, driving a very significant amount of gas toward the center of their host disks and shaping their evolution.\par

A similar conclusion could be drawn through the evidence of the modest (if any) evolution of the measured $Q_b$ ratio of the tangential force due to the non-axisymmetric perturbation over the axisymmetric radial force (see \citet{Buta_2001} for definition) for cosmic noon barred galaxies (\citet{Kalita_2026}, see also \citealt{Kim_2021}). If the relative forcing due to the bar is similar to that of local barred spirals, the loss of the gas angular momentum due to the exerted torques from the stellar component should happen in a similar fashion but scaled for the increased gas content. This will be reflected in flow rate estimated by torque modeling. In the case of G4\_38232, the measured bar strength is $Q_b\approx0.37$ (see Appendix~\ref{sec:barStrengthEstimate}), at the high end but in agreement with measurements for cosmic noon \citep{Kalita_2026} and lower redshift barred spirals \citep[e.g.,][]{Kim_2021}, further highlighting the critical role of the increased gas content in the scaling of the identified gas flow rate and, consequently, the amplified effect of the bar in the evolution of its host disk.\par

Finally, it is also useful to consider the caveats of the presented analyses. The uncertainties of some key parameters, such as the inclination and position angle of the disk of G4\_38232, which affect all aspects of our analyses, or the bar pattern speed, are significant. The fitting of an axisymmetrically rotating model to the quasi-elliptical streamlines that gas is expected to follow in the presence of a bar, could lead to discrepancies in the observed rotation curve \citep{Salibur_2026}, in turn affecting the inferred mass distribution of the galaxy. Thus, considering all caveats, our flow rate estimates are not expected to be accurate to within a few percent. Yet, the fact that a coherent picture emerges as a result of an array of independent analyses, highlights the more general and encouraging conclusion that the observed inflow rate in the case of this barred spiral is of the same order of magnitude as its integrated SFR.\par

\section{Conclusions}
\label{sec:conclusions}

In this paper, we present our deep IRAM-NOEMA CO(4-3) observation of G4\_38232, a $z\sim1.12$ massive, main sequence, barred spiral galaxy in the EGS field. Our observations, with a total on-source integration time of $\approx37$\,hrs, were carried out as part of NOEMA\textsuperscript{3D}, an ambitious GTO program focusing on molecular gas kinematics at cosmic noon \citep{Jolly_2026, Chen_2026}. Additionally, the multi-band imaging with a wavelength coverage from $\approx0.4$\,$\mu m$ to $\approx21$\,$\mu m$ provides crucial ancillary data enabling the combined study of the morphology and kinematics of this galaxy.\par

With a large number of barred spirals identified by recent studies at cosmic noon \citep[e.g.,][]{Guo_2023, Le_Conte_2024, Guo_2025, Espejo_Salcedo_2025, Le_Conte_2025} and, in some exceptional cases, even beyond \citep[e.g.,][]{Smail_2023, Tsukui_2024, Amvrosiadis_2025, Boogaard_2026}, pioneering works have provided evidence of the significant role of high-$z$ bars in shaping galaxy evolution through bar-driven gas inflows \citep{Genzel_2023, Amvrosiadis_2025, Umehata_2025, Huang_2025, Pastras_2025b}. Since detailed studies of such systems remain limited, due to the exceedingly long integration times required (a few $\times10$\,hrs of on-source integration) with state-of-the-art interferometers, we took advantage of the exquisite IRAM-NOEMA data, in order to provide well-constrained estimates of the net bar-driven flow rates in a typical, main sequence, cosmic noon barred spiral.\par

We started by constraining an array of properties for G4\_38232, showing that it is a massive ($\log(M_{\mathrm{baryons}}/M_\odot)\approx10.96$), gas-rich ($f_{\mathrm{gas}}\approx40\%$), baryon-dominated ($f_{\mathrm{dm}}(<R_e)\approx4\%$) disk galaxy. It hosts a prominent bar, which is long ($a_{\mathrm{bar}}\approx4.2$\,kpc), strong ($Q_{\mathrm{b}}\approx0.37$), and fast ($\mathcal{R}\approx1.05$), in agreement with expectations for gas-rich, baryon dominated systems \citep{Athanassoula_2003, Athanassoula_2014, Williams_2021, Beane_2023, Fragkoudi_2021, Merrow_2026}.\par

Next, we employed three methods, namely, i) the LOS velocity residual analysis (e.g., \citealt{Genzel_2023, Pastras_2025b, Jolly_2026}, C. Pulsoni et al., in prep.), ii) the torque modeling \citep[e.g.,][]{Garcia-Burillo_2005, Audibert_2019, Huang_2025}, and iii) the study of the flows parallel to the bar lanes \citep[e.g.,][]{Regan_1997, Sormani_2023}, to offer a quantification of the bar-driven gas flows for our high-$z$ barred target. For the first two analyses, we use both our tapered and naturally-weighted data products, with the former offering a higher S/N and a better recovery of the extended CO(4-3) emission, and the latter a higher resolution.\par

We find a qualitative agreement between the estimates of these methods, albeit with a large scatter, also considering the values constrained using our higher and lower resolution data products. In all cases, the inferred net gas flow rate in the bar region is inflow, with integrated values ranging from $\dot{M}_{\mathrm{taper}}\sim10$\,M$_\odot$/yr to $\dot{M}_{\mathrm{natural}}\sim45$\,M$_\odot$/yr for the tapered and naturally-weighted data products, respectively. These estimated inflow rates are of the order of the galaxy-integrated SFR, indicating efficient gas transport toward the central regions of the galaxy.\par

This transport is driven by the bar, as indicated by the overall agreement between our analyses of our highly-resolved observations focusing on its role in efficiently driving gas inflows. Our residual analysis points towards the presence of significant gas inflows, with overall patterns broadly consistent with expected gas flows in the presence of a gas-rich bar (see Appendix~\ref{sec:simulationMockObservedVelocityResiduals}). Additionally, the torque modeling, which provides an estimate of the inflow rate due to the angular momentum loss from torques exerted on the gas from the non-axisymmetric galactic potential, also supports our conclusion; with the bar shown to be the most prominent non-axisymmetric feature exerting these torques, its role in driving the observed inflows is further emphasized.\par

Moreover, the presence of a straight-line feature at the northeastern, leading side of the bar, offers compelling evidence of the presence of a dust lane shock, a known consequence of the high gas column densities at the loci of shocks driven by the potential of the bar \citep[e.g.,][]{Athanassoula_1992b, Patsis_2000, Kim_2012a, Sormani_2019b}, and typically observed in local barred spirals \citep[e.g.,][]{Stuber_2023, Sormani_2023, Schinnerer_2023}. By targeting this feature, we showed that the gas flows in its region are compatible with motions parallel to it, in the reference frame corotating with the bar, in agreement with theoretical predictions \citep{Athanassoula_1992b, Patsis_2000, Kim_2012a, Sormani_2019b, Hatchfield_2021}. Based on this assumption, we derived our third estimate of the net gas inflow rate, $\dot{M}_{\mathrm{net,\parallel~bar~lane}}\approx30$\,M$_\odot$/yr, in qualitative agreement with those derived through our other two analyses. Thus, our study offers compelling evidence of the presence of a well-defined dust lane shock in a high-$z$ barred galaxy.\par

The large gas inflows associated with such features, in addition to the high gas content of cosmic noon galaxies \citep{Tacconi_2018, Tacconi_2020, NMFS_2020}, offer strong evidence of the dominant role of these structures in the evolution of their host disks. With bars identified in a significant fraction ($\gtrsim10-20\%$) of typical, massive cosmic noon disk galaxies \citep[e.g.,][]{Le_Conte_2024, Guo_2025, Espejo_Salcedo_2025, Geron_2025, Le_Conte_2025}, our results highlight their role as crucial evolutionary drivers at the heyday of cosmic star and galaxy formation.\par

\begin{acknowledgements}
      N.M.F.S. acknowledges support, and C.B., J.C., J.M.E.S., G.T. are funded by the European Union (ERC, GALPHYS, 101055023). H.Ü. acknowledges funding by the EU (ERC APEX, 101164796). Views and opinions expressed are, however, those of the author(s) only and do not necessarily reflect those of the EU or the ERC. Neither the EU nor the granting authority can be held responsible for them. HÜ thanks the Max Planck Society for support through the Lise Meitner Excellence Program. T.N. acknowledges the support of the Deutsche Forschungsgemeinschaft (DFG, German Research Foundation) under Germany's Excellence Strategy - EXC-2094 - 390783311 of the DFG Cluster of Excellence `ORIGINS'. S.G-B. acknowledges support from the Spanish grant PID2022-138560NB-I00, funded by MCIN/AEI/10.13039/501100011033/FEDER, EU. P.A.P. acknowledges support by the Sectoral Development Program ($\rm{O\Pi\Sigma}$ 5223471) of the Greek Ministry of Education, Religious Affairs and Sports, through the National Development Program (NDP) 2021-25, as part of project 200/1025, supported by the Research Committee of the Academy of Athens. R.H-C. thanks the Max Planck Society for support under the Partner Group project `The Baryon Cycle in Galaxies' between the Max Planck for Extraterrestrial Physics and the Universidad de Concepción. R.H-C. gratefully acknowledge financial support from ANID - MILENIO - NCN2024\_112 and ANID BASAL FB210003. M.L. acknowledges support from the European Union’s Horizon Europe research and innovation programme under the Marie Skłodowska-Curie grant agreement No 101107795. L.S. acknowledges the financial support from the PhD grant funded on PNRR Funds Notice No. 3264 28-12-2021 PNRR M4C2 Reference IR0000034 STILES Investment 3.1 CUP C33C22000640006. The data analyzed in this paper are CO observations within the NOEMA\textsuperscript{3D} guaranteed time project at the Northern Extended Array for Millimeter Astronomy (NOEMA, located on the Plateau de Bure) Interferometer of the Institute for Radio Astronomy in the Millimeter Range (IRAM), Grenoble, France. IRAM is supported by INSU/CNRS (France), MPG (Germany), and IGN (Spain). This work is also based in part on observations made with the NASA/ESA/CSA James Webb Space Telescope. The data were obtained from the Mikulski Archive for Space Telescopes at the Space Telescope Science Institute, which is operated by the Association of Universities for Research in Astronomy, Inc., under NASA contract NAS 5-03127 for JWST. These observations are associated with programs \#1345, \#2234, \#3794, \#5398, and \#6434 and can be accessed via \href{https://doi.org/10.17909/6eqp-7e47}{DOI:10.17909/6eqp-7e47}. This research is also based on observations made with the NASA/ESA Hubble Space Telescope obtained from the Space Telescope Science Institute, which is operated by the Association of Universities for Research in Astronomy, Inc., under NASA contract NAS 5–26555. These observations are associated with the CANDELS Multi-Cycle Treasury Program. Some of the data products presented herein were retrieved from the Dawn JWST Archive (DJA). DJA is an initiative of the Cosmic Dawn Center (DAWN), which is funded by the Danish National Research Foundation under grant DNRF140. We also acknowledge the use of the following open-source software: \texttt{NumPy} \citep{Harris_2020}, \texttt{SciPy} \citep{Virtanen_2020}, \texttt{emcee} \citep{Foreman-Mackey_2013a, Foreman-Mackey_2013b}, \texttt{Dynesty} \citep{Speagle_2018, Speagle_2020, Koposov_2025}, \texttt{Astropy} \citep{Astropy_2022}, \texttt{Photutils} \citep{Bradley_2024}, \texttt{Matplotlib} \citep{Hunter_2007}, \texttt{CMasher} \citep{van_der_Velden_2020}, \texttt{TRILOGY} \citep{Coe_2012}, \texttt{CIGALE} \citep{Burgarella_2005, Noll_2009, Boquien_2019, Burgarella_2025}, \texttt{STPSF} \citep{Perrin_2012}, \texttt{IMFIT} \citep{Erwin_2014, Erwin_2015}, \texttt{pygad} \citep{Roettgers_2018, Roettgers_2020}, \texttt{GILDAS} \citep{GILDAS_2013} and \texttt{DysmalPy} \citep{Davies_2004a, Davies_2004b, Cresci_2009, Davies_2011, Wuyts_S_2016, Lang_2017, Price_2021, Lee_2025}.\par
\end{acknowledgements}

%


\bibliographystyle{aa} 
\bibliography{bibliography} 

\begin{appendix}




\FloatBarrier

\section{Surface density profile fitting}
\label{sec:surfaceDensityProfileFitting}

\begin{figure}[h]
    \centering
	\includegraphics[width=\columnwidth]{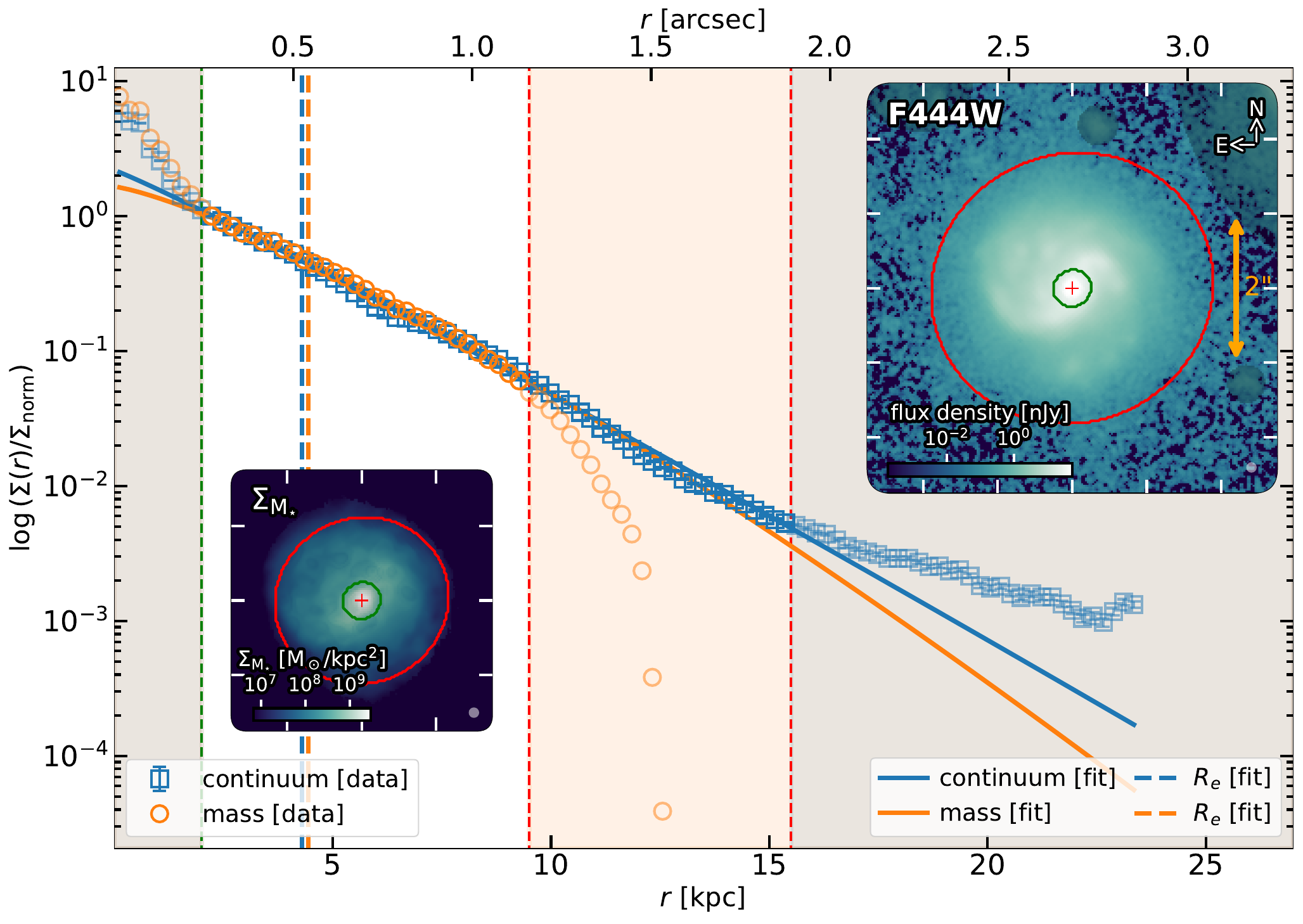}
    \caption{Overview of the fit to the surface density profiles of the NIRCam F444W image and stellar mass map: normalized flux / mass surface density profiles (main plot) and the F444W continuum (right inset) and stellar mass (left inset). As in Fig.~\ref{fig:curveOfGrowthFitting}: i) blue squares (orange circles) indicate the flux (mass) surface density profile extracted using elliptical annuli of increasing semimajor axes while the fitted profile is indicated with a solid blue (orange) line, ii) the corresponding effective radius is given with a dashed blue (orange) vertical line, and iii) the inner (outer) radius of the region considered in the fit is indicated with a green (red) dashed vertical line in the main plot and a solid green (red) ellipse in the insets.}
    \label{fig:surfaceDensityProfileFitting}
\end{figure}

We produced the surface density profiles of the NIRCam F444W continuum image and the stellar mass map of G4\_38232, using the same binning as in the curve of growth analysis of Sect.~\ref{sec:curveOfGrowthFitting}, and fit them with a S\'ersic profile. These profiles alongside the fitted ones are presented in Fig.~\ref{fig:surfaceDensityProfileFitting}. We find that the disk of the galaxy is almost exponential, with a S\'ersic index of $n_{\mathrm{s}}\approx0.9$ and an effective radius of $R_{\mathrm{e}}\approx4.3$\,kpc, in excellent agreement with the values derived in Sect.~\ref{sec:curveOfGrowthFitting}. In the central region, the observed surface density is clearly higher than the fitted profile, indicating the presence of a central mass concentration. In the outer parts of the disk, the observed F444W flux is larger than the fitted profile, stemming from the contributions of clumps in these regions, while the lower values for the stellar mass map are a result of the limited extent of this map.\par

\FloatBarrier

\section{Morphological forward modeling; a nuclear disk?}
\label{sec:morphologicalForwardModeling}

We used the photometric forward modeling code \texttt{IMFIT} \citep{Erwin_2014} to fit the morphology of G4\_38232, aiming to derive estimates for the properties of the bulge. We used two S\'ersic profiles, representing the disk and the bulge, respectively.\par

Since the morphology of G4\_38232 appears to be elongated at a very different angle compared to the kinematic $\mathrm{PA}$ (see Sects.~\ref{sec:ellipticalIsophotes} and \ref{sec:molecularGasKinematics}), we fixed the $\mathrm{PA}$ and ellipticity of the disk to the kinematically inferred values, i.e., $\mathrm{PA}_{\mathrm{disk}}\approx-66$\textdegree~and $e_{\mathrm{disk}}\approx0.06$ (corresponding to $i\approx20$\textdegree~for a razor-thin disk), respectively. We also fixed its S\'ersic index $n_{\mathrm{s,disk}}\approx0.9$ and effective radius $R_{\mathrm{e,disk}}\approx4.3$\,kpc to the values constrained in Sect.~\ref{sec:curveOfGrowthFitting}. The projection of the bulge was assumed to be axisymmetric, with an ellipticity fixed to $e_{\mathrm{bulge}}=0$. Finally, the center of both components was fixed to that determined through the 2D Gaussian fitting of Sect.~\ref{sec:ellipticalIsophotes}.\par

We first produced a model PSF for the NIRCam F444W filter which was used in the fitting, following a procedure similar to that presented in \citet{Ji_2024}, specifically: i) we determined the F444W exposures that contribute to the final image of G4\_38232 and located an ``empty" region close to the galaxy, covered by all of them, ii) we used \texttt{STPSF} \citep{Perrin_2012} to produce a mock star, in the form of a model PSF, which we added to each image at that ``empty" region, with the optical path difference (OPD) at the time of each observation and the position of the mock star on the detector taken into account in the production of the model for each exposure, and iii) we used the \texttt{CrabToolkit} to combine the exposures following the same process used for the original data. The resulting PSF model has been shown to be accurate \citep{Ji_2024}, since in addition to an accurate model PSF for each individual image, the final result includes the effects of the combination of multiple observations, with possibly different telescope orientations, i.e., PA\_V3s, as well as those of the mosaicing process.\par

\begin{figure}
    \centering
	\includegraphics[width=\columnwidth]{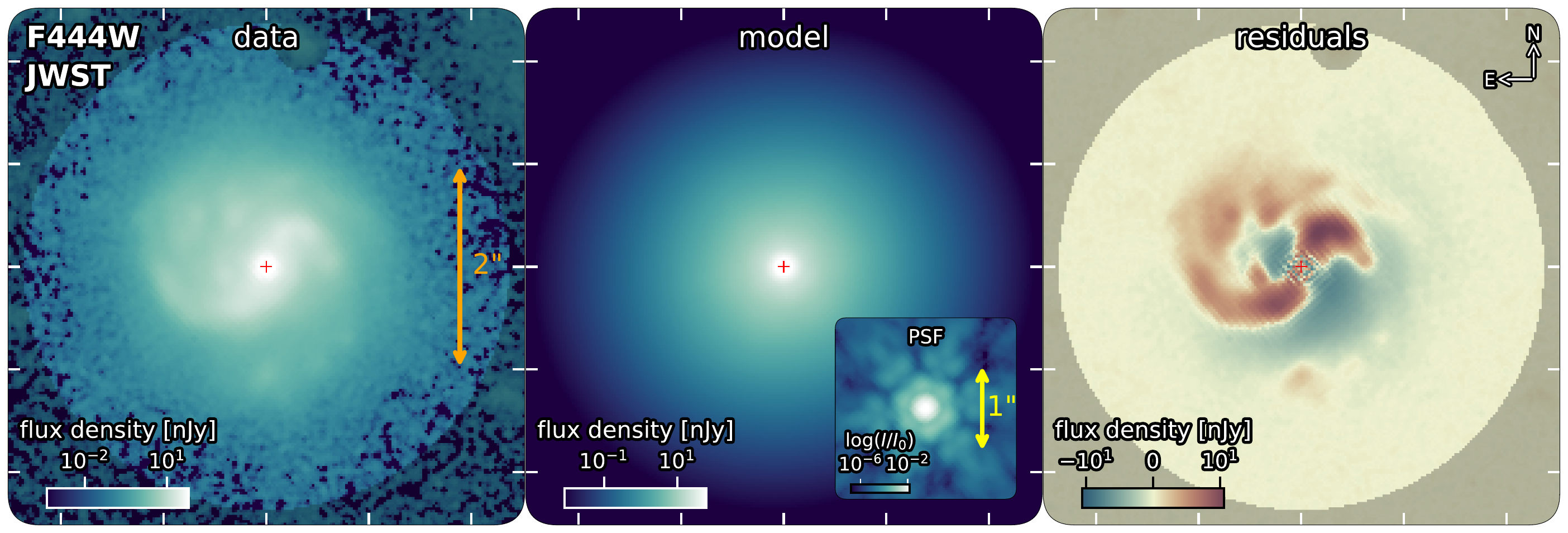}
    \caption{Overview of the morphological forward modeling: \textit{JWST} NIRCam F444W image (left), \texttt{IMFIT} bestfit model (middle), F444W PSF model (middle inset), and residuals after the subtraction of the model (right). The regions excluded from the fit are plotted with dimmer colors. In the residuals, we identify some residual flux at the northeastern part of the galaxy, with negative values at the southwestern one, as well as positive residuals along the bar and the prominent spiral arms emanating from its ends. These features respectively highlight the lopsidedness and non-axisymmetric structure of G4\_38232.}
    \label{fig:IMFITFitting}
\end{figure}

Next, we produced the maps of the uncertainties of the flux in each pixel of the F444W image using the ERR extension of our reduced products, masked the pixels corresponding to a deprojected distance larger than $\approx20$\,kpc from the center of G4\_38232, and ran \texttt{IMFIT} with our model PSF, fitting the two S\'ersic profiles for the disk and bulge. We used two fitting procedures, i) a $\chi^2$ minimization using the Differential Evolution (DE) algorithm with the Latin Hypercube sampling method, and $5000$ bootstrap resampling iterations with random masking of pixels of the F444W image for the estimation of the uncertainties, and ii) a Markov Chain Monte Carlo (MCMC) fit using $120$ MCMC chains, with $5000$ burn-in steps followed by an additional $\sim10000$ until convergence was achieved. Our DE (MCMC) fits converged to very similar values: a bulge S\'ersic index of $n_{\mathrm{s,bulge}}\approx0.87\pm0.27$~($0.92$) and an effective radius of $R_{\mathrm{e,bulge}}\approx0.42\pm0.02$~($0.41$)\,kpc, and in both cases a bulge-to-total ratio of $B/T\approx0.17$, in agreement within the uncertainties with the ratio derived in Sect.~\ref{sec:curveOfGrowthFitting}. In Fig.~\ref{fig:IMFITFitting} we present the F444W image used in the fit, alongside the \texttt{IMFIT} model and residuals.\par

It is worth noting that the central mass concentration of G4\_38232 appears to have a low S\'ersic index, hinting towards the possible presence of a rotationally-supported nuclear disk instead of a dispersion-dominated classical bulge \citep[e.g.,][]{Gadotti_2026, Le_Conte_2026}. This interpretation is also supported by the central peak in the SFR surface density map (see Appendix~\ref{sec:SEDFitting}), which shows that the central regions of G4\_38232 are actively star-forming. However, given the uncertainties in our morphological modeling as well as characterizing the central mass concentration based solely on its S\'ersic index, this result would be better described as an indication rather than a detection.\par

The size of the bulge is smaller than the resolution of the F444W image ($\mathrm{FWHM}_{\mathrm{PSF}}\approx$~0\farcs15~$\approx$\,1.2\,kpc) at the redshift of G4\_38232. Additionally, the non-axisymmetric features of the disk, namely, the bar structure and the asymmetry of the disk, are conspicuous in the residuals. These factors introduce more uncertainties in our analysis. A more sophisticated model including a bar component might be necessary for the accurate identification of the properties of the central mass concentration \citep[e.g.,][]{Gadotti_2026}. However, since the introduction of such a component, would result in a fitted model that is incompatible with our dynamical forward modeling, which includes only axisymmetric components, we chose to fit a simpler model to the photometric F444W data, considering the nonaxisymmetries as perturbations to be averaged out in the azimuthal sense.\par

\FloatBarrier

\section{Resolved SED fitting of \textit{HST} and \textit{JWST} imaging}
\label{sec:SEDFitting}

In our analysis, we used stellar mass and SFR maps derived from resolved SED fitting of the high-resolution imaging available for G4\_38232. Specifically, while the galaxy-integrated values were derived from the SED fitting of all available photometry (\citet[][Appendix~A]{Jolly_2026}, \citealt[][Sect.~2.2.2]{Chen_2026}), in this case, we used the following high-resolution \textit{HST} and \textit{JWST} images: F435W, F606W, and F814W \textit{HST} ACS filters, and F070W, F090W, F115W, F150W, F200W, F210M, F277W, F356W, F410M, F444W, and F470N \textit{JWST} NIRCam filters. A more detailed SED fitting analysis, also including \textit{JWST} MIRI filters and NOEMA dust continuum images, as obtained from our NOEMA\textsuperscript{3D} program, will be presented in a future paper (G. Tozzi et al., in prep.). Below, we summarize the main steps of our resolved SED modeling procedure, as well as the main assumptions we made.

As preliminary steps, we convolved all images to the (worst) spatial resolution of the F444W NIRCam band, and estimated the background noise by also accounting for spatial correlations between pixels, due to the data reduction, PSF matching, and instrumental features. For each cutout, we thus created a segmentation map and performed flux measurements in 100 ``empty" apertures, randomly placed, for various apertures with a variable linear size of $N$ pixels (up to $N=10$; see also \citealt{Tacchella_2015}). From the distribution of fluxes as a function of aperture size, we derived the rms within one pixel. Since the effective rms is expected to be spatially constant at fixed aperture size, we adopted it as a constant flux error for all pixels of a given cutout.

With these in hand, we modeled the SED of individual pixels using \texttt{CIGALE} \citep{Burgarella_2005, Noll_2009, Boquien_2019, Burgarella_2025}, only for those pixels with S/N\,>\,7 on the F444W band, based on our noise estimate. Such S/N threshold delivered a continuous mapping of the entire galaxy, and guaranteed a S/N\,$\gtrsim$\,2 in the bluer NIRCam filters, which indeed leads to good reduced $\chi^2$ values ($\chi^2_{{\rm red}}$\,$\approx$\,1) in the fit of individual pixels (see also \citealt{Parlanti_2025}). We then independently modeled the SED of each S/N(F444W)\,>\,7 pixel under the same set of assumptions. For all pixels, we fixed the redshift to $z=1.1159$, and assumed an exponentially declining star formation history (SFH) with an e-folding time $\tau=8000$\,Myr, approximately corresponding to a constant SFH. We adopted the stellar population synthesis models of \citep{Bruzual_2003}, with a \citep{Chabrier_2003b} initial mass function, dust attenuation law by \citep{Calzetti_2000}, and also included nebular emission. To limit the number of free parameters, we opted to fix the metallicity, trying with either solar metallicity ($Z=0.02$), or the first supersolar value allowed by \texttt{CIGALE} (i.e. $Z=0.05$). The \texttt{CIGALE} run with $Z=0.05$ overall produced better results across the entire galaxy, especially in correspondence of the clumps, where assuming a solar metallicity led to a clear (unreliable) drop in the stellar mass map compared to the surrounding bar and spiral regions.\par

\begin{figure}
    \centering
	\includegraphics[width=\columnwidth]{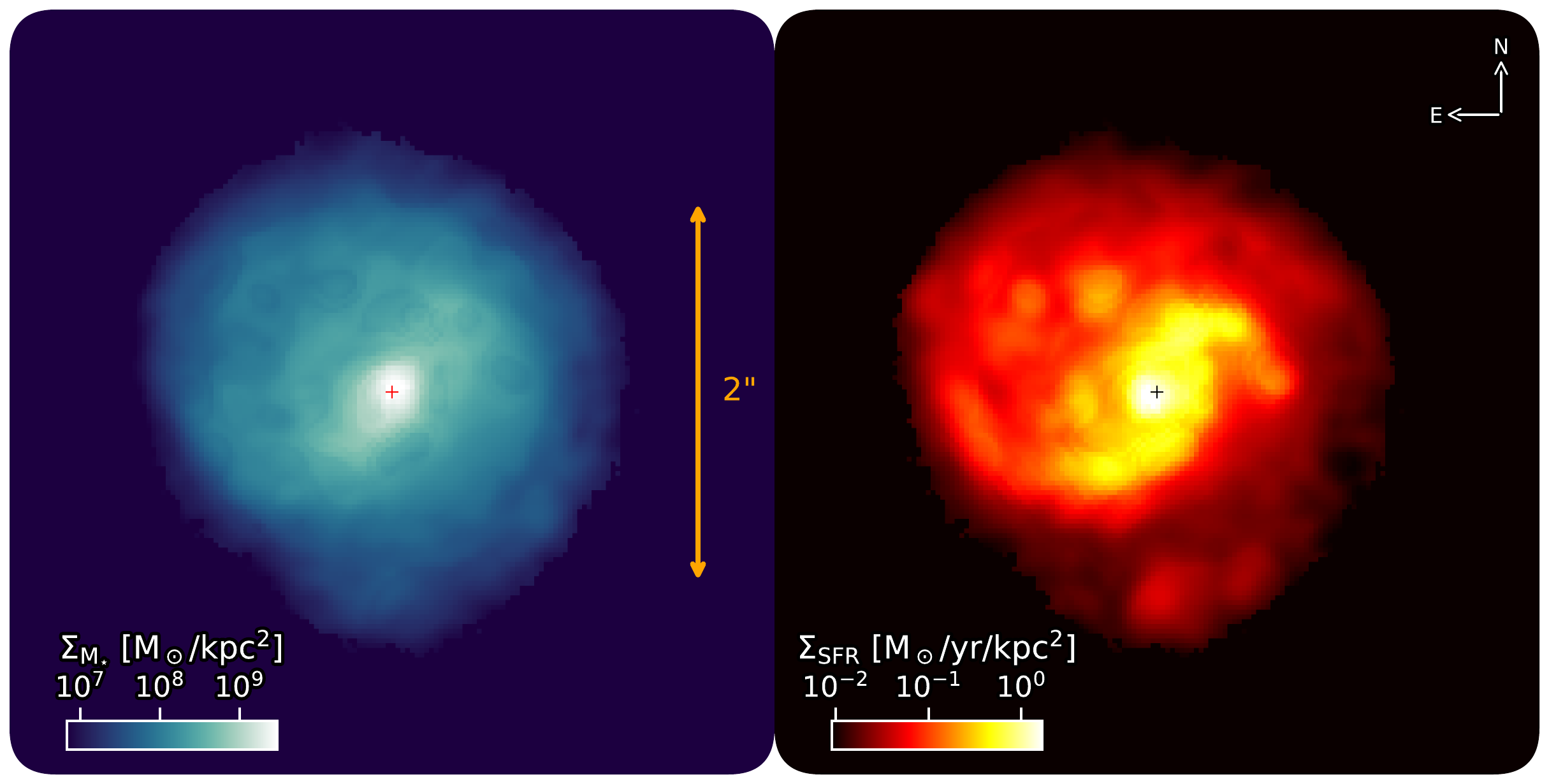}
    \caption{Surface density maps of the stellar mass (left) and SFR averaged over the last $10$\,Myr (right) used in our analysis. The stellar mass distribution of G4\_38232 appears mostly axisymmetric, with an increased stellar mass surface density at the central region and along the bar. The bulk of the star formation takes place in the bar region. We also identify a prominent peak in the SFR distribution near the center of the galaxy, with the identified gas flows possibly providing the fuel by replenishing the gas consumed in that process.}
    \label{fig:SEDFitting}
\end{figure}

In Fig.~\ref{fig:SEDFitting} we show the resulting maps of stellar mass and SFR averaged over the last $10$\,Myr. The stellar mass map was used in the curve of growth (Sect.~\ref{sec:curveOfGrowthFitting}) and surface density profile (Appendix~\ref{sec:surfaceDensityProfileFitting}) fitting analyses for deriving the properties of the disk of G4\_38232, while the SFR map was used for deriving the cumulative SFR profile offering a natural comparison baseline for our net radial flow rate estimates.\par

\FloatBarrier

\section{Stellar responses to the potential of G4\_38232}
\label{sec:stellarResponses}

\begin{sidewaysfigure*}
    \centering
    \includegraphics[width=\columnwidth]{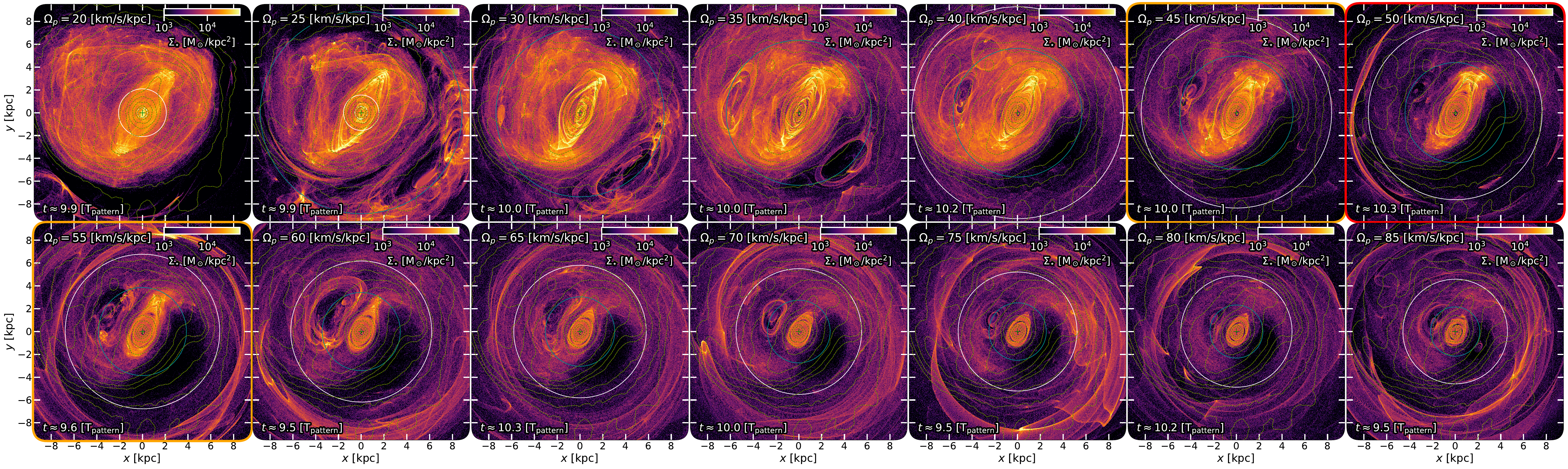}
    \caption{Stellar surface density maps of the stellar responses to the estimated potential in the midplane of G4\_38232 for pattern speeds $20\leq\Omega_{\mathrm{pattern}}\leq85$\,km/s/kpc, with overlaid contours of the NIRCam F444W image. White circles mark the radii of the inner and outer Lindblad resonance (if they exist), derived through frequency analysis on the axisymmetric potential, while the dark blue circle indicates the corotation radius. In the comparison between the observed and simulated morphologies, emphasis is put in the agreement in and around the bar region, in which the dynamics are expected to be mostly affected by the barred perturbation. The simulation with the best-matching pattern speed is marked with a red outline (top rightmost panel) and those within the estimated uncertainty marked with an orange one.}
    \label{fig:isothermalResponsesStars}
\end{sidewaysfigure*}

In Fig.~\ref{fig:isothermalResponsesStars} we present the stellar responses to the estimated potential for G4\_38232 for pattern speeds, $\Omega_{\mathrm{pattern}}$, ranging from $20$\,km/s/kpc to $85$\,km/s/kpc. The potential is introduced in the same way as in the isothermal responses of Sect.~\ref{sec:hydrodynamicFeaturesMatching}. In this case the NIRCam F444W continuum is used for the comparison between the observed and simulated structures, since this filter probes restframe near-IR wavelengths, capturing the emission of the bulk of the stellar component without being significantly affected by extinction. We find a good agreement between the observed and simulated structures for the best matching pattern speed determined through the comparison of Sect.~\ref{sec:hydrodynamicFeaturesMatching}.\par

\FloatBarrier

\section{Effective potential of G4\_38232}
\label{sec:effectivePotential}

In order to offer a possible explanation for the asymmetry in the gaseous and stellar responses to the potential derived for G4\_38232, we computed and studied the effective potential

\begin{align}
    &\Phi_{\mathrm{eff}}(r,\theta)=\Phi(r,\theta)-\frac{1}{2}\Omega_{\mathrm{pattern}}^2r^2,~\text{where}\label{eq:effectivePotential}\\
    &\Phi(r,\theta)=\Phi_0(r)+\sum_{m=1}^{6}\left[\Phi_{m,c}(r)\cos(m\theta)+\Phi_{m,s}(r)\sin(m\theta)\right]\label{eq:FourierPotential}
\end{align}

is the Fourier decomposed potential of the galaxy used in the response models and $\Omega_{\mathrm{pattern}}=50$\,km/s/kpc is the constrained bar pattern speed.\par

We first found the Lagrangian points of the effective potential. Since these are stationary points, their coordinates are solutions to the following set of equations

\begin{align}
    \vec{\nabla}\Phi_{\mathrm{eff}}=\begin{pmatrix}\frac{\partial\Phi_{\mathrm{eff}}}{\partial x}\\\frac{\partial\Phi_{\mathrm{eff}}}{\partial y}\end{pmatrix}=\vec{0}.\label{eq:LagrangianPoints}
\end{align}

We computed the partial derivatives of the effective potential $\Phi_{\mathrm{eff}}(r,\theta)$ in Cartesian coordinates, using the following conversions

\begin{align}
    \begin{split}
        \frac{\partial}{\partial x}\Phi_{\mathrm{eff}}=\left(\frac{\partial r}{\partial x}\frac{\partial}{\partial r}+\frac{\partial \phi}{\partial x}\frac{\partial}{\partial \phi}\right)\Phi_{\mathrm{eff}}=\left(\cos\phi\frac{\partial}{\partial r}-\sin\phi\frac{1}{r}\frac{\partial}{\partial\phi}\right)\Phi_{\mathrm{eff}}\\
        \frac{\partial}{\partial y}\Phi_{\mathrm{eff}}=\left(\frac{\partial r}{\partial y}\frac{\partial}{\partial r}+\frac{\partial \phi}{\partial y}\frac{\partial}{\partial \phi}\right)\Phi_{\mathrm{eff}}=\left(\sin\phi\frac{\partial}{\partial r}+\cos\phi\frac{1}{r}\frac{\partial}{\partial\phi}\right)\Phi_{\mathrm{eff}}
    \end{split}.\label{eq:cartesianDerivatives}
\end{align}

\begin{figure}
    \centering
	\includegraphics[width=\columnwidth]{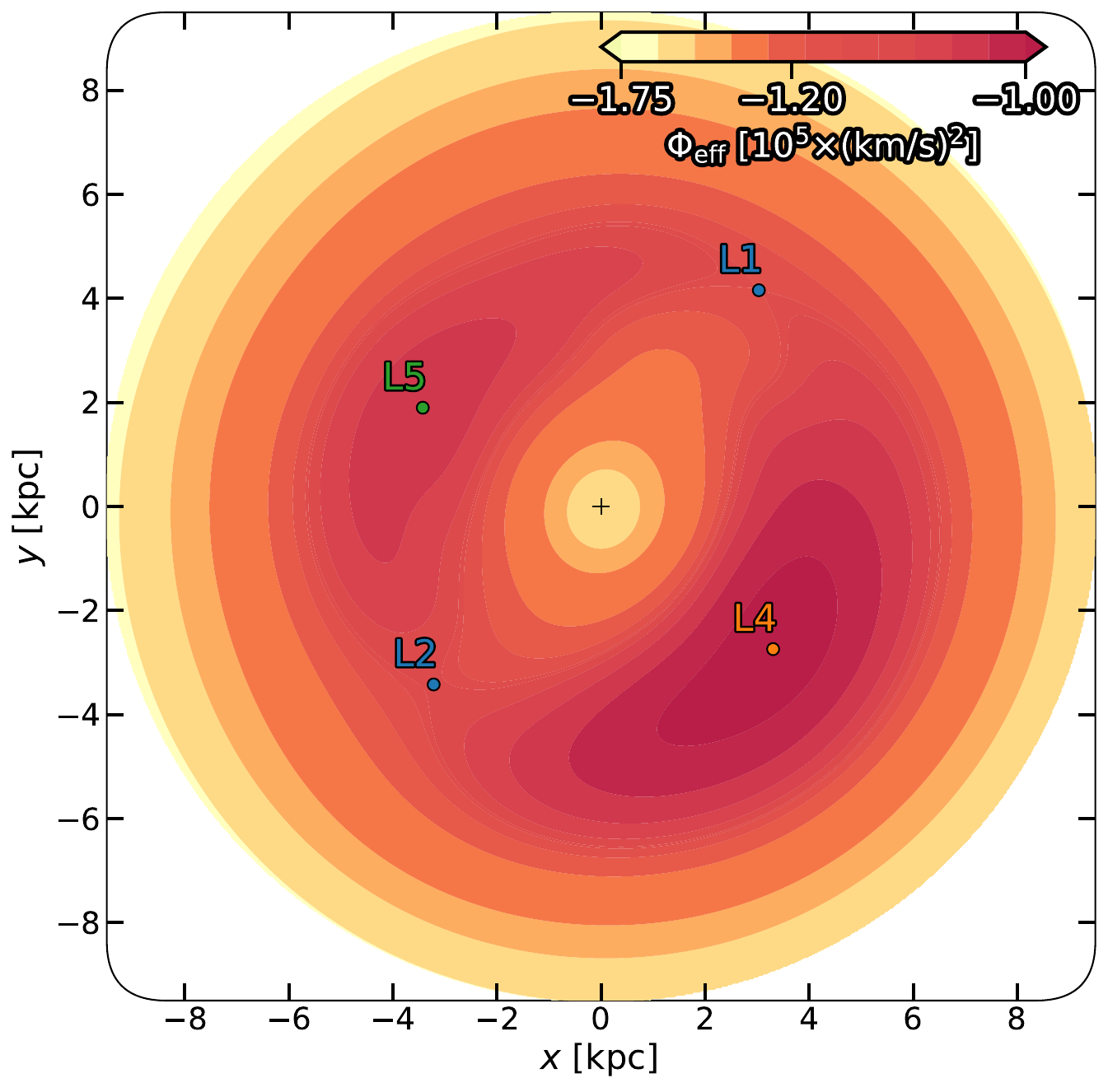}
    \caption{Contours of the effective potential of G4\_38232 for the constrained bar pattern speed of $\Omega_{\mathrm{pattern}}\approx50$\,km/s/kpc, with markers indicating the loci of the L1, L2, L4, and L5 Lagrangian points. The L1 and L2 points (blue) are unstable. The L4 (orange) point is also unstable while the L5 (green) point is stable. The difference in the stability of the L4 and L5 Lagrangian points offers an explanation for the lopsided mass distribution in the direction along the minor axis of the bar.}
    \label{fig:effectivePotential}
\end{figure}

We used as initial guesses the approximate locations of each point based on a visual inspection of the contour plot presented in Fig.~\ref{fig:effectivePotential}. In the figure, we present markers indicating the loci of the L1, L2, L4, and L5 Lagrangian points (L3, which is at the center of the galaxy, is not plotted) overlaid on the contours of the effective potential.\par

Next, we assessed the stability of the Lagrangian points by studying the linearized equations of motion of a particle in the region around each one of them. After a second order Taylor expansion of the effective potential around a Lagrangian point $L_i$, this system of equations can be written as (e.g., \citet{Romero-Gomez_2006}, \citet[][Appendix A]{Athanassoula_2009a}, see also \citealt[][Chapter~3.3.2]{Binney-Tremaine_2008})\par

\begin{align}
    \frac{d}{dt}\begin{pmatrix}\delta x\\\delta y\\\delta \dot{x}\\\delta \dot{y}\end{pmatrix}=
    \begin{bmatrix}
        0 & 0 & 1 & 0 \\
        0 & 0 & 0 & 1 \\
        -\frac{\partial^2\Phi_{\mathrm{eff}}}{\partial x^2}\big\vert_{L_i} & -\frac{\partial^2\Phi_{\mathrm{eff}}}{\partial x\partial y}\big\vert_{L_i} & 0 & 2\Omega_{\mathrm{pattern}} \\
        -\frac{\partial^2\Phi_{\mathrm{eff}}}{\partial x\partial y}\big\vert_{L_i} & -\frac{\partial^2\Phi_{\mathrm{eff}}}{\partial y^2}\big\vert_{L_i} & -2\Omega_{\mathrm{pattern}} & 0
    \end{bmatrix}\begin{pmatrix}\delta x\\\delta y\\\delta \dot{x}\\\delta \dot{y}\end{pmatrix}
\end{align}

The eigenvalues of this differential matrix determine the stability of the Lagrangian point. The imaginary part of a (generally complex) eigenvalue $\lambda$ corresponds to an oscillation. A positive (negative) real part of $\lambda$, causes the orbit of a particle in the vicinity of the point to exponentially diverge (converge). Also, if $\lambda$ is an eigenvalue of the matrix, so is $-\lambda$. Thus, in order for the Lagrangian point to be stable, all eigenvalues must be imaginary, otherwise the Lagrangian point is unstable.\par

In the effective potential of G4\_38232, the L1 and L2 Lagrangian points are saddle points and unstable, with two real (positive and negative) and two imaginary corresponding eigenvalues of the differential matrix, as is typical for barred potentials \citep[e.g.,][]{Romero-Gomez_2006, Athanassoula_2009a}. The L4 point is also unstable but with complex corresponding eigenvalues, while L5 is stable with only imaginary ones. The difference in the stability properties of the L4 and L5 Lagrangian points (as well as the associated banana-shaped periodic orbits that surround them; see, e.g., \citealt{Contopoulos_1989b}) provides a natural explanation for the asymmetry of the corresponding response models (see Figs.~\ref{fig:isothermalResponsesGas} and \ref{fig:isothermalResponsesStars}) and may also account for the lopsidedness of the galaxy itself. In a simplified picture, material tends to flow away from the vicinity of the unstable L4 region, whereas this is not the case around the stable L5 point \citep{Patsis_2017}, leading to an asymmetric mass distribution in the model under consideration. We note that, while standard analytic models typically feature two unstable L1 and L2 points and two stable L4 and L5 points, the picture in potentials derived from NIR observations of real galaxies can be considerably more complex \citep[see, e.g.,][]{Patsis_2010}.\par

Finally, it is also worth noting that the stability of the L4 and L5 Lagrangian points changes for different pattern speeds. However, in all response models with $30$\,km/s/kpc\,$\leq\Omega_{\mathrm{pattern}}\leq$\,$80$\,km/s/kpc, in which both L4 and L5 are present, L4 is unstable while L5 is either stable or less unstable than L4 (the real part of its corresponding eigenvalue is smaller than that of L4). This offers an intuitive explanation for the consistent lopsidedness of the response densities of these models.\par

\FloatBarrier

\section{Bar strength estimate}
\label{sec:barStrengthEstimate}

We estimated the strength of the bar of G4\_38232 using the $Q_b$ parameter of \citet{Buta_2001}. Starting with their definition of the strength of the non-axisymmetric perturbation $Q_T$ \citep[see also][]{Combes_1981} for the potential $\Phi(r,\theta)$ of Eq~\ref{eq:FourierPotential}, we found the local extrema of $Q_T$ near the ends of the bar by solving the following system of equations

\begin{align}
    \vec{\nabla}Q_T=\vec{\nabla}\left(\frac{\frac{1}{r}\frac{\partial \Phi}{\partial\theta}}{\frac{\partial \Phi_0}{\partial r}}\right)=\frac{1}{r\frac{\partial\Phi_0}{\partial r}}\begin{pmatrix}\frac{\partial^2\Phi}{\partial r\partial\theta}-\frac{\partial\Phi}{\partial\theta}\left(\frac{1}{r}+\frac{\partial^2\Phi_0}{\partial r^2}\big/\frac{\partial \Phi_0}{\partial r}\right)\\\frac{1}{r}\frac{\partial^2\Phi}{\partial\theta^2}\end{pmatrix}=\vec{0}.\label{eq:gradientOfQT}
\end{align}

\begin{figure}[h]
    \centering
	\includegraphics[width=\columnwidth]{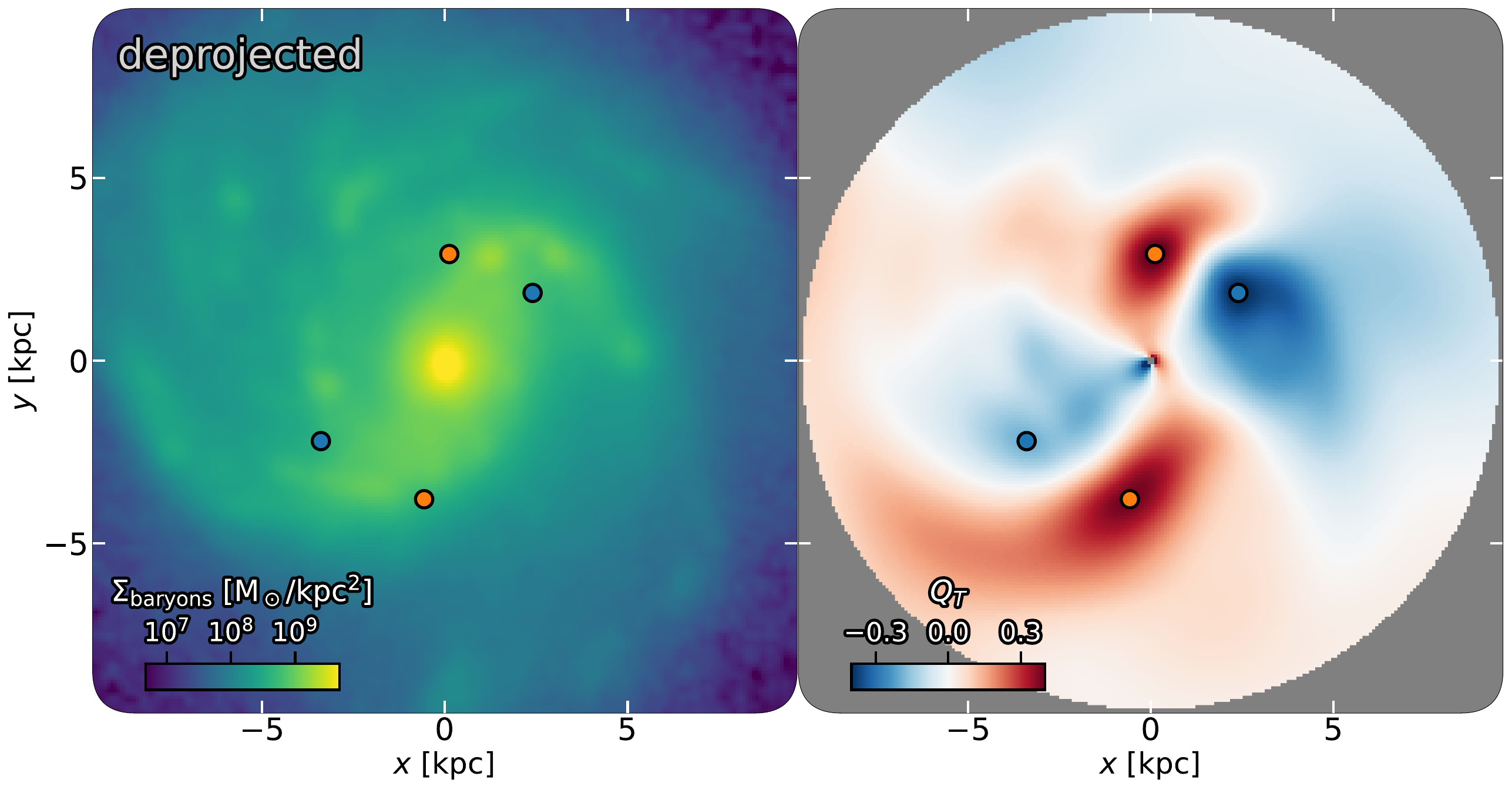}
    \caption{Deprojected baryonic surface density (left) and the ratio of the tangential force from the non-axisymmetric potential perturbation over the radial force due to the axisymmetric potential component, as a measure of the strength of the non-axisymmetric perturbation (right). We find four local maxima in this perturbation strength near the ends of the bar and use their average absolute value to estimate the strength of the bar following \citep{Buta_2001}.}
    \label{fig:barStrength}
\end{figure}

We derived the initial guesses for the coordinates of these points through a visual inspection of the non-axisymmetric potential perturbation strength presented in Fig.~\ref{fig:barStrength}. In the same figure, we have overlaid the identified extrema. The measure $Q_b$ of the strength of the bar of G4\_383232 is $Q_b\approx0.37$, computed as the average absolute value of the non-axisymmetric perturbation strength at these four points.\par

\FloatBarrier

\section{Application of observational methods to a simulated barred spiral}
\label{sec:applicationOfObervationalMethodsToASimulatedBarredSpiral}

In this section we apply an idealized version of the analyses carried out for G4\_38232 to a simulated gas-rich, barred spiral analog with similar properties. This simulation is similar to the one we presented in \citet{Pastras_2025b}, which was used as an analog to GN4\_32842, a $z\sim1.5$ massive, main sequence barred spiral also targeted by the NOEMA\textsuperscript{3D} survey. It was run using the moving mesh code \texttt{AREPO} \citep{Springel_2010, Weinberger_2020} and the \texttt{TNG} sub-grid model \citep{Weinberger_2017, Pillepich_2018a}, with similar ad-hoc modifications to those described in \citep{Pastras_2025b}: i) an adopted specific winds energy of $\overline{e_w}=7.2$, i.e., twice that of the fiducial \texttt{TNG} model, and ii) solar and one-third solar metallicities, for the gaseous disk and hot halo.\par

\begin{figure}[h]
    \centering
	\includegraphics[width=\columnwidth]{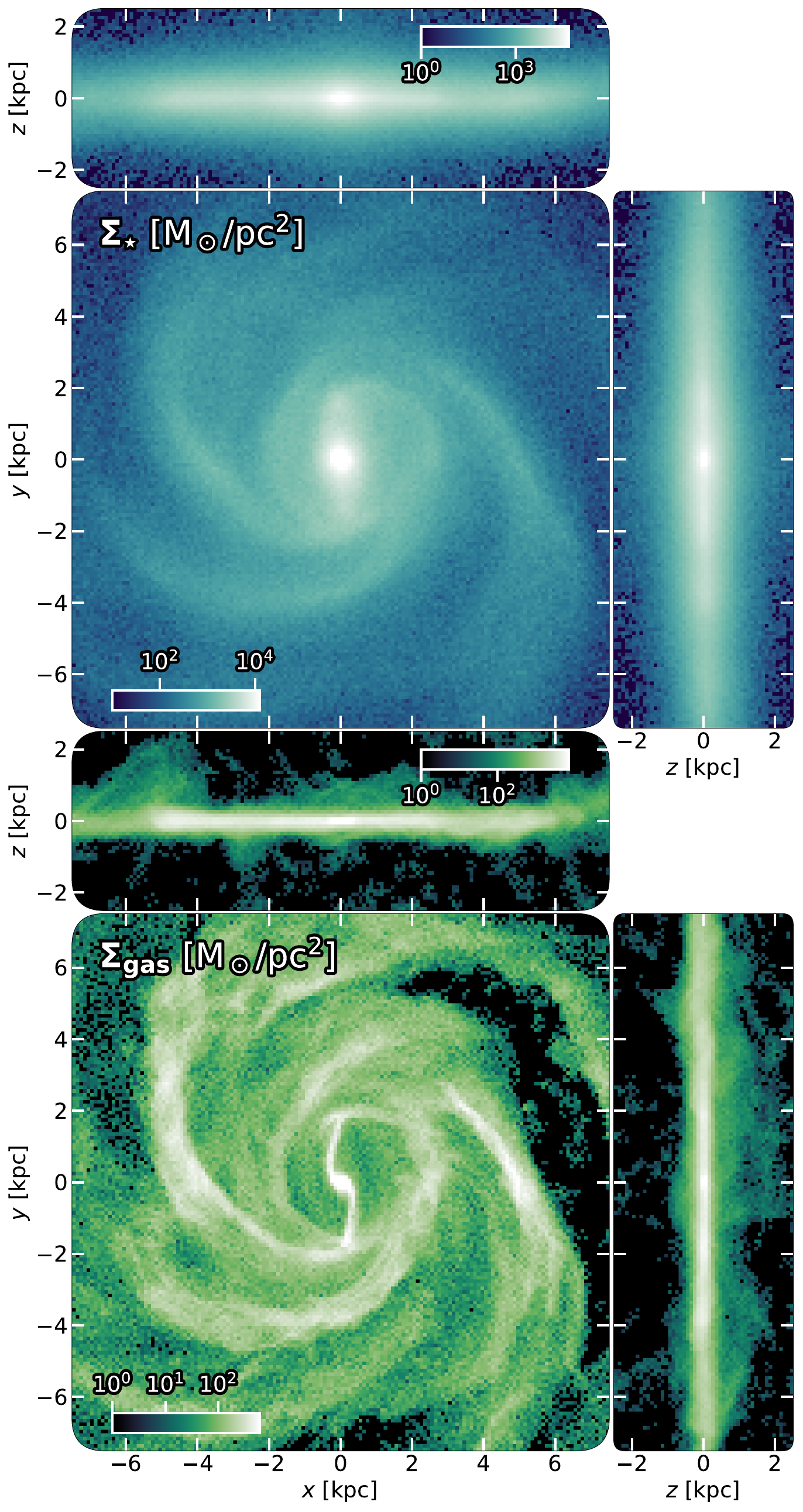}
    \caption{Overview of the stellar (top) and cold gas (bottom) components of the simulated barred spiral, with the face-on (bottom left), edge-on (top-left), and end-on (right) surface densities for each component. A central bar is identified in the stellar component as an elongated structure with the characteristic bar lanes (or dust lane shocks) clearly discernible at its leading side, with respect to the direction of the rotation. At larger radii, a multi-armed spiral pattern dominates the morphology of the disk.}
    \label{fig:simulationStellarAndGasSurfaceDensity}
\end{figure}

The initial conditions, produced using \texttt{MakeDiskGalaxy} \citep{Springel_1999, Springel_2005}, comprise a bulge, a disk, a dark matter halo and a gaseous hot halo. The properties of these components are identical to those of the \citet{Pastras_2025b} simulation, with the only difference being the spin of the hot gaseous halo, which in our case has a reduced angular momentum ratio of $\alpha\approx1.5$. Thus, our model started as a massive ($M_{\mathrm{disk}}=1.25\times10^{11}$\,M$_\odot$), extended ($R_{\mathrm{eff,disk,\star}}\approx5$\,kpc, $R_{\mathrm{eff,disk,gas}}\approx10$\,kpc), gas-rich ($f_{\mathrm{gas}}\approx20$\%) disk, with a modest ($M_{\mathrm{bulge}}=10^{10}$\,M$_\odot$) spherical bulge at its center, embedded in a $M_{\mathrm{DM}}=10^{12}$\,M$_\odot$ dark matter halo and a hot gaseous halo following a $\beta$-profile \citep{Moster_2011, Moster_2012}.\par

In the course of the simulation, the gas consumed by star formation is efficiently replenished through the cooling of the hot halo, resulting in a star-forming gas fraction of $f_{\mathrm{gas}}\approx10$\,\%, roughly constant for $\sim1$\,Gyr. The integrated SFR is also approximately constant throughout this period at a value of $\mathrm{SFR}\approx70$\,M$_\odot$/yr. A bar forms very early in this simulated galaxy, allowing the use of a snapshot at $t\approx440$\,Myr after the start of the simulation for the comparison with G4\_38232. An overview of the face-on stellar and gas surface densities of this showcase snapshot is presented in Fig.~\ref{fig:simulationStellarAndGasSurfaceDensity}.\par

A short ($a_{\mathrm{bar}}\approx2$\,kpc) bar is clearly discernible in the surface density map of the stellar component, with the bar lanes (or dust lane shocks) at the leading side of the bar, with respect to the direction of rotation, being the most prominent feature of the gaseous component, in agreement with expectations for the gas morphology in the presence of a bar \citep[e.g.,][]{Athanassoula_1992b}. In the outskirts of the disk, we identify a multi-armed pattern in both the stellar and gaseous components, as is the case for G4\_38232. Thus, in the following sections we studied this snapshot of a gas-rich barred spiral theoretically motivating our observational analyses and assessing, in an idealized fashion, their efficiency in recovering the known kinematical features of this model.\par

\FloatBarrier

\subsection{Non-axisymmetric structure; a gas-rich bar}
\label{sec:simulationnon-axisymmetricStructure}

We started by estimating the potential in the midplane of our simulated galaxy using a methodology inspired by our observational techniques. First, we split the components comprising our galaxy into spherical (bulge, dark matter halo, and hot gaseous halo) and disky (old and new stellar disk and gaseous disk). We estimated the potential of each spherical component by splitting it into $50$ spherical shells for $R\leq10$\,kpc and deriving their total potential as

\begin{equation}
    \Phi(r)=-G\left[\sum_{i~:~r_i\leq r}\left(\frac{M_i}{r}\right)+\sum_{i~:~r_i>r}\left(\frac{M_i}{r_i}\right)\right],
        \label{eq:potentialOfSphericallySymmetricAnnuli}
\end{equation}

where $G$ is the gravitational constant, $i$ the index of a spherical shell, $r_i$ the radius at its center, and $M_i$ its mass.

For each disky component, we first determined its scale-height, by splitting its particles with $-10$\,kpc\,$\leq x,y\leq$\,$10$\,kpc into $25$ horizontal slabs between $z=\pm5$\,kpc, computing the total mass of the component within each slab, effectively producing its vertical profile, normalizing it, and fitting it with a normalized hyperbolic secant squared function, i.e., $\rho(z)=\left[1/\left(4z_0\right)\right]\mathrm{sech}^2\left[z/(2z_0)\right]$. Next, we produced the vertically-integrated surface density map of the component with a bin size of $\approx0.2$\,kpc along the $x$ and $y$ axes, and used the methodology of \citet{Quillen_1994} (as presented in Sect.~\ref{sec:potentialEstimation}) to estimate its contribution to the potential at the midplane of the disk. Finally, we summed the contribution of all components, deriving the total potential at the disk midplane.\par

\begin{figure}[h]
    \centering
	\includegraphics[width=\columnwidth]{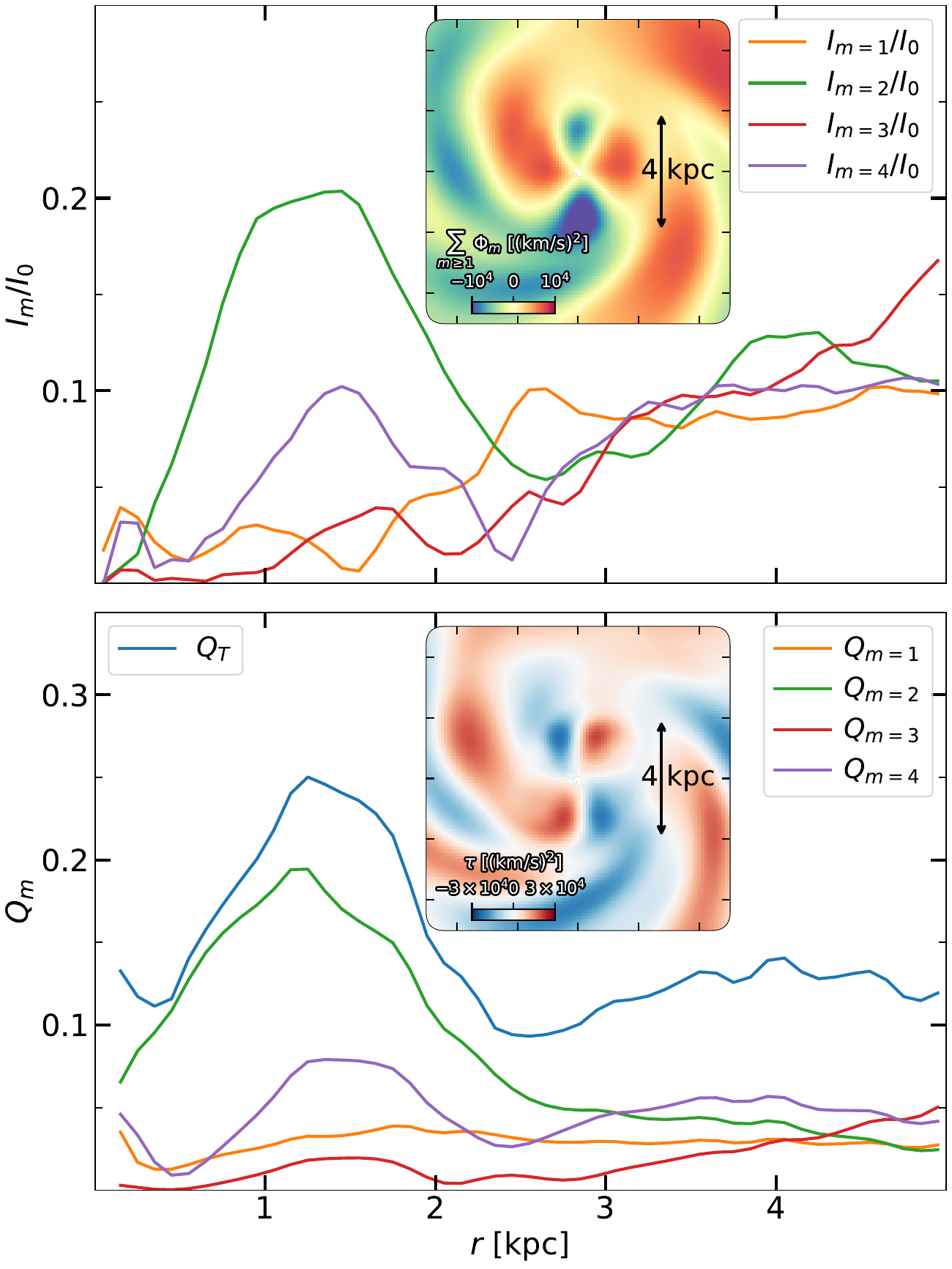}
    \caption{Normalized Fourier mode amplitudes of the deprojected stellar surface density of the simulated bar (top main) and strength of the non-axisymmetric potential perturbation (bottom main), along with a map of the non-axisymmetric perturbation of the potential (top inset) and the resulting torque map (bottom inset). A clear peak of the normalized $m=2$ mode amplitude is identified in the Fourier decomposition and the strength of the $m=2$ potential perturbation as expected in the presence of a bar. The alignment of the potential minimum, in the azimuthal direction, with the axis of the bar and the resulting quadrupole torque pattern, are also characteristic of this structure.}
    \label{fig:simulationnon-axisymmetricPerturbation}
\end{figure}

In a similar fashion as for G4\_38232 (Sect.~\ref{sec:potentialEstimation}), we assessed the strength of the bar, which is the most prominent non-axisymmetric structure, using the Fourier decomposition of the stellar surface density and the strength of the non-axisymmetric perturbation of the potential \citep{Combes_1981}. The results of this analysis are presented in Fig.~\ref{fig:simulationnon-axisymmetricPerturbation}.\par

In the figure, we identify the characteristic signatures of the presence of a bar: i) the normalized $m=2$ amplitude peak, ii) the peak in the ratio $Q_{\mathrm{m=2}}$ of the maximum tangential force due to the $m=2$ perturbation over the axisymmetric radial force \citep{Combes_1981}, iii) the alignment of the potential minimum, in the azimuthal sense, with the major axis of the bar, and iv) the resulting quadrupole pattern in the map of the gravitational torques. It is also worth mentioning that the maximum normalized $m=2$ amplitude, $I_\mathrm{m=2}/I_0$, is effectively consistent with the threshold value of $I_\mathrm{m=2}/I_0\approx0.2$, routinely used in the literature as a threshold for the identification of a bar \citep[e.g.,][]{Fujii_2018, Fragkoudi_2020, Fragkoudi_2021, Rosas-Guevara_2022, Rosas-Guevara_2024, Pastras_2025b}, indicating that this is a weak, young bar, which has not had enough time to grow stronger at the time of this showcase snapshot.\par

\begin{figure*}[h]
    \centering
	\includegraphics[width=2.0\columnwidth]{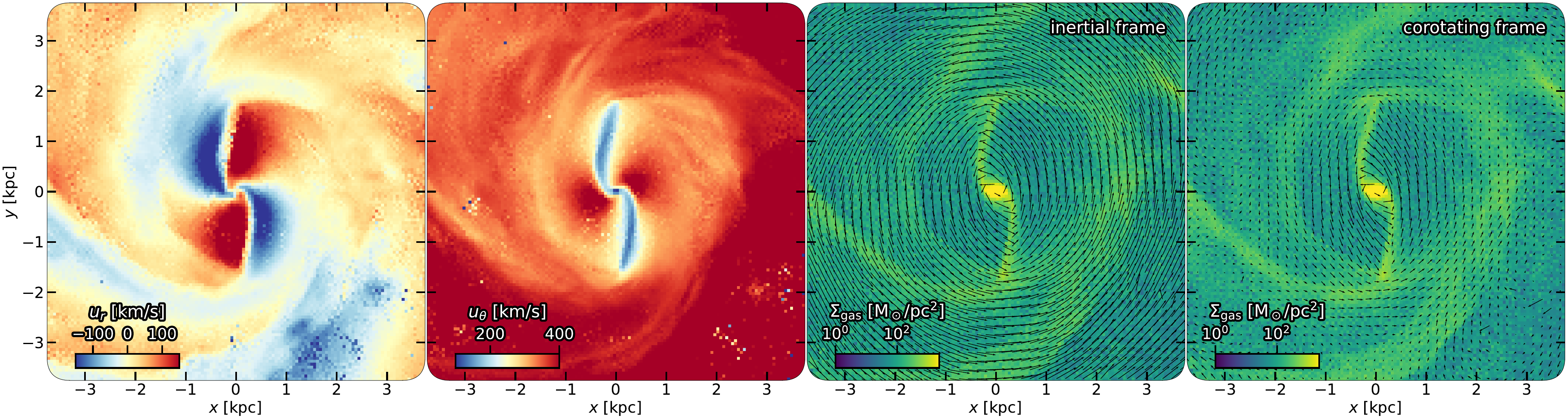}
    \caption{Overview of the gas flow velocities in the plane of the simulated galaxy: the radial (left) and tangential (middle left) components of the in-plane velocities, alongside the velocity field vectors in the inertial frame (middle right) and a frame corotating with the bar (right) overlaid on the gas surface density maps. In the radial velocity field, a quadrupole pattern is identified with negative (local inflows) and positive (local outflows) radial velocities identified at the leading and trailing side of the bar, respectively. In the azimuthal direction, the most prominent features are the strong dust lane shocks identified as regions of significantly reduced tangential velocities, indicating the presence of strong shocks. Both of these features can be inferred from the velocity vectors in the inertial frame, in which a quasi elliptical velocity streaming pattern is identified. However, since the bar also rotates in the disk plane, the most coherent picture of the flow patterns arises in the frame of reference corotating with the bar, in which gas streams into the dust lane shocks, loses angular momentum, and is subsequently funneled parallel to them toward the central region.}
    \label{fig:simulationInPlaneVelocity}
\end{figure*}

Next, we determined the bar pattern speed directly from the simulation through the Fourier method presented in \citep{Dehnen_2023} (but see also \citealt{Pfenniger_2023}), recovering a angular velocity of $\Omega_{\mathrm{pattern}}\approx110$\,km/s/kpc. A well-constrained pattern speed is critical in the interpretation of the gas flows in a barred spiral since the most coherent view of the flow patterns emerges in the reference frame that corotates with the bar (see Fig.~\ref{fig:simulationInPlaneVelocity}).\par

\FloatBarrier

\subsection{Gas flows in a simulated gas-rich barred spiral}
\label{sec:simulationGasFlows}

An overview of the kinematics of the star-forming gas in the showcase snapshot of the simulated barred spiral is presented in Fig.~\ref{fig:simulationInPlaneVelocity}. Splitting the in-plane velocities into their radial and tangential components aids the identification of the main features of the gas flows. We identify: i) negative (positive) radial velocities at the leading (trailing) side of the bar, corresponding to local inflows (outflows), ii) strong shocks along the bar lanes, with significantly reduced tangential velocities, and faster azimuthal streaming along the minor axis of the bar. This picture is compatible with gas streaming in an elliptical-like manner in the bar region, with the instantaneous velocity field in the inertial frame showing clearly a similar flow pattern.\par

However, since the bar rotates at its own pattern speed, the most coherent picture of the gas flow is identified in a frame of reference corotating with the bar. In that frame, the quasi elliptical streamlines become yet clearer, while the gas flows parallel to the bar lanes towards the central regions. When it reaches the nuclear region, some is accreted while the rest overshoots into the bar lane at the opposite side. This flow pattern, which has been identified in both simulations \citep[e.g.,][]{Athanassoula_1992b, Patsis_2000, Sormani_2019b, Hatchfield_2021} and observations \citep[e.g.,][]{Quillen_1995, Regan_1997, Sormani_2019a, Sormani_2023}, allows for the estimation of the gas flow rate through the study of the gas streaming parallel to the bar lanes, toward the central region.\par

\subsubsection{Radial motions}
\label{sec:simulationRadialMotions}

\begin{figure}[h]
    \centering
    \includegraphics[width=\columnwidth]{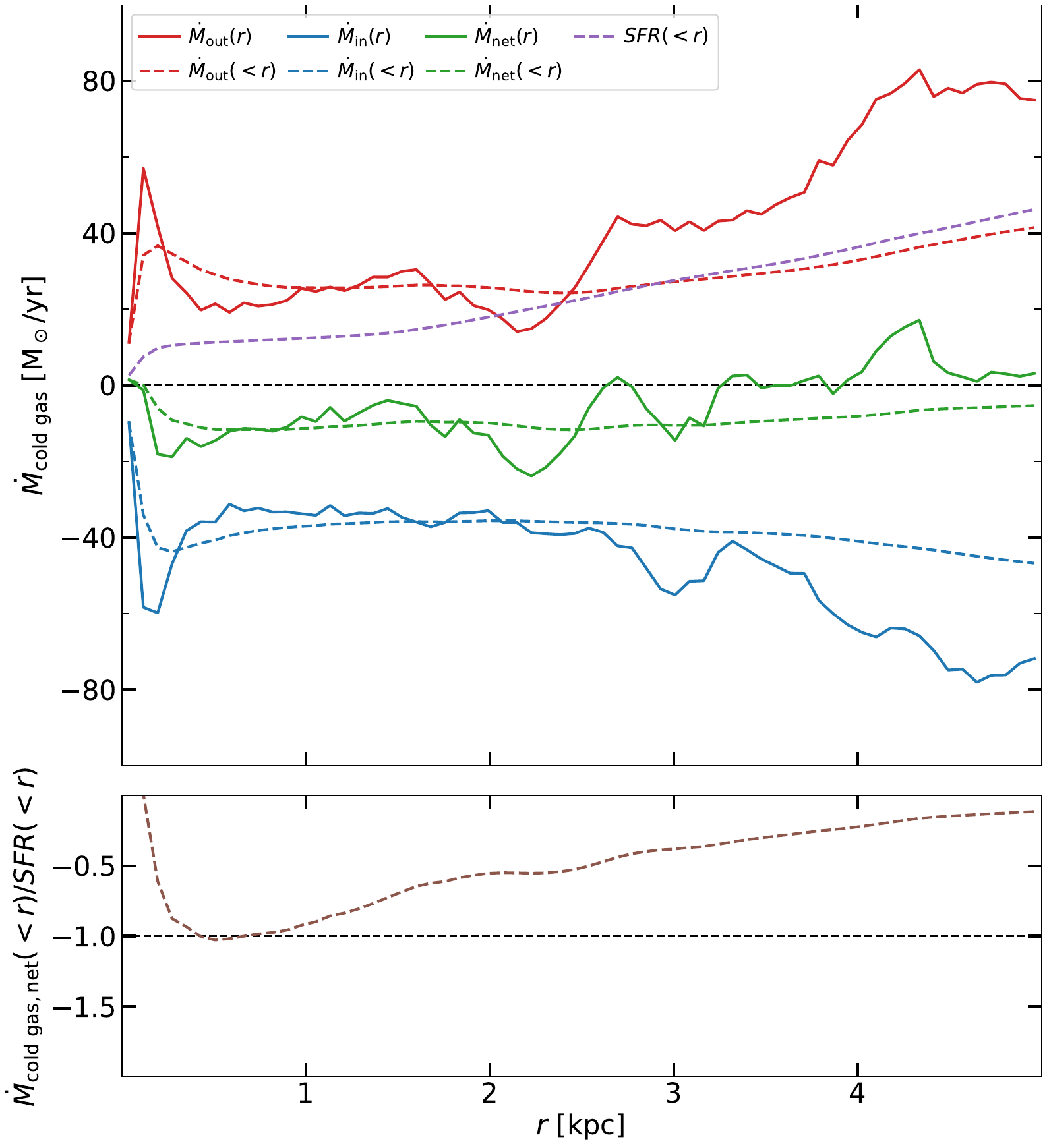}
    \caption{Radial gas flow rate and SFR profiles (top) and comparison between the net flow rate and the SFR as a function of radius (bottom) for the simulation snapshot. The solid lines show the outflow (red), inflow (blue) and net (green) flow rate as a function of radius. The dashed lines indicate the average radial flow rates and cumulative SFR (purple) from the center up to each radius. A comparison of the net gas flow rate and the integrated SFR highlights the similar order of the two, with an increased efficiency of radial gas transport in the bar region, i.e., $r\leq$\,2\,kpc.}
    \label{fig:simulationRadialFlowAndSFRates}
\end{figure}

We estimated the net gas flow rate in the plane of the simulated barred spiral by using the radial velocities of the gas in a similar way as in Appendix~C of \citet{Pastras_2025b}, under the assumption that the flow pattern is established and does not vary significantly in short timescales. The resulting gas flow rate profiles are presented in Fig.~\ref{fig:simulationRadialFlowAndSFRates}. We identify the presence of significant radial flows throughout the extent of the disk, with a consistent prevalence of negative radial motions, i.e., inflows. The estimated gas inflow rate resulting from this radial streaming up to each radius is of the order of the star formation rate, with an increased efficiency of the gas flows in the bar region, decreasing towards the outer regions where gas from the hot halo settles.\par

\FloatBarrier

\subsubsection{Torque modeling}
\label{sec:simulationTorqueModeling}
\begin{figure}[h]
    \centering
	\includegraphics[width=\columnwidth]{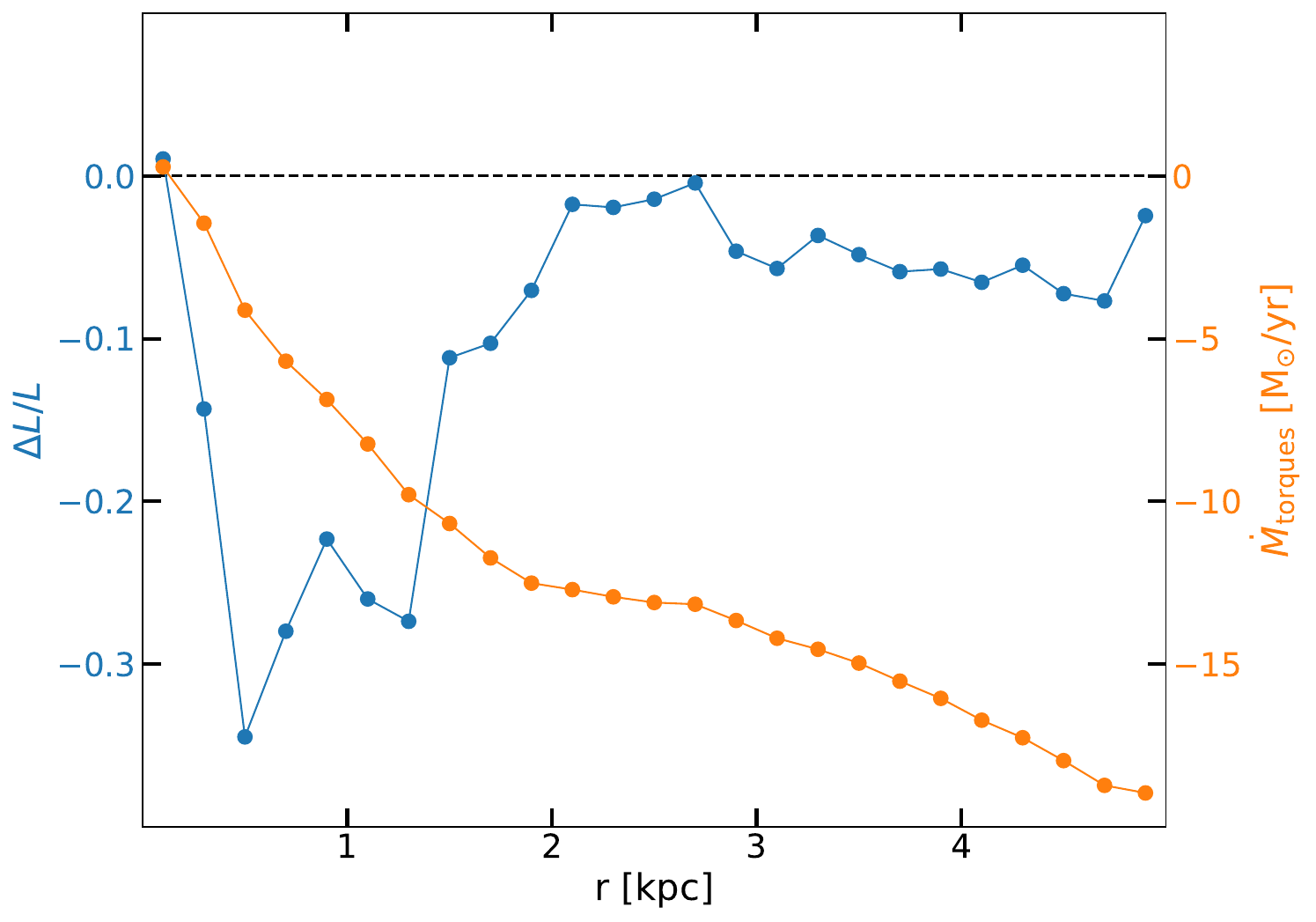}
    \caption{Angular momentum loss efficiency (blue) and resulting gas inflow rate (orange) profiles for the simulated barred spiral. We find that the angular momentum is efficiently lost in the region of the bar ($r\leq2$\,kpc), while at larger radii gas still loses angular momentum but with reduced efficiency. The corresponding gas flow rate is negative throughout the extent of the disk (excluding only the very central region), with values of a similar order as the integrated SFR up to each radius.}
    \label{fig:simulationTorqueModeling}
\end{figure}

We used the potential in the midplane of the simulated galaxy estimated in Sect.~\ref{sec:simulationnon-axisymmetricStructure} to model the torques exerted on the gaseous component, as we did for G4\_38232. The resulting angular momentum loss efficiency and gas flow rate profiles are presented in Fig.~\ref{fig:simulationTorqueModeling}. We find that the gas loses angular momentum efficiently in the bar region, resulting in an expected net inflow rate of $\dot{M}\approx20$\,M$_\odot$/yr. This inflow rate is larger but in agreement within the uncertainties with that derived through the study of the radial motions of Sect.~\ref{sec:simulationRadialMotions}, highlighting the role of the torques exerted by the non-axisymmetric structure, with the bar being the most prominent such feature, in driving the gas inflows in this simulated barred spiral. It is also worth noting that this estimate of the net gas flow rate is of the same order as the SFR integrated up to each radius.\par

\FloatBarrier

\subsubsection{Gas funneling parallel to the bar lanes}
\label{sec:simulationGasFunnelingParallelToTheBarLanes}

\begin{figure}
    \centering
	\includegraphics[width=\columnwidth]{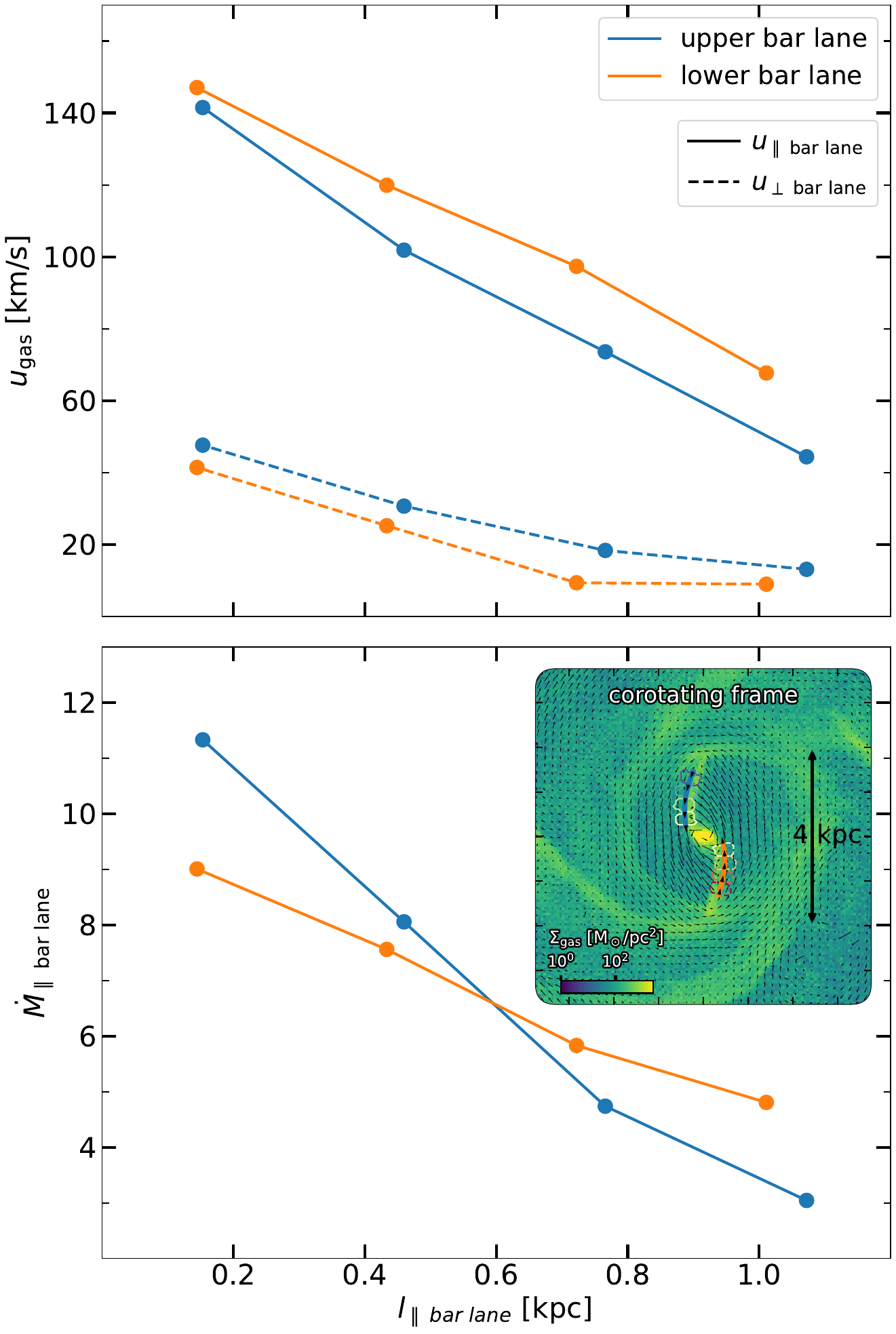}
    \caption{Gas streaming velocities parallel and perpendicular to the bar lanes in the corotating frame of reference (top) and resulting gas flow rate parallel to them (bottom) based on the extracted kinematics from apertures placed parallel to the bar lanes (bottom inset). The velocities are positive for streaming towards the center ($u_{\mathrm{\parallel}}>0$) and in the direction of the rotation ($u_{\mathrm{\perp}}>0$). We find that the streaming velocities perpendicular to the shocks are much smaller than those parallel to them, with the latter reaching values of the order of $\sim100$\,km/s. Assuming an accretion efficiency of $\approx30\%$, the resulting net inflow rate due to both bar lanes is estimated at $\sim5$\,M$_\odot$/yr.}
    \label{fig:simulationInflowParallelToDustLanes}
\end{figure}

Taking advantage of the coherent gas streaming pattern in the region of the bar lanes in the corotating frame, we place a set of apertures parallel to each dust lane shock and use the extracted gas kinematics to derive a gas flow rate estimate in a similar fashion as that presented in Sect.~\ref{sec:gasFunnelingParallelToTheBarLanes}. In Fig.~\ref{fig:simulationInflowParallelToDustLanes}, we present the velocity of the cold gas parallel and perpendicular to the shocks as well as the estimated gas flow rate parallel to it. We find that the streaming motions perpendicular to the shock are limited, with streaming parallel to the bar lanes being the most prominent flow pattern, with velocity amplitudes of the order of $\sim100$\,km/s. The resulting gas flow rate parallel to the bar lanes is also significant, resulting in an expected net inflow rate of $\sim5$\,M$_\odot$/yr, given the typically assumed accretion efficiency of $30$\% \citep{Regan_1997, Hatchfield_2021}.\par

\FloatBarrier

\subsubsection{Mock-observed velocity residuals}
\label{sec:simulationMockObservedVelocityResiduals}

\begin{figure*}
    \centering
	\includegraphics[width=2.0\columnwidth]{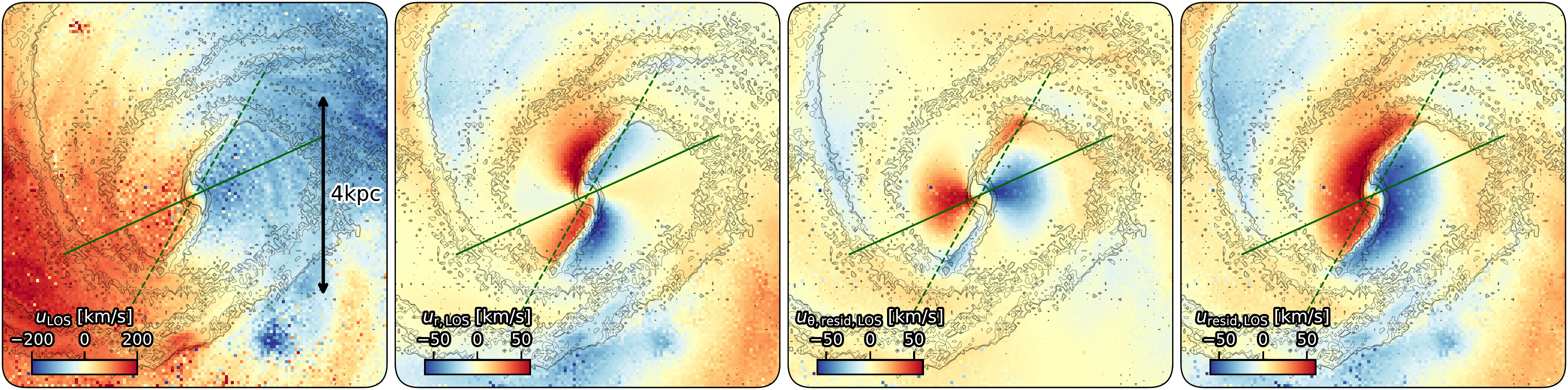}
    \caption{Overview of the expected contributions of noncircular motions to the LOS velocity residuals with overlaid contours of the projected cold gas surface density. From left to right: total LOS velocities (left), LOS contributions of radial motions (middle left), LOS contributions of tangential noncircular motions (middle right), and total LOS velocity contributions of noncircular motions (right). We find that the observed velocity residual pattern in the northern side of the bar of G4\_38232 is in overall agreement with expectations based on our simulation. However, the negative residuals identified in the southern part of the simulated bar are not observed in G4\_38232, possibly due to the presence of a weak or not well-defined dust lane shock. The strong positive residual pattern observed along the northeastern arm of G4\_38232 is also qualitatively reproduced but with a lower amplitude.}
    \label{fig:simulationIdealizedVelocityResiduals}
\end{figure*}

Following \citep{Pastras_2025b}, we oriented the simulated galaxy in a similar fashion as G4\_38232 and derived the expected LOS velocity residuals due to planar noncircular motions, with the results presented in Fig.~\ref{fig:simulationIdealizedVelocityResiduals}. We estimated the axisymmetric tangential velocity by placing a $10$\,kpc long and 0\farcs3 wide, when projected and at $z\sim1.1159$, slit along the major kinematic axis, sampled with $40$ bins, and computing the SFR-weighted average tangential velocity of the particles falling within the bounds of each pixel. The resulting rotational profile was used to estimate the contribution of the axisymmetric rotation to the tangential velocity of each particle.\par

In Fig.~\ref{fig:simulationIdealizedVelocityResiduals}, we present the LOS velocity component of the cold gas alongside the estimated LOS contributions of the noncircular motions. We find a significant contribution from radial flows in the region of the disk minor axis, with radial motions close to the bar lanes being the most prominent feature. In the tangential direction, the slower rotation along the bar lanes and the faster rotation along the minor axis of the bar are responsible for the most prominent features in the LOS residuals, with the contribution of the former being significantly dumped in the regions close to the minor axis, where azimuthal streaming does not contribute to the LOS velocities. Finally, a coherent pattern emerges as the sum of the previous two components in the expected residuals. This pattern can be described as an apparent inflow-outflow pattern along the bar and can be compared to the observed LOS velocity residuals in G4\_38232 (see Fig.~\ref{fig:residualVelocityMaps}).\par

This comparison shows that the positive residuals identified at the northern leading side of the bar of G4\_38232 could stem from either negative in-plane radial velocities (local inflows) or slower rotation (shocks) or both. At the central southern bar region, the identified positive residuals likely stem from positive radial velocities (local outflows). The expected negative residuals due to shocks and slower rotation at the southern leading side of the bar predicted by our model are not observed in G4\_38232, possibly due to the reduced strength of the southern bar lane (see also Sect.~\ref{sec:gasFunnelingParallelToTheBarLanes}). Finally, the prominent positive residuals observed along the northeastern arm of G4\_38323, interpreted as local inflows, are also identified in the case of our simulation along the spiral arm at the northeastern projected side of the bar, but in this case with a lower amplitude.\par

\FloatBarrier 
\clearpage

\end{appendix}

\end{document}